%% file: Template.tex
\documentclass[12pt, draftclsnofoot, onecolumn]{IEEEtran}

\ifCLASSINFOpdf

\fi

\usepackage[subrefformat=parens,labelformat=parens,caption=false,font=footnotesize]{subfig}
\usepackage{array}
\usepackage{amsmath}
\usepackage[printonlyused]{acronym}
\usepackage{graphicx,wrapfig,lipsum}
\usepackage{rotating}
\usepackage{amsmath, amssymb, amsthm}
\usepackage{stmaryrd}
\usepackage{tikz}
\usepackage{verbatim}
\usepackage{booktabs}
\usepackage{siunitx}
\usepackage{arydshln}
\usepackage{url}
\usepackage[utf8]{inputenc}
\usepackage[english]{babel}

\usepackage{multicol}
\usepackage{xr-hyper}

\usepackage{scalerel}
\usepackage{multicol}

\usepackage[noadjust]{cite}
\usepackage{bbm}
\usepackage[normalem]{ulem}
\usepackage{color}
\usepackage{dsfont}
\usepackage{bm}
\usepackage[short]{optidef}
\usepackage{diagbox}
\usepackage{enumitem}
\usepackage{slashbox}
\usepackage[ruled]{algorithm2e}
\setlist{nosep}

\usepackage{mathtools}

\DeclarePairedDelimiter{\flcl}{\lfloor}{\rceil}

\begin{document}
\title{A Survey on 5G Radio Access Network Energy Efficiency: Massive MIMO, Lean Carrier Design, Sleep Modes, and Machine Learning}

\author{David L\'opez-P\'erez, Antonio De Domenico, Nicola Piovesan, Geng Xinli, Harvey Bao, Song Qitao and M\'erouane Debbah}

\maketitle

\begin{abstract}

Cellular networks have changed the world we are living in,
and the fifth generation (5G) of radio technology is expected to further revolutionise our everyday lives,
by enabling a high degree of automation, 
through its larger capacity, massive connectivity, and ultra-reliable low-latency communications. 
In addition,
the third generation partnership project (3GPP) new radio (NR) specification also provides tools to significantly decrease the energy consumption and the green house emissions of next generations networks,
thus contributing towards information and communication technology (ICT) sustainability targets. 
In this survey paper,
we thoroughly review the state-of-the-art on current energy efficiency research. 
We first categorise and carefully analyse the different power consumption models and energy efficiency metrics,
which have helped to make progress on the understanding of green networks.
Then,
as a main contribution, 
we survey in detail
---from a theoretical and a practical viewpoint--- 
the main energy efficiency enabling technologies that 3GPP NR provides,
together with their main benefits and challenges.
Special attention is paid to four key enabling technologies,
i.e., massive multiple-input multiple-output (MIMO), lean carrier design, and advanced idle modes,
together with the role of artificial intelligence capabilities.
We dive into their implementation and operational details,
and thoroughly discuss their optimal operation points and theoretical-trade-offs from an energy consumption perspective.
This will help the reader to grasp the fundamentals of 
---and the status on--- 
green networking.
Finally, the areas of research where more effort is needed to make future networks greener are also discussed. 
\end{abstract}


\section{Introduction}
\label{sec:intro}
\input{sections/Introduction}


\section{3GPP NR Energy Efficiency Related Enabling Technologies}
\label{sec:energyEfficiencyEnabler}
\input{sections/energyEfficiencyEnabler}


\section{Power Consumption Models}
\label{sec:PC_Models}
\input{sections/PC_models}


\section{Energy Efficiency Metrics}
\label{sec:energyEfficiencyMetrics}
\input{sections/Metrics}


\section{Theoretical Understanding of Energy Efficiency: Massive MIMO}
\label{sec:MMIMO}
\input{sections/mMIMO}


\section{Time-domain (symbol shutdown) energy saving-based solutions}
\label{sec:TimeD}
\input{sections/ON_OFF_symbol}


\section{Carrier-domain (carrier shutdown) energy saving-based solutions}
\label{sec:CarrierD}
\input{sections/ON_OFF_carrier}


\section{Antenna-domain (channel shutdown) energy saving-based solutions}
\label{sec:AntennaD}
\input{sections/ON_OFF_channel}


\section{Machine Learning and Data-driven Energy Efficiency Optimization}
\label{sec:ML}
\input{sections/ML}


\section{Open Research Directions}
\label{sec:OpenresearchDirections}
\input{sections/OpenResearchDirections}

\section{Conclusions}
\label{sec:conclusions}

In this paper, 
we have provided an overview of the state-of-the-art on the fundamental understanding and practical considerations of the energy efficiency challenge in \ac{5G} networks. 
We have surveyed in detail the available \ac{BS} power consumption models and metrics for the optimization of the energy efficiency. 
We have also reviewed the impact on energy efficiency of four \ac{3GPP} \ac{NR} key enabling technologies,
i.e., \ac{mMIMO}, the lean carrier design, \acp{ASM}, and \ac{ML},
presenting the findings in the literature with respect to their available bounds, trade-offs and/or practical achievable energy savings. 
Importantly, 
we have highlighted the need for adapting the network resources to meet the end-users' \ac{QoS} demands, 
while minimizing network power consumption,
and have surveyed the related research differentiating among different algorithm time scales and classes 
(i.e. micro-sleeps, carrier and channel shutdown).
We have also stressed the role that spatio-temporal predictions and online optimisation via \ac{ML} will play in the previous network power consumption minimization task,
and discussed state-of-the-art \ac{ML} related approaches in such energy efficiency field.
To conclude, 
we have also provided discussion around the lines of research that need further work to make \ac{5G} networks greener.

As a final note,
given the enabling effect of that telecommunications systems and the impact that they can have in meeting the requirements for a sustainable development,
we encourage the research community to continue making progress toward a sustainable communication system.  


\section*{List of Acronyms}
\begin{acronym}[AAAAAAAAA]  
\acro{2D}{two-dimensional}
 \acro{3D}{three-dimensional}
 \acro{3G}{third generation}
 \acro{3GPP}{third generation partnership project}
 \acro{4G}{fourth generation}
 \acro{5G}{fifth generation}
 \acro{5GC}{5G core network}
 \acro{AAA}{authentication, authorisation and accounting}
 \acro{ABS}{almost blank subframe}
 \acro{AC}{alternating current}
 \acro{ACIR}{adjacent channel interference rejection ratio}
 \acro{ACK}{acknowledgment}
 \acro{ACL}{allowed CSG list}
 \acro{ACLR}{adjacent channel leakage ratio}
 \acro{ACPR}{adjacent channel power ratio}
 \acro{ACS}{adjacent channel selectivity}
 \acro{ADC}{analog-to-digital converter}
 \acro{ADSL}{asymmetric digital subscriber line}
 \acro{AEE}{area energy efficiency}
 \acro{AF}{amplify-and-forward}
 \acro{AGCH}{access grant channel}
 \acro{AGG}{aggressor cell}
 \acro{AH}{authentication header}
 \acro{AI}{artificial intelligence}
 \acro{AKA}{authentication and key agreement}
 \acro{AMC}{adaptive modulation and coding}
 \acro{ANN}{artificial neural network}
 \acro{ANR}{automatic neighbour relation}
 \acro{AoA}{angle of arrival}
 \acro{AoD}{angle of departure}
 \acro{APC}{area power consumption}
 \acro{API}{application programming interface}
 \acro{APP}{a posteriori probability}
 \acro{AR}{augmented reality}
 \acro{ARIMA}{autoregressive integrated moving average}
 \acro{ARQ}{automatic repeat request}
 \acro{AS}{access stratum}
 \acro{ASE}{area spectral efficiency}
 \acro{ASM}{advanced sleep mode}
 \acro{ASN}{access service network}
 \acro{ATM}{asynchronous transfer mode}
 \acro{ATSC}{Advanced Television Systems Committee}
 \acro{AUC}{authentication centre}
 \acro{AWGN}{additive white gaussian noise}
 \acro{BB}{baseband}
 \acro{BBU}{baseband unit}
 \acro{BCCH}{broadcast control channel}
 \acro{BCH}{broadcast channel}
 \acro{BCJR}{Bahl-Cocke-Jelinek-Raviv} 
 \acro{BE}{best effort}
 \acro{BER}{bit error rate}
 \acro{BLER}{block error rate}
 \acro{BPSK}{binary phase-shift keying}
 \acro{BR}{bit rate}
 \acro{BS}{base station}
 \acro{BSC}{base station controller}
 \acro{BSIC}{base station identity code}
 \acro{BSP}{binary space partitioning}
 \acro{BSS}{blind source separation}
 \acro{BTS}{base transceiver station}
 \acro{BWP}{bandwidth part}
 \acro{CA}{carrier aggregation}
 \acro{CAC}{call admission control}
 \acro{CaCo}{carrier component}
 \acro{CAPEX}{capital expenditure}
 \acro{capex}{capital expenses}
 \acro{CAS}{cluster angular spread}
 \acro{CATV}{community antenna television}
 \acro{CAZAC}{constant amplitude zero auto-correlation}
 \acro{CC}{component carrier}
 \acro{CCCH}{common control channel}
 \acro{CCDF}{complementary cumulative distribution function}
 \acro{CCE}{control channel element}
 \acro{CCO}{coverage and capacity optimisation}
 \acro{CCPCH}{common control physical channel}
 \acro{CCRS}{coordinated and cooperative relay system}
 \acro{CCTrCH}{coded composite transport channel}
 \acro{CDF}{cumulative distribution function}
 \acro{CDMA}{code division multiple access}
 \acro{CDS}{cluster delay spread}
 \acro{CESM}{capacity effective SINR mapping}
 \acro{CO$_{2e}$}{carbon dioxide equivalent}
 \acro{CFI}{control format indicator}
 \acro{CFL}{Courant-Friedrichs-Lewy}
 \acro{CGI}{cell global identity}
 \acro{CID}{connection identifier}
 \acro{CIF}{carrier indicator field}
 \acro{CIO}{cell individual offset}
 \acro{CIR}{channel impulse response}
 \acro{CNN}{convolutional neural network}
 \acro{CMF}{cumulative mass function}
 \acro{C-MIMO}{cooperative MIMO}
 \acro{CN}{core network}
 \acro{COC}{cell outage compensation}
 \acro{COD}{cell outage detection}
 \acro{CoMP}{coordinated multi-point}
 \acro{ConvLSTM}{convolutional LSTM}
 \acro{CP}{cycle prefix}
 \acro{CPC}{cognitive pilot channel}
 \acro{CPCH}{common packet channel}
 \acro{CPE}{customer premises equipment}
 \acro{CPICH}{common pilot channel}
 \acro{CPRI}{common public radio interface}
 \acro{CPU}{central processing unit}
 \acro{CQI}{channel quality indicator}
 \acro{CR}{cognitive radio}
 \acro{CRAN}{centralized radio access network} 
 \acro{C-RAN}{cloud radio access network} 
 \acro{CRC}{cyclic redundancy check}
 \acro{CRE}{cell range expansion}
 \acro{C-RNTI}{cell radio network temporary identifier}
 \acro{CRP}{cell re-selection parameter}
 \acro{CRS}{cell-specific reference symbol}
 \acro{CRT}{cell re-selection threshold}
 \acro{CSCC}{common spectrum coordination channel}
 \acro{CSG ID}{closed subscriber group ID}
 \acro{CSG}{closed subscriber group}
 \acro{CSI}{channel state information}
 \acro{CSIR}{receiver-side channel state information}
 \acro{CSI-RS}{channel state information-reference signals}
 \acro{CSO}{cell selection offset}
 \acro{CTCH}{common traffic channel}
 \acro{CTS}{clear-to-send} 
 \acro{CU}{central unit}
 \acro{CV}{cross-validation}
 \acro{CWiND}{Centre for Wireless Network Design}
 \acro{D2D}{device to device}
 \acro{DAB}{digital audio broadcasting}
 \acro{DAC}{digital-to-analog converter}
 \acro{DAS}{distributed antenna system}
 \acro{dB}{decibel}
 \acro{dBi}{isotropic-decibel}
 \acro{DC}{direct current}
 \acro{DCCH}{dedicated control channel}
 \acro{DCF}{decode-and-forward}
 \acro{DCH}{dedicated channel}
 \acro{DC-HSPA}{dual-carrier high speed packet access}
 \acro{DCI}{downlink control information}
 \acro{DCM}{directional channel model}
 \acro{DCP}{dirty-paper coding}
 \acro{DCS}{digital communication system}
 \acro{DECT}{digital enhanced cordless telecommunication}
 \acro{DeNB}{donor eNodeB}
 \acro{DFP}{dynamic frequency planning}
 \acro{DFS}{dynamic frequency selection}
 \acro{DFT}{discrete Fourier transform}
 \acro{DFTS}{discrete Fourier transform spread}
 \acro{DHCP}{dynamic host control protocol}
 \acro{DL}{downlink}
 \acro{DMC}{dense multi-path components}
 \acro{DMF}{demodulate-and-forward}
 \acro{DMT}{diversity and multiplexing tradeoff}
  \acro{DNN}{deep neural network} 
 \acro{DoA}{direction-of-arrival}
 \acro{DoD}{direction-of-departure}
 \acro{DoS}{denial of service}
 \acro{DPCCH}{dedicated physical control channel}
 \acro{DPDCH}{dedicated physical data channel}
 \acro{D-QDCR}{distributed QoS-based dynamic channel reservation}
 \acro{DQL}{deep Q-learning}
  \acro{DRAN}{distributed radio access network}
 \acro{DRS}{discovery reference signal}
 \acro{DRL}{deep reinforcement learning}
 \acro{DRX}{discontinuous reception}
 \acro{DS}{down stream}
 \acro{DSA}{dynamic spectrum access}
 \acro{DSCH}{downlink shared channel}
 \acro{DSL}{digital subscriber line}
 \acro{DSLAM}{digital subscriber line access multiplexer}
 \acro{DSP}{digital signal processor}
 \acro{DT}{decision tree}
 \acro{DTCH}{dedicated traffic channel}
 \acro{DTX}{discontinuous transmission}
   \acro{DU}{distributed unit}
 \acro{DVB}{digital video broadcasting}
 \acro{DXF}{drawing interchange format}
 \acro{E2E}{end-to-end}
 \acro{EAGCH}{enhanced uplink absolute grant channel}
 \acro{EA}{evolutionary algorithm}
 \acro{EAP}{extensible authentication protocol}
 \acro{EC}{evolutionary computing}
 \acro{ECGI}{evolved cell global identifier}
 \acro{ECR}{energy consumption ratio}
 \acro{ECRM}{effective code rate map}
 \acro{EDCH}{enhanced dedicated channel}
 \acro{EE}{energy efficiency}
 \acro{EESM}{exponential effective SINR mapping}
 \acro{EF}{estimate-and-forward}
 \acro{EGC}{equal gain combining}
 \acro{EHICH}{EDCH HARQ indicator channel}
 \acro{eICIC}{enhanced intercell interference coordination}
 \acro{EIR}{equipment identity register}
 \acro{EIRP}{effective isotropic radiated power}
 \acro{ELF}{evolutionary learning of fuzzy rules}
 \acro{eMBB}{enhanced mobile broadband}
  \acro{EMR}{Electromagnetic-Radiation}
 \acro{EMS}{enhanced messaging service}
 \acro{eNB}{evolved NodeB}
 \acro{eNodeB}{evolved NodeB}
 \acro{EoA}{elevation of arrival}
 \acro{EoD}{elevation of departure}
 \acro{EPB}{equal path-loss boundary}
 \acro{EPC}{evolved packet core}
 \acro{EPDCCH}{enhanced physical downlink control channel}
 \acro{EPLMN}{equivalent PLMN}
 \acro{EPS}{evolved packet system}
 \acro{ERAB}{eUTRAN radio access bearer}
 \acro{ERGC}{enhanced uplink relative grant channel}
 \acro{ERTPS}{extended real time polling service}
 \acro{ESB}{equal downlink receive signal strength boundary}
 \acro{ESF}{even subframe}
 \acro{ESP}{encapsulating security payload}
 \acro{ETSI}{European telecommunications standards institute}
 \acro{E-UTRA}{evolved UTRA}
 \acro{EU}{European Union}
 \acro{EUTRAN}{evolved UTRAN}
 \acro{EVDO}{evolution-data optimised}
 \acro{FACCH}{fast associated control channel}
 \acro{FACH}{forward access channel}
 \acro{FAP}{femtocell access point}
 \acro{FARL}{fuzzy assisted reinforcement learning}
 \acro{FCC}{Federal Communications Commission}
 \acro{FCCH}{frequency-correlation channel}
 \acro{FCFS}{first-come first-served}
 \acro{FCH}{frame control header}
 \acro{FCI}{failure cell ID}
 \acro{FD}{frequency-domain}
 \acro{FDD}{frequency division duplexing}
 \acro{FDM}{frequency division multiplexing}
 \acro{FDMA}{frequency division multiple access}
 \acro{FDTD}{finite-difference time-domain}
 \acro{FE}{front-end}
 \acro{FeMBMS}{further evolved multimedia broadcast multicast service}
 \acro{FER}{frame error rate}
 \acro{FFR}{fractional frequency reuse}
 \acro{FFRS}{fractional frequency reuse scheme}
 \acro{FFT}{fast Fourier transform}
 \acro{FFU}{flexible frequency usage}
 \acro{FGW}{femtocell gateway}
 \acro{FIFO}{first-in first-out}
 \acro{FIS}{fuzzy inference system}
 \acro{FMC}{fixed mobile convergence}
 \acro{FPC}{fractional power control}
 \acro{FPGA}{field-programmable gate array}
 \acro{FRS}{frequency reuse scheme}
 \acro{FTP}{file transfer protocol}
 \acro{FTTx}{fiber to the x}
 \acro{FUSC}{full usage of subchannels}
 \acro{GA}{genetic algorithm}
 \acro{GAN} {generic access network}
 \acro{GANC}{generic access network controller}
 \acro{GBR}{guaranteed bitrate}
 \acro{GCI}{global cell identity}
 \acro{GERAN}{GSM edge radio access network}
 \acro{GGSN}{gateway GPRS support node}
 \acro{GHG}{greenhouse gas}
 \acro{GMSC}{gateway mobile switching centre}
 \acro{gNB}{next generation NodeB}
 \acro{GNN}{graph neural network}
 \acro{GNSS}{global navigation satellite system}
 \acro{GP}{genetic programming}
 \acro{GPON}{Gigabit passive optical network}
 \acro{GPP}{general purpose processor}
 \acro{GPRS}{general packet radio service}
 \acro{GPS}{global positioning system}
 \acro{GPU}{graphics processing unit}
 \acro{GRU}{gated recurrent unit}
 \acro{GSCM}{geometry-based stochastic channel models}
 \acro{GSM}{global system for mobile communication}
 \acro{GTD}{geometry theory of diffraction}
 \acro{GTP}{GPRS tunnel protocol}
 \acro{GTP-U}{GPRS tunnel protocol - user plane}
 \acro{HA}{hybrid access}
 \acro{HARQ}{hybrid automatic repeat request}
 \acro{HBS}{home base station}
 \acro{HCN}{heterogeneous cellular network}
 \acro{HCS}{hierarchical cell structure}
  \acro{HD}{high definition}
 \acro{HDFP}{horizontal dynamic frequency planning}
 \acro{HeNB}{home eNodeB}
 \acro{HeNodeB}{home eNodeB}
 \acro{HetNet}{heterogeneous network}
 \acro{HiFi}{high fidelity}
 \acro{HII}{high interference indicator}
 \acro{HLR}{home location register}
 \acro{HNB}{home NodeB}
 \acro{HNBAP}{home NodeB application protocol}
 \acro{HNBGW}{home NodeB gateway}
 \acro{HNodeB}{home NodeB}
 \acro{HO}{handover}
 \acro{HOF}{handover failure}
 \acro{HOM}{handover hysteresis margin}
 \acro{HPBW}{half power beam width}
 \acro{HPLMN}{home PLMN}
 \acro{HPPP}{homogeneous Poison point process}
 \acro{HRD}{horizontal reflection diffraction}
 \acro{HSB}{hot spot boundary}
 \acro{HSDPA}{high speed downlink packet access}
 \acro{HSDSCH}{high-speed DSCH}
 \acro{HSPA}{high speed packet access}
 \acro{HSS}{home subscriber server}
 \acro{HSUPA}{high speed uplink packet access}
 \acro{HUA}{home user agent}
 \acro{HUE}{home user equipment}
 \acro{HVAC}{heating, ventilating, and air conditioning}
 \acro{IC}{interference cancellation}
 \acro{ICI}{inter-carrier interference}
 \acro{ICIC}{intercell interference coordination}
 \acro{ICNIRP}{International Commission on Non-Ionising Radiation Protection}
 \acro{ICS}{IMS centralised service}
 \acro{ICT}{information and communication technology}
 \acro{ID}{identifier}
 \acro{IDFT}{inverse discrete Fourier transform}
 \acro{IE}{information element}
 \acro{IEEE}{institute of electrical and electronics engineers}
 \acro{IETF}{internet engineering task force}
 \acro{IFA}{inverted-F-antennas}
 \acro{IFFT}{inverse fast Fourier transform}
 \acro{i.i.d.}{independent and identical distributed}
 \acro{IIR}{infinite impulse response}
 \acro{IKE}{Internet key exchange}
 \acro{IKEv2}{Internet key exchange version 2}
 \acro{ILP}{integer linear programming}
 \acro{IMEI}{international mobile equipment identity}
 \acro{IMS}{IP multimedia subsystem}
 \acro{IMSI}{international mobile subscriber identity}
 \acro{IMT}{international mobile telecommunications}
 \acro{INH}{indoor hotspot}
 \acro{IOI}{interference overload indicator}
 \acro{IoT}{Internet of things}
 \acro{IP}{Internet protocol}
 \acro{IPSEC}{Internet protocol security}
 \acro{IR}{incremental redundancy}
 \acro{IRC}{interference rejection combining}
 \acro{ISD}{inter site distance}
 \acro{ISI}{inter symbol interference}
 \acro{ITU}{international telecommunication union}
 \acro{Iub}{UMTS interface between RNC and NodeB}
 \acro{IWF}{IMS interworking function}
 \acro{JFI}{Jain's fairness index}
 \acro{KPI}{key performance indicator}
 \acro{KNN}{$k$-nearest neighbours}
 \acro{L1}{layer one}
 \acro{L2}{layer two}
 \acro{L3}{layer three}
 \acro{LA}{location area}
 \acro{LAA}{licensed Assisted Access}
 \acro{LAC}{location area code}
 \acro{LAI}{location area identity}
 \acro{LAU}{location area update}
 \acro{LDA}{linear discriminant analysis} 
 \acro{LIDAR}{laser imaging detection and ranging}
 \acro{LLR}{log-likelihood ratio}
 \acro{LLS}{link-level simulation}
 \acro{LMDS}{local multipoint distribution service}
 \acro{LMMSE}{linear minimum mean-square-error}
 \acro{LoS}{line-of-sight}
 \acro{LPC}{logical PDCCH candidate}
 \acro{LPN}{low power node}
 \acro{LR}{likelihood ratio}
 \acro{LSAS}{large-scale antenna system}
 \acro{LSP}{large-scale parameter}
 \acro{LSTM}{long short term memory cell}
 \acro{LTE/SAE}{long term evolution/system architecture evolution}
 \acro{LTE}{long term evolution}
 \acro{LTE-A}{long term evolution advanced}
 \acro{LUT}{look up table}
 \acro{MAC}{medium access control}
 \acro{MaCe}{macro cell}
 \acro{MAE}{Mean Absolute Error}
 \acro{MAP}{media access protocol}
 \acro{MAXI}{maximum insertion}
 \acro{MAXR}{maximum removal}
 \acro{MBMS}{multicast broadcast multimedia service} 
 \acro{MBS}{macrocell base station}
 \acro{MBSFN}{multicast-broadcast single-frequency network}
 \acro{MC}{modulation and coding}
 \acro{MCB}{main circuit board}
 \acro{MCM}{multi-carrier modulation}
 \acro{MCP}{multi-cell processing}
 \acro{MCS}{modulation and coding scheme}
 \acro{MCSR}{multi-carrier soft reuse}
 \acro{MDAF}{management data analytics function}
 \acro{MDP}{markov decision process }
 \acro{MDT}{minimisation of drive tests}
 \acro{MEA}{multi-element antenna}
 \acro{MeNodeB}{Master eNodeB}
 \acro{MGW}{media gateway}
 \acro{MIB}{master information block}
 \acro{MIC}{mean instantaneous capacity}
 \acro{MIESM}{mutual information effective SINR mapping}
 \acro{MIMO}{multiple-input multiple-output}
 \acro{MINI}{minimum insertion}
 \acro{MINR}{minimum removal}
 \acro{MIP}{mixed integer program}
 \acro{MISO}{multiple-input single-output}
 \acro{ML}{machine learning}
 \acro{MLB}{mobility load balancing}
 \acro{MLB}{mobility load balancing}
 \acro{MM}{mobility management}
 \acro{MME}{mobility management entity}
 \acro{mMIMO}{massive multiple-input multiple-output}
 \acro{MMSE}{minimum mean square error}
 \acro{mMTC}{massive machine type communication}
 \acro{MNC}{mobile network code}
 \acro{MNO}{mobile network operator}
 \acro{MOS}{mean opinion score}
 \acro{MPC}{multi-path component}
 \acro{MR}{measurement report}
 \acro{MRC}{maximal ratio combining}
 \acro{MR-FDPF}{multi-resolution frequency-domain parflow}
 \acro{MRO}{mobility robustness optimisation}
 \acro{MRT}{maximum ratio transmission}
 \acro{MS}{mobile station}
 \acro{MSC}{mobile switching centre}
 \acro{MSE}{mean square error}
 \acro{MSISDN}{mobile subscriber integrated services digital network number}
 \acro{MUE}{macrocell user equipment}
 \acro{MU-MIMO}{multi-user MIMO}
 \acro{MVNO}{mobile virtual network operators}
 \acro{NACK}{negative acknowledgment}
 \acro{NAS}{non access stratum}
 \acro{NAV}{network allocation vector}
 \acro{NB}{Naive Bayes}   
 \acro{NCL}{neighbour cell list}
 \acro{NEE}{network energy efficiency}
  \acro{NF}{network function}
 \acro{NFV}{network functions virtualization}
 \acro{NG}{next generation}
 \acro{NGMN}{next generation mobile networks}
 \acro{NG-RAN}{next generation radio access network} 
 \acro{NIR}{non ionisation radiation}
 \acro{NLoS}{non-line-of-sight}
 \acro{NN}{nearest neighbour} 
 \acro{NR}{new radio}
 \acro{NRMSE}{normalised root mean square error}
 \acro{NRTPS}{non-real-time polling service}
 \acro{NSS}{network switching subsystem}
 \acro{NTP}{network time protocol}
 \acro{NWG}{network working group}
 \acro{NWDAF}{network data analytics function} 
 \acro{OA}{open access}
 \acro{OAM}{operation, administration and maintenance}
 \acro{OC}{optimum combining}
 \acro{OCXO}{oven controlled oscillator}
 \acro{ODA}{omdi-directional antenna} 
 \acro{ODU}{optical distribution unit}
 \acro{OFDM}{orthogonal frequency division multiplexing}
 \acro{OFDMA}{orthogonal frequency division multiple access}
 \acro{OFS}{orthogonally-filled subframe}
 \acro{OLT}{optical line termination}
 \acro{ONT}{optical network terminal}
 \acro{OPEX}{operational expenditure}
 \acro{OSF}{odd subframe}
 \acro{OSI}{open systems interconnection}
 \acro{OSS}{operation support subsystem}
 \acro{OTT}{over the top}
 \acro{P2MP}{point to multi-point}
 \acro{P2P}{point to point}
 \acro{PAPR}{peak-to-average power ratio}
 \acro{PA}{power amplifier}
 \acro{PBCH}{physical broadcast channel}
 \acro{PC}{power control}
 \acro{PCB}{printed circuit board}
 \acro{PCC}{primary carrier component}
 \acro{PCCH}{paging control channel}
 \acro{PCCPCH}{primary common control physical channel}
 \acro{PCell}{primary cell}
 \acro{PCFICH}{physical control format indicator channel}
 \acro{PCH}{paging channel}
 \acro{PCI}{physical layer cell identity}
 \acro{PCPICH}{primary common pilot channel}
 \acro{PCPPH}{physical common packet channel}
 \acro{PDCCH}{physical downlink control channel}
 \acro{PDCP}{packet data convergence protocol}
 \acro{PDF}{probability density function}
 \acro{PDSCH}{physical downlink shared channel}
 \acro{PDU}{packet data unit}
 \acro{PeNB}{pico eNodeB}
 \acro{PeNodeB}{pico eNodeB}
 \acro{PF}{proportional fair}
 \acro{PGW}{packet data network gateway}
 \acro{PGFL}{probability generating functional}
 \acro{PhD}{doctor of philosophy}
 \acro{PHICH}{physical HARQ indicator channel}
 \acro{PHY}{physical layer}
 \acro{PIC}{parallel interference cancellation}
 \acro{PKI}{public key infrastructure}
 \acro{PL}{path loss}
 \acro{PMI}{precoding-matrix indicator}
 \acro{PLMN ID}{public land mobile network identity}
 \acro{PLMN}{public land mobile network}
 \acro{PML}{perfectly matched layer}
 \acro{PMF}{probability mass function}
 \acro{PMP}{point to multi-point}
 \acro{PN}{pseudorandom noise}
 \acro{POI}{point of interest}
 \acro{PON}{passive optical network}
 \acro{POP}{point of presence}
 \acro{PP}{point process}
 \acro{PPP}{Poisson point process}
 \acro{PPT}{PCI planning tools}
 \acro{PRACH}{physical random access channel}
 \acro{PRB}{physical resource block}
 \acro{PSC}{primary scrambling code}
 \acro{PSD}{power spectral density}
 \acro{PSS}{primary synchronisation channel}
 \acro{PSTN}{public switched telephone network}
 \acro{PTP}{point to point}
 \acro{PUCCH}{Physical Uplink Control Channel}
 \acro{PUE}{picocell user equipment}
 \acro{PUSC}{partial usage of subchannels}
 \acro{PUSCH}{physical uplink shared channel}
 \acro{QAM}{quadrature amplitude modulation}
 \acro{QCI}{QoS class identifier}
 \acro{QoE}{quality of experience}
 \acro{QoS}{quality of service}
 \acro{QPSK}{quadrature phase-shift keying}
 \acro{RAB}{radio access bearer}
 \acro{RACH}{random access channel}
 \acro{RADIUS}{remote authentication dial-in user services}
 \acro{RAN}{radio access network}
 \acro{RANAP}{radio access network application part}
 \acro{RAT}{radio access technology}
 \acro{RAU}{remote antenna unit}
 \acro{RAXN}{relay-aided x network}
 \acro{RB}{resource block}
 \acro{RCI}{re-establish cell id}
 \acro{RE}{resource efficiency}
 \acro{REB}{range expansion bias}
 \acro{REG}{resource element group}
 \acro{RF}{radio frequency}
  \acro{RFID}{radio frequency identification}
 \acro{RFP}{radio frequency planning}
 \acro{RI}{rank indicator}
 \acro{RL}{reinforcement learning}
 \acro{RLC}{radio link control}
 \acro{RLF}{radio link failure}
 \acro{RLM}{radio link monitoring}
 \acro{RMA}{rural macrocell}
 \acro{RMS}{root mean square}
 \acro{RMSE}{root mean square error}
 \acro{RN}{relay node}
 \acro{RNC}{radio network controller}
 \acro{RNL}{radio network layer}
 \acro{RNN}{recurrent neural network}
 \acro{RNP}{radio network planning}
 \acro{RNS}{radio network subsystem}
 \acro{RNTI}{radio network temporary identifier}
 \acro{RNTP}{relative narrowband transmit power}
 \acro{RPLMN}{registered PLMN}
 \acro{RPSF}{reduced-power subframes}
 \acro{RR}{round robin}
 \acro{RRC}{radio resource control}
 \acro{RRH}{remote radio head}
 \acro{RRM}{radio resource management}
 \acro{RS}{reference signal}
 \acro{RSC}{recursive systematic convolutional}
 \acro{RS-CS}{resource-specific cell-selection}
 \acro{RSQ}{reference signal quality}
 \acro{RSRP}{reference signal received power}
 \acro{RSRQ}{reference signal received quality}
 \acro{RSS}{reference signal strength}
 \acro{RSSI}{receive signal strength indicator}
 \acro{RTP}{real time transport}
 \acro{RTPS}{real-time polling service}
 \acro{RTS}{request-to-send}
 \acro{RTT}{round trip time}
  \acro{RU}{remote unit}
  \acro{RV}{random variable}
 \acro{RX}{receive}
 \acro{S1-AP}{s1 application protocol}
 \acro{S1-MME}{s1 for the control plane}
 \acro{S1-U}{s1 for the user plane}
 \acro{SA}{simulated annealing}
 \acro{SACCH}{slow associated control channel}
 \acro{SAE}{system architecture evolution}
 \acro{SAEGW}{system architecture evolution gateway}
 \acro{SAIC}{single antenna interference cancellation}
 \acro{SAP}{service access point}
 \acro{SAR}{specific absorption rate}
 \acro{SARIMA}{seasonal autoregressive integrated moving average}
 \acro{SAS}{spectrum allocation server}
 \acro{SBS}{super base station}
 \acro{SCC}{standards coordinating committee}
 \acro{SCCPCH}{secondary common control physical channel}
 \acro{SCell}{secondary cell}
 \acro{SCFDMA}{single carrier FDMA}
 \acro{SCH}{synchronisation channel}
 \acro{SCM}{spatial channel model}
 \acro{SCN}{small cell network}
 \acro{SCOFDM}{single carrier orthogonal frequency division multiplexing}
 \acro{SCP}{single cell processing}
 \acro{SCTP}{stream control transmission protocol}
 \acro{SDCCH}{standalone dedicated control channel}
 \acro{SDMA}{space-division multiple-access}
  \acro{SDO}{standard development organization}
 \acro{SDR}{software defined radio}
 \acro{SDU}{service data unit}
 \acro{SE}{spectral efficiency}
 \acro{SeNodeB}{secondary eNodeB}
 \acro{SFBC}{space frequency block coding}
 \acro{SFID}{service flow ID}
 \acro{SG}{signalling gateway}
 \acro{SGSN}{serving GPRS support node}
 \acro{SGW}{serving gateway}
 \acro{SI}{system information}
 \acro{SIB}{system information block}
 \acro{SIB1}{systeminformationblocktype1}
 \acro{SIB4}{systeminformationblocktype4}
 \acro{SIC}{successive interference cancellation}
 \acro{SIGTRAN}{signalling transport}
 \acro{SIM}{subscriber identity module}
 \acro{SIMO}{single input multiple output}
 \acro{SINR}{signal to interference plus noise ratio}
 \acro{SIP}{session initiated protocol}
 \acro{SIR}{signal to interference ratio}
 \acro{SISO}{single input single output}
 \acro{SLAC}{stochastic local area channel}
 \acro{SLL}{secondary lobe level}
 \acro{SLNR}{signal to leakage interference and noise ratio}
 \acro{SLS}{system-level simulation}
  \acro{SMB}{small and medium-sized businesses}
 \acro{SmCe}{small cell}
 \acro{SMS}{short message service}
 \acro{SN}{serial number}
 \acro{SNMP}{simple network management protocol}
 \acro{SNR}{signal to noise ratio}
 \acro{SOCP}{second-order cone programming}
 \acro{SOHO}{small office/home office}
 \acro{SON}{self-organising network}
 \acro{son}{self-organising networks}
 \acro{SOT}{saving of transmissions}
 \acro{SPS}{spectrum policy server}
 \acro{SRS}{sounding reference signals}
 \acro{SS}{synchronization signal}
 \acro{SSB}{synchronization Signal/PBCH block}
 \acro{SIB1}{system information block 1} 
 \acro{SSL}{secure socket layer}
 \acro{SSMA}{spread spectrum multiple access}
 \acro{SSS}{secondary synchronisation channel}
 \acro{STA}{steepest ascent}
 \acro{STBC}{space-time block coding}
 \acro{SUI}{stanford university interim}
 \acro{SVR}{support vector regression}
 \acro{TA}{timing advance}
 \acro{TAC}{tracking area code}
 \acro{TAI}{tracking area identity}
 \acro{TAS}{transmit antenna selection}
 \acro{TAU}{tracking area update}
 \acro{TCH}{traffic channel}
 \acro{TCO}{total cost of ownership}
 \acro{TCP}{transmission control protocol}
 \acro{TCXO}{temperature controlled oscillator}
 \acro{TD}{temporal difference}
 \acro{TDD}{time division duplexing}
 \acro{TDM}{time division multiplexing}
 \acro{TDMA}{time division multiple access}
  \acro{TDoA}{time difference of arrival}
 \acro{TEID}{tunnel endpoint identifier}
 \acro{TLS}{transport layer security}
 \acro{TNL}{transport network layer}
  \acro{ToA}{time of arrival}
 \acro{TP}{throughput}
 \acro{TPC}{transmit power control}
 \acro{TPM}{trusted platform module}
 \acro{TR}{transition region}
 \acro{TS}{tabu search}
 \acro{TSG}{technical specification group}
 \acro{TTG}{transmit/receive transition gap}
 \acro{TTI}{transmission time interval}
 \acro{TTT}{time-to-trigger}
 \acro{TU}{typical urban}
 \acro{TV}{television}
 \acro{TWXN}{two-way exchange network}
 \acro{TX}{transmit}
 \acro{UARFCN}{UTRA absolute radio frequency channel number}
 \acro{UAV}{unmanned aerial vehicle}
 \acro{UCI}{uplink control information}
 \acro{UDP}{user datagram protocol}
 \acro{UDN}{ultra-dense network}
 \acro{UE}{user equipment}
 \acro{UGS}{unsolicited grant service}
 \acro{UICC}{universal integrated circuit card}
 \acro{UK}{united kingdom}
 \acro{UL}{uplink}
 \acro{UMA}{unlicensed mobile access}
 \acro{UMi}{urban micro}
 \acro{UMTS}{universal mobile telecommunication system}
 \acro{UN}{United Nations}
 \acro{URLLC}{ultra-reliable low-latency communication}
 \acro{US}{upstream}
 \acro{USIM}{universal subscriber identity module}
 \acro{UTD}{theory of diffraction}
 \acro{UTRA}{UMTS terrestrial radio access}
 \acro{UTRAN}{UMTS terrestrial radio access network}
 \acro{UWB}{ultra wide band}
 \acro{VD}{vertical diffraction}
 \acro{VDFP}{vertical dynamic frequency planning}
 \acro{VDSL}{very-high-bit-rate digital subscriber line}
 \acro{VeNB}{virtual eNB}
 \acro{VeNodeB}{virtual eNodeB}
 \acro{VIC}{victim cell}
 \acro{VLR}{visitor location register}
 \acro{VNF}{virtual network function}
 \acro{VoIP}{voice over IP}
 \acro{VoLTE}{voice over LTE}
 \acro{VPLMN}{visited PLMN}
 \acro{VR}{visibility region}
  \acro{VRAN}{virtualized radio access network}
 \acro{WCDMA}{wideband code division multiple access}
 \acro{WEP}{wired equivalent privacy}
 \acro{WG}{working group}
 \acro{WHO}{world health organisation}
 \acro{Wi-Fi}{Wi-Fi}
 \acro{WiMAX}{wireless interoperability for microwave access}
 \acro{WiSE}{wireless system engineering}
 \acro{WLAN}{wireless local area network}
 \acro{WMAN}{wireless metropolitan area network}
 \acro{WNC}{wireless network coding}
 \acro{WRAN}{wireless regional area network}
 \acro{WSEE}{weighted sum of the energy efficiencies}
 \acro{WPEE}{weighted product of the energy efficiencies}
 \acro{WMEE}{weighted minimum of the energy efficiencies}
 \acro{X2}{x2}
 \acro{X2-AP}{x2 application protocol}
 \acro{ZF}{zero forcing}
\end{acronym}

\bibliographystyle{IEEEtran}
\bibliography{reference.bib}

\end{document}

%% file: sections/Introduction.tex


The industrial revolution and the automation of labour greatly accelerated the natural pace of evolution, 
and since then, 
we have significantly transformed the world we are living in~\cite{Rifkin2011}.
After the Second World War, 
technology has developed faster than ever before. 
Automation is making our lives easier,
and until very recently,
it felt like nothing could stop us.
However, we are starting to see the consequences of our unsustainable progress now~\cite{Weart2008}.

From 1937 to 2019, 
the world population has grown from 2.3 to 7.7\,billions~\cite{Worldometer2020},
and our modern agricultural, manufacturing, transport and life styles have sharply increased energy consumption.
As a result,
the amount of carbon in the atmosphere raised from 280 to 409 parts per millions in such time period~\cite{NASA2020}, 
and the levels of \ac{GHG} emissions reached the historical record of 37.5\,gigatones of \ac{CO$_{2e}$} in 2018,
---a 1.5\,\% increase with respect to 2008~\cite{UN2019}. 

Importantly, 
the consequences of such high levels of \ac{GHG} emissions are already tangible today. 
The increase of \ac{GHG} emissions, 
combined with current trends on deforestation, 
have contributed to global warming
---the rise of the average Earth surface temperature~\cite{etde_22221318}.
Between 1906 and 2005,
the planet temperature rose 0.6 to 0.9 degrees Celsius due to global warming, 
and the rate of temperature increase has nearly doubled in the last 50 years~\cite{GISTEMP2020}~\cite{Lenssen2019}. 

If our societies do not significantly change the manner in which we consume energy,
the consequences may be catastrophic~\cite{UNTrade2019}. 
To put the above numbers in perspective, 
it should be noted that an increase in the Earth’s temperature of 1.5 to 2.0 degrees Celsius,
above pre-industrial temperatures, 
has been estimated to be \emph{the limit} 
---a threat to most natural ecosystems on Earth today,
and thus to our planet and everyday lives~\cite{Cazenave2006,Seager2007,Bonan2008,Clement2009}.

To address this challenge,
international policymakers are targeting a dramatic increase in energy efficiency, 
and a sharp shift from fossil fuels to renewable sources of energy,
such as solar, wind and water. 
This will entail a completely new approach to the generation and use of energy, 
which must be adopted by every government, industry, business, and individual. 

In the following subsections,
we first motivate the need for the 5G of mobile technology to enable a green environment,
reviewing the important role that 5G will play to revert the global unsustainable energy consumption trend.
Then,
in the rest of the manuscript, 
we survey the technical innovations that 5G \acp{RAN} bring in terms of energy efficiency with respect to previous technology generations
---from a theoretical and practical viewpoint--- 
to enable a greener cellular operation, 
and in turn,
support such environmental change.

\subsection{The Enabling Role of \ac{5G}}

\begin{figure}[t]
    \centering
    \includegraphics[width=14cm]{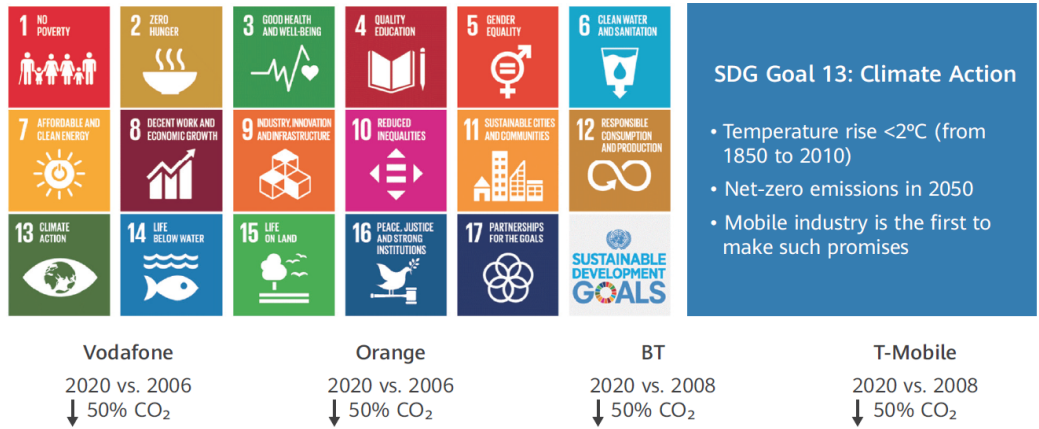}
    \caption{Wireless industry client action goals~\cite{GSMA2019_2}.}
    \label{fig:wirelessIndustryGoals}
\end{figure}

Governments and industries have set 
---or are setting--- 
ambitious targets to reduce their \ac{GHG} emissions and deal with global warming.
To date,
77 major economies have already established a net-zero \ac{GHG} emission target by 2050,
and their industries are accordingly (re)defining their energy efficiency and consumption road maps.
In this regard,
the telecommunication sector has taken the lead 
---and an set exemplary role---, 
setting stringent requirements for both the energy efficiency and consumption of their networks,
together with clear plans to meet them~\cite{GSMA2019_2}
(see Fig.~\ref{fig:wirelessIndustryGoals}).

Enhancing the energy efficiency and reducing the energy consumption of \ac{5G} networks will help reducing \ac{GHG} emissions
---no question about it. 
Their enabling effect, however, will be
---without a doubt--- 
the most important contribution of the mobile industry to address the current climate change~\cite{GSMA2020}.

\begin{figure}[t]
    \centering
    \includegraphics[width=9cm]{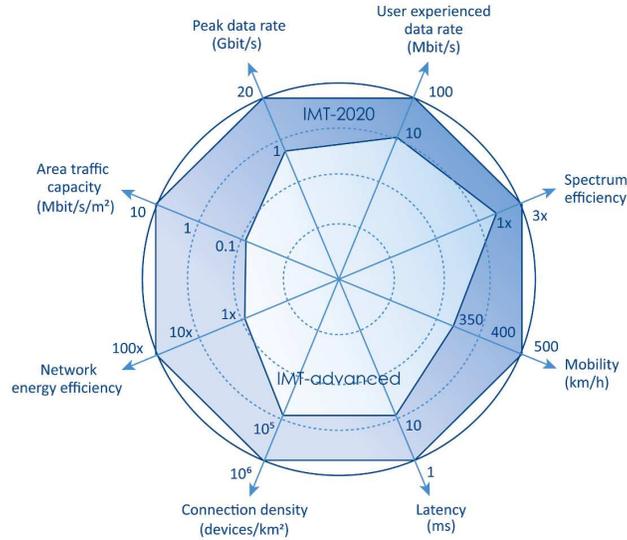}
    \caption{Comparison of key capabilities of \ac{IMT}-Advanced (\ac{4G}) with \ac{IMT}-2020 (\ac{5G}) according to~\cite{ITU-RM.2083}.}
    \label{fig:imtRequirements}
\end{figure}

At a macro-scale level,
the new \ac{5G} technology enables a new type of networking capability able to connect for the first time both everyone and everything together, including machines, objects, and devices,
thanks to its higher capacity, lower latency, improved reliability, and larger number of supported connections.
Please refer to Fig.~\ref{fig:imtRequirements} for a comparison between \ac{IMT}-Advanced (\ac{4G}) and \ac{IMT}-2020 (\ac{5G}) specification capabilities. 
Importantly, 
these new \ac{5G} communication capabilities are already helping governments, current industries, and new forms of businesses to implement novel processes with improved effectiveness,
by supporting a more flexible, tailored, and efficient use of resources.

As a result,
\ac{5G} has already become an integral part of governmental and industrial energy efficiency and consumption programs,
as it is envisioned that an intelligent exploitation of resources will enable a significant decrease of \ac{GHG} emissions through different avenues,
for example 
\emph{i)} an improved support for smart city and building energy management, 
\emph{ii)} reduced requirements for office space and business travel, and
\emph{iii)} efficient just-in-time supply chains, 
enabled by predictive analytics,
to cite a few~\cite{Huawei2020}.

To give some idea of the magnitude and importance of this \ac{5G} enabling effect,
it should be noted that,
according to the \ac{ITU} SMART 2020 report~\cite{ITU-T2008},
the enabling effect of mobile communications alone was estimated to be around 2,135 million tones of \ac{CO$_{2e}$} in 2018,
and that according to~\cite{ITU-T2012},
the scale of it will increase in the \ac{5G} era,
where the enabling effect across all the \ac{ICT} sector was predicted to be equivalent to 15\,\% of all global emissions by the end of 2020.

Recent reports indicate that industries, 
such as transportation, health care, and manufacturing, 
are already significantly benefiting from such \ac{5G} enabling effect. 
For example, 
through smart city programs and \ac{5G}-related innovations,
London, Berlin, and Madrid have already reduced \ac{GHG} emissions of motor vehicles by 30\,\% each from their peak rates, 
and Copenhagen by 61\,\%~\cite{C402019}.

To further assess the of breath and depth of the enabling effect of \ac{5G} networks,
interested readers are refereed to~\cite{GSMA2020},
and references there in. 

\subsection{The \ac{5G} Energy Efficiency Challenge}

Unfortunately,
the \ac{5G} enabling effect is no free lunch.
In fact, 
it comes at the expense of a tremendous challenge for the telecommunication sector in terms of both carried data and energy consumption.

Allowing governments, industries, businesses, and individuals in general to increase their energy efficiencies and reduce their energy consumption through more flexible, tailored, and efficient operations via a telecommunication network entails 
\begin{itemize}
    \item 
    a dramatic growth of data usage in some scenarios,
    and
    \item
    the need for more sophisticated networking to meet the required low-latency, high-reliability, and/or large volume of data connections in some others.
\end{itemize} 
Quantifying such challenge, 
recent studies already indicate that,
by 2030,
the number of connected devices is expected to grow to 100 billion~\cite{SMARTer2030},
and that \ac{5G} networks may be supporting up to 1,000$\,$times more data than \ac{4G} ones did in 2018~\cite{Huawei2020}.

Importantly,
\begin{itemize}
    \item
    to reach their established energy efficiency and consumption targets, 
    and 
    \item
    reduce their energy bills, 
    up to 40$\,\%$ by 2030~\cite{Huawei2020},
    to make their businesses profitable,
\end{itemize}
\acp{MNO} will need to meet the aforementioned more challenging traffic demands and requirements with significantly reduced \ac{GHG} emissions with respect to those of today's \ac{4G} networks.    

Considering the aforementioned two aspects, 
the \ac{3GPP} stakeholders have already called for a 90$\,\%$ reduction in energy consumption of \ac{3GPP} \ac{NR} compared to \ac{3GPP} \ac{LTE}~\cite{GSMA2019}.
However, whether these gains can be realised or not in practical networks will not only depend on what the new specification can do and/or the energy performance of a single site,
but also on how the actual network is deployed and operated as a whole.

In fact, 
to support the growing use of \ac{5G} connectivity and its more stringent requirements, 
while reducing energy consumption on a per-bit basis through an intelligent use of the network,
changes are needed at all levels of it to achieve the maximum holistic effect. 
\acp{MNO} must thus embrace new approaches to network planning, deployment, management, and optimisation that have energy efficiency at heart, 
and are implemented end-to-end.
Without energy efficiency driving future deployments, 
the study in~\cite{Abrol2016} indicates that a \ac{5G} network,
despite of its enhanced energy efficiency in bits per Joule due to its larger bandwidth and better spatial multiplexing capabilities, 
could typically consume over 140\,\% more energy than a 4G one,
with a similar coverage area. 
This unwanted energy consumption arises from \ac{5G}'s greater density of \acp{BS}, antennas, cloud infrastructure, and \ac{UE},
among others. 

\begin{figure}[t]
\centering
\includegraphics[width=9cm]{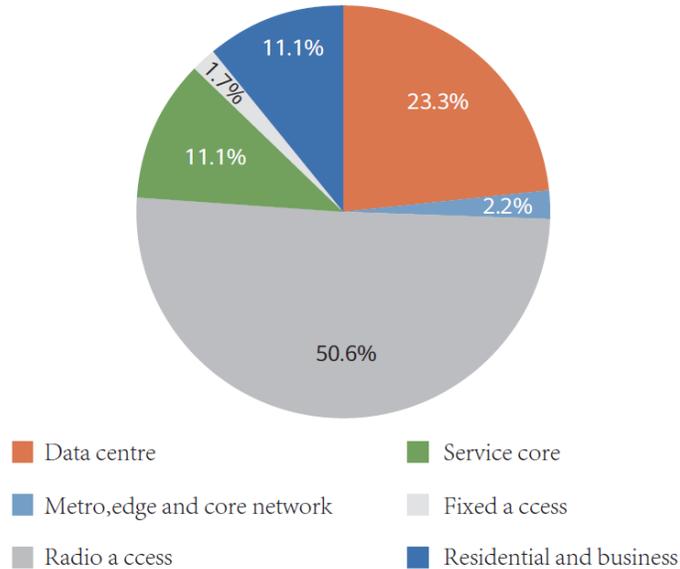}
\caption{Energy consumption breakdown by network element in 2025~\cite{Lorincz2019}.}
\label{fig:ranConsumption}
\end{figure}

To address this challenge,
and better understand where the energy consumption could be meaningfully reduced in a \ac{5G} network,
thus helping \acp{MNO} to take educated decisions,
the authors of~\cite{Huawei2020} put together an interesting analysis,
reporting that the most consuming part is the \ac{RAN},
and in particular, the \acp{BS}.
They currently account for about 57\,\% of the total network energy consumption~\cite{Usama2019}.
By 2025, 
that figure should be lower, 
as \ac{5G} becomes more prevalent, 
but the \ac{RAN} will still be the biggest consumer of energy in the network,
a 50.6\,\% according to~\cite{Lorincz2019}  
(see Fig.~\ref{fig:ranConsumption}).

Within a \ac{BS} itself, 
the \ac{RF} equipment, 
i.e., the power amplifier plus the transceivers and cables, 
have been identified as the largest energy consumer,
typically using about 65$\,\%$ of the total \ac{BS} energy. 
The cooling system, the digital signal and base band processing as well as the \ac{AC}-\ac{DC} converters follow with an energy consumption of around 17.5$\,\%$, 10$\,\%$, and 7.5$\,\%$, respectively.

In this line, 
\ac{5G} can be ---and has been--- improved for a better energy efficiency.
More efficient power amplifiers have been developed,
renewable energy sources for powering on-grid and off-grid sites, 
including solar power,
are starting to be widely adopted. 
Moreover, 
smart lithium batteries are becoming an integral part of any \ac{5G} site to enhance energy management,
and liquid cooling is being implemented to reduce the need for air conditioning~\cite{Huawei2020}.

Importantly,
since the \acp{BS} 
---and their \ac{RF} equipment--- 
consume most of the energy at a \ac{5G} network,
a judicious networking is of imperative importance~\cite{Cao2018}.
\acp{BS} and/or its most relevant components must thus only be active 
---and consume energy--- 
when handling actual data.
Plainly speaking, 
the Joules consumed in a \ac{5G} \ac{RAN} should 'follow' the bits transmitted/received. 
As a result,
and in contrast to \ac{4G},
the amount of always-on signalling in next generation \acp{RAN} must be greatly minimised at all cost. 
It is also equally necessary that data 
---and its related signalling--- 
are transmitted to/received from the intended \acp{UE}, 
while consuming the minimum possible energy to meet the end-users' \ac{QoS} demands~\cite{Pierucci2015}.
Avoiding the resource waste occasioned by the uncontrolled over-provisioning of end-users' \ac{QoS} is essential to significantly reduce the \ac{GHG} emissions of \ac{5G} \acp{RAN}.

To realise such network-wide energy efficient operation, 
\ac{5G} \acp{RAN} in general and the \ac{3GPP} \ac{NR} specification~\cite{3GPPNR} in particular
\emph{i)} have been redesigned and developed using a new and more flexible user-centric principle~\cite{Book_Dahlman2018}, 
and \emph{ii)} provide support for new ---or further enhance existing--- enabling technologies that help reducing energy consumption at the \ac{RAN} level.
In this line, 
three technologies stand out,
i.e. \ac{mMIMO}, the lean carrier design, and \ac{5G} sleep modes.
Moreover, it is important to note that \ac{5G} \ac{RAN} optimization is already benefiting from the latest advancements in the \ac{ML} field,
which in terms of energy efficiency enable a more accurate \ac{RAN} modeling and the use of more sophisticated optimization techniques.
Such over-the-top \ac{ML}-based optimization frameworks are also supported by new \ac{3GPP} \ac{NR} enhancements, 
which facilitate, among others, statistics collection and prediction capabilities.

The good understanding of these \ac{5G} technologies, how they enable energy efficiency and their optimal operation points in terms of energy efficiency are the main focus of this survey.

\subsection{Comparison to previous \ac{5G} Energy Efficiency Surveys}

Several surveys have been already published on energy efficiency and related aspects during the last 10 years. 
In the following, 
we describe the most relevant ones.

In~\cite{Wu2018}, 
the authors have surveyed the goals set by the \ac{UN} for sustainable development,
where among others,
urgent actions to combat climate change are called upon. 
Importantly, 
the wide variety of opportunities in the \ac{ICT} sector for enabling energy efficiency have been highlighted,
but although of relevance to understand the current scene,
this survey did not touch on the particularities of wireless networks. 
The survey in~\cite{Mahapatra2016}, instead, has focused on green wireless networks,
and has explained at a high level 
---and an introductory form--- 
how different metrics such as energy efficiency, spectral efficiency, throughput, and delay relate to each other. 
The discussion has been organised around the \ac{OSI} layers,
and has presented an exhaustive summary of the different frameworks and techniques that can be used to achieve optimal networking trade-offs in practical \acp{RAN}. 
Descriptions, however, are only qualitative,
and lack of detail. 
For example, the paper has considered a simple power consumption model in most explanations,
which only accounts for the transmit power, 
and has neglected the power consumption of equipment hardware.
With a similar scope, 
the authors in~\cite{Chen2011} have surveyed the literature around the understanding of fundamental energy performance trade-offs,
i.e., energy and deployment efficiency, energy and spectral efficiency, bandwidth and power as well as delay and power,
this time, 
however, 
from a more focused cellular perspective, 
covering both \ac{3GPP} \ac{UMTS} and \ac{LTE} aspects.
Importantly,
this paper has brought the attention to the importance of accurate \ac{BS} power consumption models. 
Based on this work,
the authors in~\cite{Li2011} have further surveyed techniques to optimise the above mentioned four trade-offs,
while putting the spotlight on the potential benefits of two enabling technologies,
i.e., \ac{MIMO} and relay.
With a more \ac{3GPP} \ac{LTE} standard focus,
the survey in~\cite{Hasan2011} has provided an in-depth review on green aspects,
discussing advancements in power amplifier technology, \ac{3GPP} \ac{LTE} protocols for symbol and carrier shutdown, as well as the potential benefits of small cell \acp{RAN}.
On a forward looking note,
the authors have also put forward cognitive radio and cooperative relaying as energy saving enabling technologies,
and share the latest developments in these areas. 
Complementarily,
the authors in~\cite{Feng2013} have surveyed the efforts of different academic and industry projects on energy efficiency,
emphasizing how to make use of daily network traffic variations to save energy at the \ac{RAN} level. 
Several deployment strategies, 
such as cell-size, heterogeneous networks, cooperative communications and network coding, 
have been also discussed,
together with their merits and challenges.
On a similar note,
in~\cite{Domenico2014},
a complete survey on \ac{3GPP} metrics and power consumption models for energy efficiency analysis can be found,
with a detailed formulation of the most relevant energy efficiency trade-offs.
An exhaustive 
---and critical discussion--- 
on standardised enhancements for energy-aware management in \ac{3GPP} \ac{LTE} cellular \acp{RAN} from a broadband networking point of view has been also provided.
In~\cite{Budzisz2014},
the authors have turned the focus on the survey of efficient resource management schemes, 
which are capable of controlling how much of the \ac{RAN} infrastructure is actually needed in a given space-time and which parts can be temporarily powered off to save energy.
The survey includes both cellular- and Wi-Fi-based algorithms. 
Focusing on more modern applications, 
the work in~\cite{Cao2018} has overviewed the challenges brought by the big data era,
describing issues and solutions around energy efficient data acquisition, communication, storage, computation, and analytics.
Discussion on the necessity of avoiding radio resource waste to reduce energy consumption in \acp{RAN} has been at the core of the survey,
and four types of schemes have been surveyed in this line, 
i.e., power control, time-domain scheduling, spatial resource allocation, and spectrum sharing.

Unfortunately, 
it should be highlighted that none of the above mentioned surveys has gone into the specifics of \ac{3GPP} \ac{NR}.
They are either generic or \ac{3GPP} \ac{LTE} focus,
and as a result, 
they did not survey the assets of this new technology generation to harvest energy savings. 

Giving a more related \ac{5G} perspective,
the authors in \cite{Lin2014} have provided guidelines for the development of energy efficient enabling technologies in \ac{3GPP} \ac{NR},
and surveyed the literature around the user-centric concept of \emph{no more cells}.
This networking paradigm would enable energy efficiency through a flexible network comprised of heterogeneous cells, decoupled signal- and data-planes as well as downlink and uplink.
A dynamic \ac{C-RAN}-based configuration has also been proposed to handle spatial and temporal mobile traffic variations without energy over-provisioning.  
In this work,
the potential of both \ac{mMIMO} and full duplex in \ac{5G} for power savings has been also surveyed,
while considering hardware issues.
In the same line,
the work in~\cite{Wu2017} has provided an overview of the latest research on green \ac{5G} techniques. 
The authors have explored ultra-dense sub-6GHz and millimetre wave networks, unlicensed spectrum as well as \ac{D2D} and \ac{mMIMO} communications,
while analysing their potential energy efficiency improvements,
as well as circuit power consumption issues. 
As a main contribution, 
energy harvesting has been presented as fundamental to meet \ac{5G} green requirements,
and the different lines of work in this have been discussed 
(e.g., renewable, \ac{RF} energy harvesting).
Taking a more theoretical and systematic approach,
the authors in~\cite{Zhang2017_2} have extended their work in~\cite{Chen2011},
surveying energy efficient solutions for \ac{5G} \acp{RAN} using the aforementioned fundamental green trade-offs as a driver. 
This overview has been around three main pillars,
i.e., non-orthogonal access, \ac{mMIMO}, and heterogeneous networks.
Importantly,
the paper concludes that \ac{mMIMO} is the most effective approach to enable high energy efficiencies,
provided that issues around \ac{CSI} acquisition, transceiver hardware impairments, and power inefficient components are addressed.  
A large number of references has also been provided around carrier and channel (antenna) shutdown techniques for dense small cell networks,
and the implications of centralised versus distributed \ac{RAN} architectures have been discussed. 
With a more practical 
---but still vanilla---
\ac{5G} perspective,
the authors in~\cite{Mughees2020} provide a comprehensive survey on how \ac{ML} can be used to address the energy efficiency challenges encountered in generic \ac{5G} \acp{RAN}.
Finally,
the recent survey in~\cite{Li2020} has provided the most up-to-date overview on power saving techniques supported by the \ac{3GPP} \ac{NR} standard,
covering developments in Release 15~and~16, 
and the potential upcoming ones in Release~17.
Such overview, however, has mainly focused on 
---and evaluated---
\ac{UE} power saving mechanisms, 
such as bandwidth parts, \ac{RRC} inactive state, \ac{DRX} mechanism, wake up signaling, cross-slot scheduling, and \ac{MIMO} layer adaptation.
At the \ac{RAN} side,
the lean carrier and \ac{DTX} concepts 
together with that of dormant cells, 
have been only briefly touched upon.

As it can be derived from the previous summary,
most of the existing energy efficiency surveys were written in a pre-\ac{5G} era before the \ac{3GPP} \ac{NR} existed/matured, 
or have a strong focus on the \ac{UE} 
---and not on the \ac{RAN}--- 
side. 

For completeness, Table~\ref{tab:surveys} provides a summary of the contributions and gaps observed in the aforementioned surveys.

\begin{table}
\caption{Summary and comparison of the most relevant energy efficiency surveys.}
\label{tab:surveys}
\centering
\scalebox{0.85}{
\small
\begin{tabular}{ m{1cm} | m{1cm}| m{1.4cm} |  m{9.4cm} |  m{3.2cm} } \toprule
    {\bf Ref.} & {\bf Year} & {\bf Tech. era} & {\bf Contribution/techniques} & {\bf Gaps} \\ \midrule
    
    \cite{Wu2018}  
    & 2018 
    & ICT 
    & Surveys \ac{UN} frameworks for sustainable development 
    & No cellular or \ac{ML} related.  \\ \midrule
    
    \cite{Mahapatra2016}  
    & 2016 
    & General wireless  
    & Surveys green metrics and performance trade-offs. 
    & No 5G or \ac{ML} related.   \\ \midrule
    
    \cite{Chen2011}  
    & 2011 
    & 3G/4G  
    & Surveys green metrics and performance trade-off. 
    & No 5G or \ac{ML} related.   \\ \hdashline
    
    \cite{Li2011}  
    & 2011 
    & 4G  
    & Surveys green performance trade-off optimization. Explores MIMO and relay technologies.
    & No 5G or \ac{ML} related.   \\ \hdashline
    
    \cite{Hasan2011}  
    & 2011 
    & 4G  
    & Surveys \ac{3GPP} \ac{LTE} green protocols. Focuses on symbol and carrier shutdown technologies. Explores cognitive radio, and cooperative relaying. 
    & No 5G or \ac{ML} related.    \\ \hdashline
    
    \cite{Feng2013}   
    & 2013 
    & 4G  
    & Surveys green academic and industry projects, and energy minimization under end-users' \ac{QoS} demands. Explores heterogeneous networks, cooperative communications, and network coding technologies.   
    & No 5G or \ac{ML} related.    \\ \hdashline
    
    \cite{Domenico2014} 
    & 2014    
    & 4G 
    & Surveys \ac{3GPP} green metrics, power consumption models, and \ac{3GPP} \ac{LTE} protocols. Detailed formulation of energy efficiency trade-offs.
    & No 5G or \ac{ML} related.    \\ \hdashline
    
    \cite{Budzisz2014} 
    & 2014
    & 4G and Wi-Fi 
    & Surveys energy efficient resource management schemes (with focus on Network and MAC layers).  
    & No 5G or \ac{ML} related.    \\ \hdashline
    
    \cite{Cao2018}  
    & 2018  
    & Big data era and 4G  
    & Surveys the challenges brought by the big data era. With respect to cellular, surveys MAC layer power control, time-domain scheduling, spatial resource allocation, and spectrum sharing green protocols.
    & No 5G or \ac{ML} related.    \\ \midrule
    
    \cite{Lin2014} 
    & 2014 
    & Pre-5G   
    & Surveys potential 5G energy efficiency enabling technologies. Explores the user-centric concept, downlink and uplink split \ac{C-RAN}, \ac{mMIMO}, and full duplex technologies.
    & Guesses what 5G could be. No \ac{3GPP} \ac{NR} or \ac{ML} related.  \\ \hdashline
    
    \cite{Wu2017} 
    & 2017 
    & Pre-5G  
    & Surveys research on green \ac{5G} techniques. Explores  ultra-dense sub-6GHz and millimetre wave networks, unlicensed spectrum as well as \ac{D2D}, \ac{mMIMO}, and energy harvesting technologies.
    & Guesses what 5G could be. No \ac{3GPP} \ac{NR} or \ac{ML} related.  \\ \hdashline 
    
    \cite{Zhang2017_2} 
    & 2017 
    & Pre-5G   
    & Surveys green metrics and performance trade-offs. Explores non-orthogonal access, \ac{mMIMO}, and heterogeneous networks. 
    & Guesses what 5G could be. No \ac{3GPP} \ac{NR} or \ac{ML} related.  \\ \hdashline
    
    \cite{Mughees2020} 
    & 2020 
    & 5G generic  
    & Surveys how \ac{ML} can be used to address 5G energy efficiency challenges. 
    & No \ac{3GPP} \ac{NR} focus. Lack of detail.   \\ \hdashline
    
    \cite{Li2020} 
    & 2020 
    & 5G  
    & Surveys power saving techniques supported by the \ac{3GPP} \ac{NR} standard (Release 15/16) with focus on \ac{UE}
    & Does not cover network aspects or \ac{ML} techniques. \\ \bottomrule
\end{tabular}
\normalsize
}
\end{table}

\subsection{Objective and Structure of this Survey}

In this survey,
contrary to the previous ones,
which speculatively discussed what \ac{5G} \acp{RAN} could be in terms of energy efficiency, 
we provide for the first time a detailed, up-to-date overview of the most relevant technologies that a \ac{5G} \ac{RAN} can currently use 
---or can further leverage--- 
to increase its energy efficiency in sub-6\,GHz deployments,
from both a theoretical and practical perspective.
These technologies are \ac{mMIMO}, the lean carrier design, \ac{5G} sleep modes, and \ac{ML}.
The structure of this survey is built around these technologies.
Note that \ac{5G} core networks aspects are not covered in this survey.

Importantly,
and for motivation purposes, 
we should highlight that the potential of \ac{mMIMO} to significantly enhance energy efficiency and/or decrease transmit power has been already shown in \ac{5G} deployments.
Moreover, the lean carrier design together with the \ac{5G} sleep modes have also taken significant \ac{3GPP} specification work, 
and are critical to allow the adaptation of the \ac{RAN} capabilities to varying traffic loads at both small and large time scales with symbol as well as carrier and channel (antenna) shutdown mechanisms, respectively. 
\ac{ML} frameworks are also having a major impact now on the design and optimization of the aforementioned energy efficiency techniques in practical \ac{5G} \acp{RAN}. 

It should also be noted that, 
differently than other surveys, 
this one provides much more detailed descriptions, 
including, for example, the formulation of the most relevant \ac{BS} power consumption models and energy efficiency metrics currently used for energy efficiency optimisation.
It also presents the most important energy efficiency bounds, trade-offs, and optimal operation points derived in the literature.
This makes this survey a self-contained manuscript. 
This survey also clearly distinguishes main practical concepts around the analysed enabling technologies already existing in previous technology generations from the latest \ac{3GPP} \ac{NR} specified (supporting) energy efficiency enhancements.

Given that much improvements in terms of energy efficiency are still required in practical \ac{5G} \acp{RAN},
this survey also points out and provides a fresh overview on such challenges,
and the potential of other new technologies.

\begin{figure}[t]
\includegraphics[width=16cm]{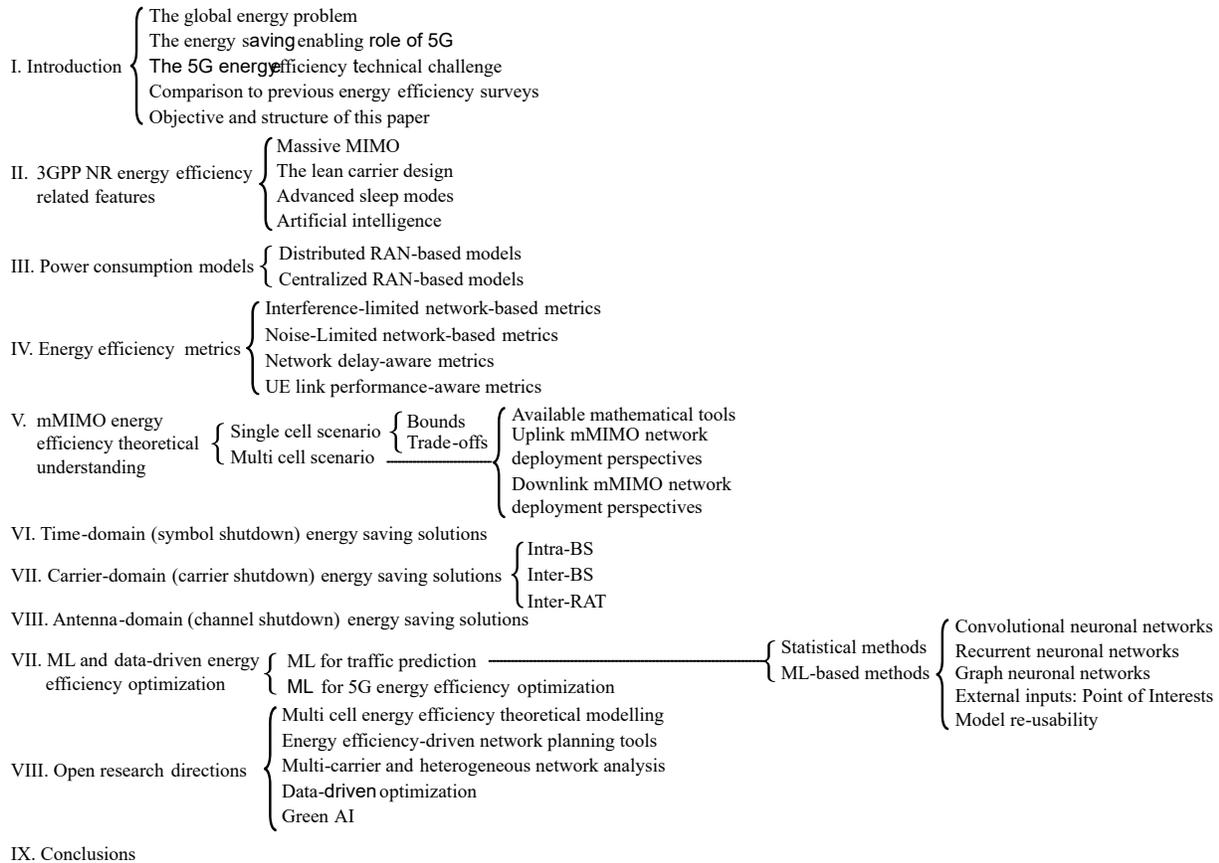}
\caption{Outline and structure of this survey.}
\label{fig:outline}
\end{figure}

With this in mind,
the rest of this survey is organised as follows 
(see Fig.~\ref{fig:outline}):
\begin{itemize}
    \item 
In Section~\ref{sec:energyEfficiencyEnabler},
we introduce \ac{mMIMO}, the lean carrier design, \ac{5G} sleep modes, and \ac{ML} as energy efficiency enabling technologies in \ac{5G} \acp{RAN};
    \item 
In Section~\ref{sec:PC_Models} and Section~\ref{sec:energyEfficiencyMetrics}, 
we present and discuss in detail existing \ac{BS} power consumption models and metrics used for energy efficiency optimisation,
respectively;
    \item 
In Section~\ref{sec:MMIMO}\footnote{
Since \ac{3GPP} \ac{NR} \ac{mMIMO} specification work mostly relates to control signalling and protocols,
and not to energy efficiency \textit{per se},
and because the amount of valuable practical work on the energy efficiency of realistic \ac{mMIMO}-based \acp{RAN} is limited,
with regard to both detailed system-level simulations as well as measurements campaigns
---the latter owing to 3GPP NR mMIMO deployments being quite new---,
in this section,
we focus on well-established theoretical aspects.},
we overview the current and well-established theoretical understanding of \ac{mMIMO} in terms of energy efficiency from a single and a multi-cell perspective and from an uplink and a downlink viewpoint. 
Important bounds and trade-offs with other key performance indicators are formulated and explained, 
and the most relevant optimisation frameworks to enhance energy efficiency via \ac{mMIMO} tuning are presented.
    \item
In Sections~\ref{sec:TimeD},~\ref{sec:CarrierD}, and~\ref{sec:AntennaD}\footnote{
Since the body of work on the theoretical understanding of symbol, carrier, and channel (antenna) shutdown methods is limited, 
given the complexity of their theoretical modelling,
as the performance of these schemes highly depends on the details of the \ac{3GPP} physical layer specification as well as scenario, deployment, traffic as well as system characteristics and dynamics,
in this section,
we focus on the \ac{3GPP} specification work related these techniques,
and the valuable work around their performance analysis and optimization.}
we dive into the details of the lean carrier design, 
and highlight the importance of sleeping modes and their optimisation at different levels 
(i.e., symbol, carrier, and channel (antenna) shutdown);
    \item 
In Section~\ref{sec:ML}, 
we highlight the potential of spatio-temporal traffic predictions and \ac{ML} approaches to maximize energy efficiency,
and overview the research in this area;
    \item 
Finally, 
in Section~\ref{sec:OpenresearchDirections} and \ref{sec:conclusions}, we discuss
future research directions, 
and draw the conclusions,
respectively.
\end{itemize}

%% file: sections/EnergyEfficiencyEnabler.tex
To reach the ambitious targets of \ac{5G} networks in terms of capacity, latency, reliability, number of supported connections, and energy efficiency, 
the \ac{3GPP} \ac{NR} specification presents a paradigm shift with respect to any preceding cellular technology~\cite{Book_Dahlman2018}.

In comparison with \ac{3GPP} \ac{LTE} and with regard to energy efficiency,
\ac{3GPP} \ac{NR} introduces a new beam-centric ---or \ac{mMIMO}-centric--- design,
which enables both 
\begin{itemize}
   \item 
    an extensive use of beamforming through a massive number of antenna elements,
    not only for data transmissions,
    but also for control-plane procedures, 
    such as the initial access~\cite{Nadeem2019},
    and
    \item 
    a larger spatial multiplexing of information on a given time-frequency resource.
\end{itemize}
Such beamforming and multiplexing gains allow for a reduced transmit power to reach a given targeted distance and/or meet a given end-user \ac{QoS} demand,
with the potential consequent benefits in terms of interference mitigation and energy savings.
Similarly, they also allow for larger data rates given a fixed transmit power,
which enable a lager \ac{BS} deactivation time,
and thus further energy savings~\cite{Rusek2013}.

Moreover, \ac{3GPP} \ac{NR} follows a new ultra-lean design principle,
in which control signals are not consistently transmitted in every radio frame,
but on demand and more sparsely, 
based on traffic requirements~\cite{Lin2019}.
This ultra-lean design allows for a more efficient operation of the \ac{mMIMO}-centric design,
and facilitates \ac{5G} sleep modes support,
i.e., \ac{BS} (de)activation,
including sophisticated symbol, carrier, or channel (antenna) shutdown mechanisms,
which can significantly reduce energy consumption~\cite{Fuentes2020}.

\ac{3GPP} \ac{NR} enhancements also allow for a more distributed network architecture,
which facilitates the usage of \ac{ML} to derive optimum network-wide energy efficient operation policies through a centralised data gathering and processing~\cite{Senzafili2019}\cite{Nguyen2020}. 
In this line,
the current \ac{3GPP} \ac{NR} architecture introduces new functions in the core and the management domains, 
i.e., the \ac{NWDAF} and the \ac{MDAF}, 
which can either run analytics on collected data or enhance the already supported network functions with statistics collection and prediction capabilities~\cite{Pateromichelakis2019}\cite{Ferrus2020}.

\bigskip 

In the following,
we introduce the main concept behind these \ac{5G} enabling technologies,
and the possibilities they enable,
whose proper optimisation will be key to minimise the energy consumption of \ac{5G} networks.
We will further elaborate on their details and potential energy saving abilities in the rest of this survey.

\subsection{Massive \ac{MIMO}}
\label{sec:energyEfficiencyMetrics:mMIMO}

Full dimension \ac{MIMO}
---generally refereed to as \ac{mMIMO}--- 
is probably one of the most important developments in \ac{5G}~\cite{Nadeem2019}.
Plainly speaking,
\ac{mMIMO} refers to a technology where \acp{BS} are equipped with antenna arrays comprised of a large number of antenna elements~\cite{Marzetta2010}.
At higher frequency bands,
due to their more challenging propagation conditions,
the large number of antenna elements are primarily used for beamforming to extend coverage.
At lower frequency bands,
the focus of this survey,
in addition to leveraging beamforming gains,
the large number of antenna elements is readily used to enable extensive spatial multiplexing and interference mitigation by spatial separation~\cite{Marzetta2016}~\cite{Bjornson2017}.
In more detail,
the large \ac{mMIMO} antenna array can excite a plurality of channel sub-spaces to support multiple simultaneous transmissions to 
---or receptions from--- several \acp{UE}. 
In this way, 
the network capacity can potentially linearly grow with the number of spatial streams multiplexed.

Despite its significant benefits, 
\ac{mMIMO} also presents a number of optimization challenges. 
Importantly,
\ac{mMIMO} requires \ac{CSI} at the \ac{BS} to realise the necessary multi-user uplink signal detection operations and downlink precoding ones. 
Most practical \ac{mMIMO} implementations take advantage of the channel reciprocity of \ac{TDD} systems to acquire such \ac{CSI} with the minimum possible overhead~\cite{4176578}. 
However, the number of orthogonal uplink pilots in a given time-frequency resource is finite,
which limits the number of \acp{UE} over which a \ac{BS} can perform \ac{CSI} acquisition at once. 
Thus, such \ac{UL} pilots have to be reused across neighbouring \acp{BS}, 
and \acp{UE} allocated to the same \ac{UL} pilot in neighbouring cells will interfere with each other in channel estimation phase. 
This effect is known as pilot contamination, 
and can significantly affect the performance of a \ac{mMIMO} system~\cite{5898372}.
Given that pilot contamination imposes a fundamental limit on what can be achieved with
a non-cooperative \ac{mMIMO} system,
coordinated uplink pilot reuse among neighboring cells has become an important area of research and optimization.

Embracing the advantages and challenges of \ac{mMIMO},
\ac{3GPP} \ac{NR} provides extensive support to realise such \ac{mMIMO} operation,
and the basis for its optimization.
Channels and signals, 
specially those used for control and synchronization,
have been re-designed with respect to \ac{3GPP} \ac{LTE} to natively use beamforming. 
This has been a major specification work. 
The acquisition of \ac{CSI} for the large number of antenna elements in a \ac{mMIMO} \ac{BS} is also now supported through
\emph{i)} new \ac{UE} \ac{CSI} reports estimated over \ac{CSI-RS} in the downlink for \ac{FDD} systems, 
and \emph{ii)} new channel measurements over \acp{SRS} in the uplink, 
exploiting channel reciprocity, 
for \ac{TDD} systems.
Among others,
\ac{3GPP} \ac{NR} is also providing new functionalities to support analog beam-forming as well as digital precoding~\cite{Book_Holma2020}.

\smallskip 

In Section~\ref{sec:MMIMO}, 
we provide a detailed survey and analysis of the energy efficiency benefits and trade-offs of this important \ac{5G} technology
---\ac{mMIMO}---
in both single- and multi-cell setups,
using well established theoretical results.  

\subsection{The Lean Carrier Design}
\label{sec:energyEfficiencyMetrics:leanCarrier}

In previous technology generations, 
signals for \ac{BS} detection, broadcast of system information, and channel estimation were always-active or transmitted very frequently over the air,
regardless of whether the \ac{BS} was serving \acp{UE} or not~\cite{Book_Dahlman2013}. 
It is important to note that,
while these always-active signals facilitate \ac{UE} operations 
---as \acp{UE} always have signals to rely on---,
they also
\begin{itemize}
    \item 
    result in a large overhead in dense deployments, 
    \item
    introduce inter-cell interference to other cells, thus reducing the achievable throughput,
    \item
    reduce the battery lifetime of the \ac{UE}, 
    and
    \item
    worsen the energy efficiency,
\end{itemize}
thus becoming a burden to efficient network operation. 

In \ac{3GPP} \ac{NR}, 
the transmission of such control signalling and the related procedures have been revisited,
following a new lean carrier design~\cite{Lin2019},
to enable larger sleep ratios and longer sleep duration.

The logic behind is as follows. 
When the traffic load of a cell 
---or group of them--- 
is low,
or the mobility conditions of the \acp{UE} allow,
larger signaling cycles could be selected to make the carrier leaner,
and thus reduce overhead, mitigate interference, and more importantly save energy,
if the \ac{BS} hardware is accordingly shutdown.

With respect to symbol switch-off in \ac{3GPP} \ac{LTE},
\ac{3GPP} \ac{NR} allows for longer and deeper sleep periods up to 160\,ms,
where more and more \ac{BS} hardware is progressively shutdown.  
To accommodate for such sparser signalling cycles, 
cell search, (re)selection as well as \ac{CSI} related signalling and procedures,
such as the transmission of \ac{SSB}, \ac{SIB1} and paging, 
have been accordingly redesigned in the new specification~\cite{3GPPTR37.910}. 

\smallskip 

In Section~\ref{sec:TimeD}, 
we survey and discuss how the \ac{5G} lean carrier design benefits energy efficiency. 

\subsection{\ac{5G} Sleep Modes}
\label{sec:energyEfficiencyMetrics:idleMode}

Cellular networks are usually planned and deployed to meet certain peak-hour requirements, 
which leads to an over-dimensioning of the network for the less challenging traffic loads during the day time~\cite{Blume2010}. 
As the traffic demands fluctuates over both time and space, 
underutilized \ac{BS} resources could be dynamically switched off to save energy.
The more network components that are shutdown and the longer the time that they are shutdown,
the more energy can be saved~\cite{Mishra2018}.

Importantly,
as mentioned earlier,
the less required always-active signalling resulting from the lean carrier design allows for longer micro-sleeps of up to up to 160\,ms in the presence of bursty traffic~\cite{3GPPTR37.910}.
In addition, 
when coupled with traffic shaping techniques and an efficient hardware at the \ac{BS} able to power up and down in fractions of a millisecond,
these times with no transmission can be leveraged by \acp{ASM} to allow deeper sleep modes,
which can progressively switch off more circuitry depending on such time length~\cite{Frenger2019}. 

Most of the improvements that a network can achieve by appropriate resource management, however, may not lay on such micro-sleep space,
as these sleep periods are usually only opportunistic and generally short.
To enable longer \ac{BS} resource deactivation times,
---macro-sleeps---
an appropriate management of the carriers and channels (antennas) of the \ac{BS} according to traffic distributions and demands at a macro-time, 
e.g. minutes or even hours, 
can avoid the resource waste emanating from the over-dimensioning of the network to meet peak hour requirements~\cite{Niu2011}~\cite{Oh2013}.

In \ac{3GPP} \ac{NR}, 
new functionality to allow large macrocells to (de)activate smaller overlapping ones in an ad-hoc manner has been specified. 
This signalling, for example, includes messaging for cell activation/deactivation request over $\rm X_n$/$\rm X_2$/$\rm F_1$ interface, 
as well as for confirming activation/deactivation actions~\cite{3GPPTR37.816}.
 
\smallskip 

In Sections~\ref{sec:TimeD},~\ref{sec:CarrierD}, and~\ref{sec:AntennaD} 
we survey and discuss how the \ac{5G} sleep modes\footnote{
It is important to note that, 
even if most of these 5G sleep mode concepts, 
i.e. symbol, carrier, and channel (antenna) shutdown, 
already existed in \ac{4G}, 
the \ac{5G} advancements on \ac{mMIMO} and the lean carrier design, 
together with other new \ac{5G} complexities, 
such as multi-\ac{RAT} \ac{4G} and \ac{5G} deployments, 
demand for novel, more complex sleep modes in \ac{5G} networks, 
and thus new specification and/or algorithmic work.} 
can operate carriers and channels (antennas) to enhance energy efficiency.

\subsection{Machine Learning}
\label{sec:energyEfficiencyMetrics:AI}

The heterogeneous and stringent service requirements of \ac{5G} networks,
together with their increasing complexity
---a pinch of which has been depicted in previous sections---
are making traditional approaches to network operation and optimization no longer adequate. 
Such methods use a significant level of expert knowledge and theoretical assumptions to characterize real environments. 
Thus, they do not scale well,
and cannot handle the complexity of real scenarios with their many parameters and imperfections as well as stochastic and non-linear processes.
To bridge this gap,
and provide \ac{5G} networks with the intelligence required to strike optimum operation points, 
equipment vendors and \acp{MNO} have started to equip their products with \ac{ML}-based functionalities~\cite{Cao2018}~\cite{Nguyen2020}. 

Fed by network measurements,
supervised and unsupervised learning tools~\cite{Luong2019}, 
two different branches of \ac{ML},
are being extensively used nowadays to model \ac{5G} network behaviour first,
and subsequently, 
take educated decisions and/or make predictions on complex scenarios~\cite{Hu2014}.
This is particularly relevant to energy efficiency.
As one can infer from the previous discussions,
minimising \ac{5G} energy consumption is a large-scale network problem,
which highly depends on complex \ac{BS} and \ac{UE} distributions, varying traffic demands and wireless channels as well as hidden network trade-offs.
Thus, understanding and predicting \ac{UE} behaviours and requirements, 
as well as their evolution in time and space,
is critical  to tailor the \ac{5G} network configuration
---\ac{mMIMO}, lean carrier, and \ac{5G} sleep modes---,
and address \ac{UE} specific communications needs with the minimum possible energy consumption. 
Specifically, the current trend is to replace rule-based heuristics and associated thresholds with e.g., optimal parameters configured trough the knowledge acquired by machine learning models.
\begin{figure}[t]
    \centering
    \includegraphics[width=12cm]{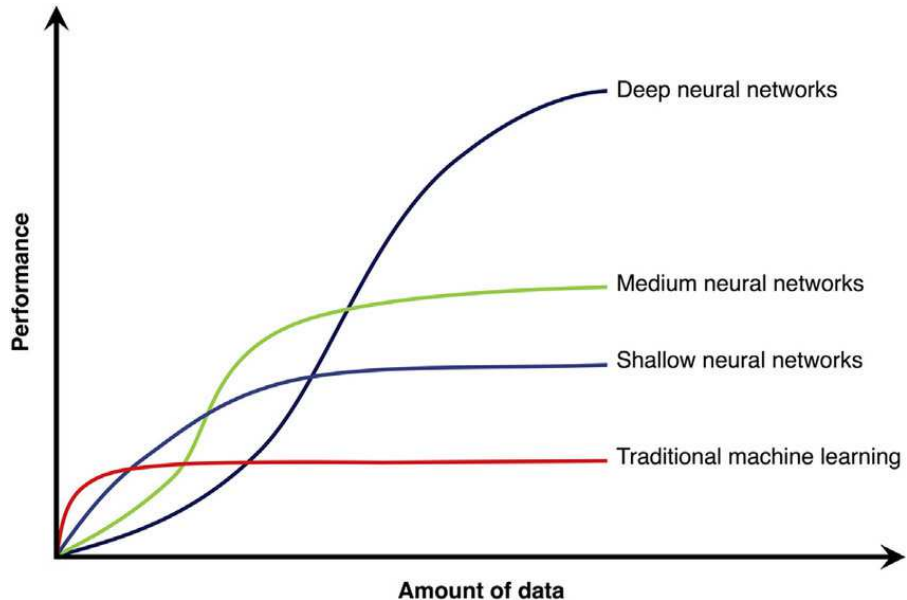}
    \caption{Relation between available training data and ML model performance \cite{tang2018canadian}.}
    \label{fig:data_ML}
\end{figure}

Additionally, 
due to the dynamic nature of wireless networks,
and the lack of network measurements data for all network procedures and on all the possible configurations they can adopt,
\ac{RL}~\cite{Sutton2018} is also being widely explored to optimise \ac{5G} network performance in general,
and energy efficiency in particular. 
For example,
shutting down network elements is a combinatorial problem with a large number of variables.
\ac{RL} agents may be used to let the network interact with the environment,
and learn optimum resource shutdown policies to minimise the total network energy consumption.
In addition, it is expected that current promising machine learning results will be improved with time, 
while the 5G ecosystem collects and makes available larger data sets related to the different problems. 
In fact, the amount of training data has a notable impact on the performance of ML algorithms,
i.e. adding data generally improves performance, 
as shown in Fig. \ref{fig:data_ML}, 
particularly for the most recent models.

However, such learning may come at the expense of both \emph{i)} an undesirably long exploration phase, 
where \ac{5G} network performance may be highly suboptimal,
and \emph{ii)} large computing powers and storage capabilities~\cite{thompson2020computational}.
Moreover, 
continuously adapting an optimal policy, 
derived from and for a limited set of specific system configurations,
to variations of network settings is a challenge,
which may drive the need for data-driven model-based approaches. 

\smallskip 

In Section \ref{sec:ML}, 
we review how \ac{ML} is being used to tackle the energy efficiency problem in \ac{5G} networks,
and discuss both main its benefits and challenges.

%% file: sections/PC_models.tex
To assess the impact of the different enabling technologies presented earlier on the energy efficiency of a \ac{5G} network, 
it is necessary to define models that provide a good estimation of their energy consumption.
Importantly,
such energy consumption models need to offer the right balance between accuracy and tractability,
while embracing different network and \ac{BS} architectures,
to empower \ac{5G} system performance characterisation and optimisation.

\begin{figure}
\centering
    \subfloat[]
    {\includegraphics[width = 7cm]{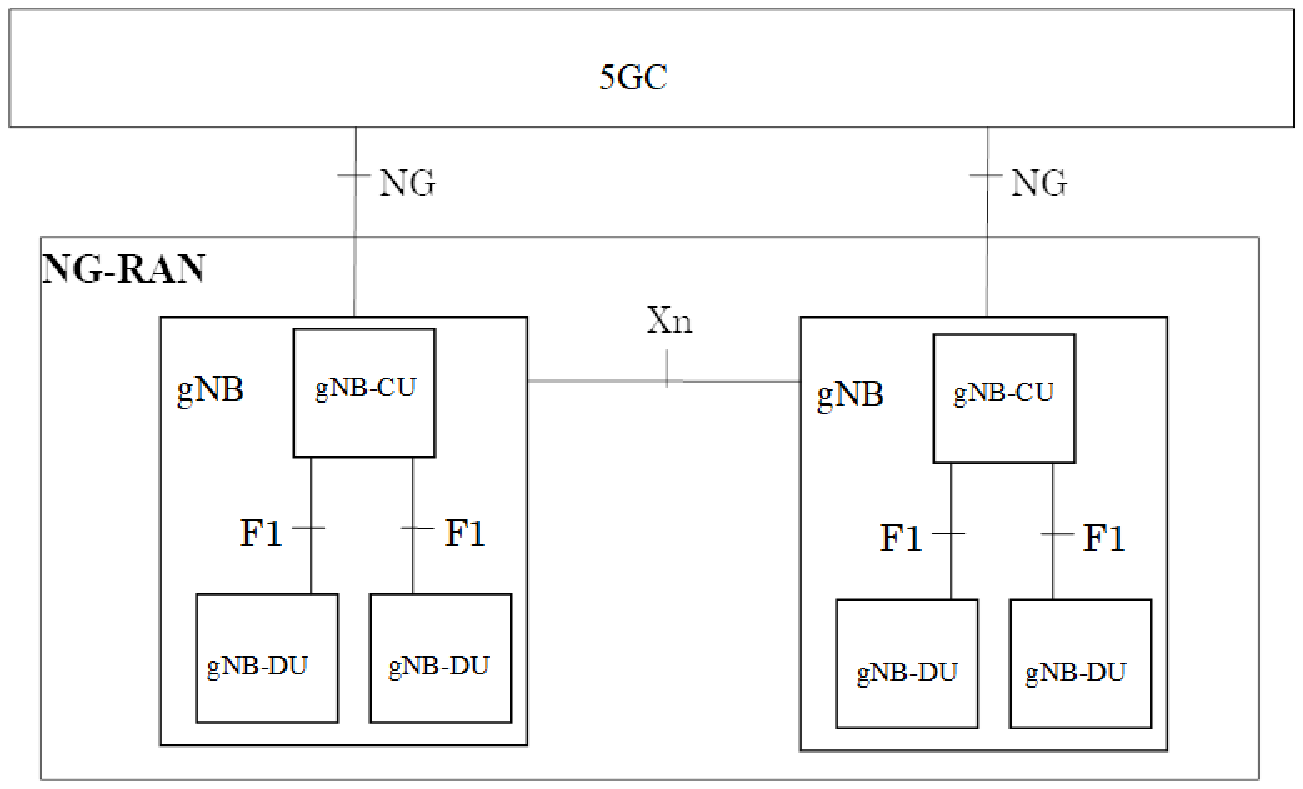}
    \label{fig:5GRAN}}
    \subfloat[]
    {\includegraphics[width = 4.5cm]{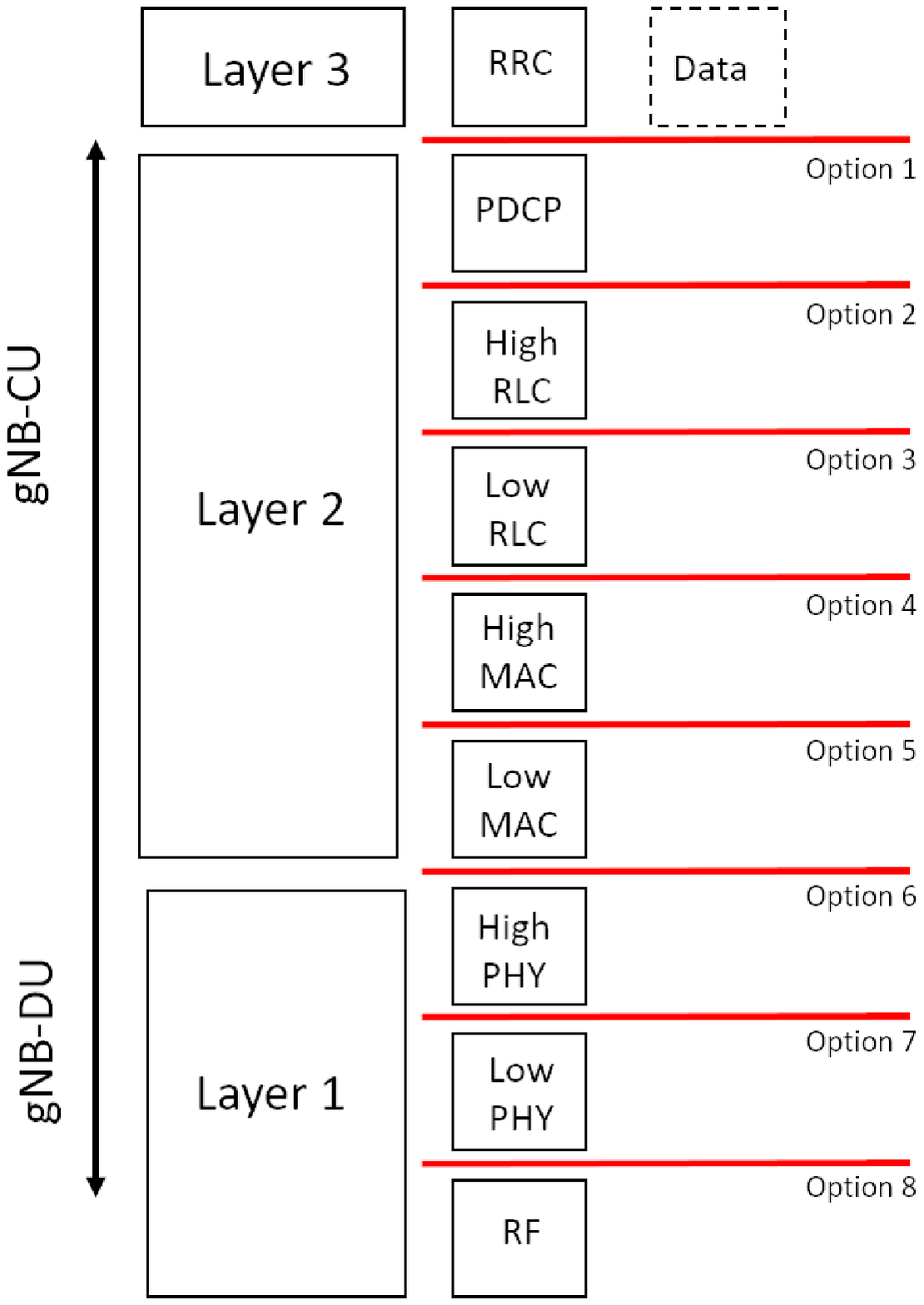}
    \label{fig:split}}
    \caption{(a) \ac{NG-RAN} overall architecture~\cite{3GPPTTR32.972}; 
    (b) \ac{3GPP} options for the function split between \ac{gNB}-\acp{DU} and \ac{gNB}-\ac{CU}~\cite{3GPPTR38.801}.}
\end{figure}

Fig. \ref{fig:5GRAN} shows the \ac{3GPP} \ac{NR} \ac{RAN} logical architecture. 
This architecture, 
denoted as \ac{NG-RAN} architecture, 
consists of a set of \acp{gNB} 
---the \ac{3GPP} \ac{NR} \acp{BS} in the \ac{3GPP} terminology--- 
connected \emph{i)} amongst them through the $\mbox{X}_{\mbox{n}}$ interface and \emph{ii)} to the \ac{5GC} through the \ac{NG} interface. 
To take advantage of virtualization technologies
and provide more implementation flexibility, 
a \ac{gNB} may also consist of a \ac{CU} and multiple \acp{DU}, 
connected to each other through the $\mbox{F1}$ interface. 
The \ac{3GPP}  has studied eight functional split options between \ac{CU} and \ac{DU} (see Fig. \ref{fig:split}), 
and current \ac{RAN} implementations are focusing on option 2~\cite{3GPPTR38.801}, in which \ac{RRC} and \ac{PDCP} are in the \ac{CU}, while \ac{RLC}, \ac{MAC}, \ac{PHY}, and \ac{RF} are in the \ac{DU}.
  
From an implementation perspective, 
each functional split corresponds to a distinct deployment option, 
which needs an appropriate power consumption characterisation. 
Specifically, 
it is necessary to take into consideration both the power consumption of the sites where the \acp{DU} and \acp{CU} are deployed, 
as well as the transport network, 
also referred to as fronthaul, 
that connects the \acp{DU} and the \ac{CU}. 
In addition, 
with the advent of \ac{NFV}, 
\acp{DU} and \acp{CU} functions can be implemented either through standard dedicated hardware or \acp{VNF} in a network cloud. 
Therefore, the overall \ac{RAN} power consumption model may need to include the contribution of the cloud server, 
where the \ac{RAN} \acp{VNF} are deployed. 
Accordingly, 
we can generally express the aggregated \ac{RAN} power consumption of a \ac{5G} network as follows:
\begin{equation}
    P_{\mbox{\scriptsize{RAN}}}=\sum_i P_{\mbox{\scriptsize{BS}}_{i}}+\sum_j P_{\mbox{\scriptsize{FH}}_j}+\sum_k P_{\mbox{\scriptsize{VBBU}}_k},
    \label{eq:RANPC}
\end{equation}
where 
$P_{\mbox{\scriptsize{BS}}_{i}}$, $P_{\mbox{\scriptsize{FH}}_j}$, and $P_{\mbox{\scriptsize{VBBU}}_k}$ 
are the power consumption of the $i$-th \ac{gNB}, the power consumption of the $j$-th fronthaul, and the power consumption of the $k$-th virtualized \ac{BBU}, 
respectively. 
Depending on the specific \ac{RAN} architecture, distributed or centralised, some of these components may not be considered. 

\bigskip

In the following,
we survey the most relevant power consumption models for both distributed and centralised \ac{RAN},
while considering their most relevant characteristics.

\subsection{Power Consumption Model for Distributed \ac{RAN}}
\label{sec:PC_RAN}

In case of a fully \ac{DRAN}, 
the \ac{RAN} power consumption can be modelled by taking into account the contributions of all \acp{BS}
as in this architecture there are no fronthaul links and virtualized \acp{BBU}, 
whose power consumption contributions correspond to the second and third terms in \eqref{eq:RANPC}, 
respectively. 

\begin{figure}[t]
    \centering
    \includegraphics[width=9cm]{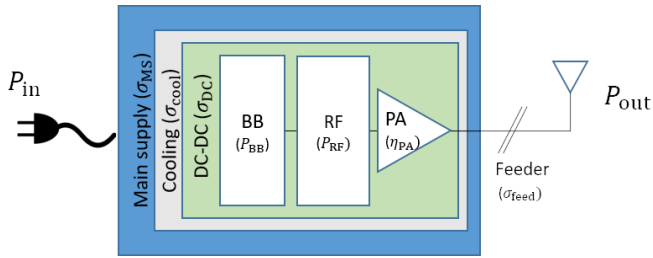}
    \caption{Power consumption block diagram of a BS RF transceiver.}
    \label{fig:BS_block}
\end{figure}

Fig.~\ref{fig:BS_block} shows the block diagram of the widely used BS power consumption model defined in~\cite{Auer2011},
where the power consumption of a non-\ac{mMIMO} \ac{BS} is computed as a function of the power consumption of all its antennas,
each one including an \ac{RF} transceiver module with its \ac{PA}, 
plus that of the \ac{BBU} associated to them, the \ac{DC}-\ac{DC} power supply, the active cooling system and the \ac{AC}-\ac{DC}C unit for connection to the electrical power grid.
Such model is formulated as follows:
\begin{equation}
    P_{\mbox{\scriptsize{BS}}}= N_{\mbox{\scriptsize{TRX}}}\frac{\frac{P_{\mbox{\scriptsize{out}}}}{\eta_{\mbox{\scriptsize{PA}}}}\left(1-\sigma_{\mbox{\scriptsize{feed}}}\right)+P_{\mbox{\scriptsize{RF}}}+P_{\mbox{\scriptsize{BB}}}}{\left(1-\sigma_{\mbox{\scriptsize{DC}}}\right)\left(1-\sigma_{\mbox{\scriptsize{MS}}}\right)\left(1-\sigma_{\mbox{\scriptsize{cool}}}\right)},
    \label{eq:PC}
\end{equation}
where  
$N_{\mbox{\scriptsize{TRX}}}$ is the overall number of \ac{RF} transceiver modules in a \ac{BS}, 
and $P_{\mbox{\scriptsize{in}}}$ is the input power of each transceiver.
$P_{\mbox{\scriptsize{out}}}$ is the transmit power, 
$P_{\mbox{\scriptsize{RF }}}$ and $P_{\mbox{\scriptsize{BB }}}$ are the \ac{RF} transceiver module and the \ac{BBU} power consumption,
$\eta_{\mbox{\scriptsize{PA}}}$ is the \ac{PA} power efficiency,
and $\sigma_{\mbox{\scriptsize{feed}}}$, $\sigma_{\mbox{\scriptsize{DC}}}$, $\sigma_{\mbox{\scriptsize{MS}}}$, and $\sigma_{\mbox{\scriptsize{cool}}}$ 
are the power losses in the feeder, \ac{DC}-\ac{DC} power supply, mains supply, and active cooling, 
respectively.

It should be noted that the model in eq.~\eqref{eq:PC} is widely represented in the literature by a simplified version of it,
which explicitly shows the linear relation between the \ac{BS} power consumption, $P_{\mbox{\scriptsize{BS}}}$, and the transmit power, $P_{\mbox{\scriptsize{out}}}$, as follows:
\begin{equation}
    P_{\mbox{\scriptsize{BS}}}= N_{\mbox{\scriptsize{TRX}}}\cdot\left(P_{\mbox{\scriptsize{0}}}+ \delta_{\mbox{\scriptsize{p}}}P_{\mbox{\scriptsize{out}}}\right),
    \label{eq:linPC}
\end{equation}
where 
$P_{\mbox{\scriptsize{0}}}$ and $\delta_{\mbox{\scriptsize{p}}}$ are cell-type dependent parameters, 
which indicate the power consumption at the minimum non-zero output power and the slope of the load-dependent power consumption, 
respectively. 
Note that this model is general, 
and accommodates to macro, micro and small cells. 
For example,
the parameters of eq.~\eqref{eq:linPC} for different types of small cells are provided in~\cite{Auer2011}.

Importantly,
in the last decade, 
in addition to appearance of the aforementioned smaller cells~\cite{David2015},
two other main solutions have emerged as key enablers to boost the mobile network capacity,
i.e., \ac{CA} (see Section \ref{sec:CarrierD}) and \ac{mMIMO} (see Section \ref{sec:MMIMO}).
Accordingly,
a number of works have evolved the previous presented model to capture the impact of these technologies on the \ac{BS} power consumption. 

\subsubsection{Carrier Aggregation Power Consumption Model}
\label{sec:PC_RAN_CA}

\ac{CA} is a \ac{3GPP} flagship feature primarily introduced to increase the cell throughput in \ac{3GPP} \ac{LTE-A}. 
The first version of \ac{CA} allowed to aggregate up to 5 \acp{CC} of up to 20~MHz,
and currently, 
in \ac{3GPP} \ac{NR},
it has been extended to support up to 16 \acp{CC} and 1\,GHz of bandwidth.
The \ac{CA} framework is flexible by design.
It enables to use continuous or discontinuous intra-band \acp{CC} as well discontinuous inter-band \acp{CC}, 
which can be characterized by different bandwidth or coverage.
It is also a key technology to enable \ac{LAA} \cite{Chen2017}, \ac{HetNet} deployments \cite{Wang2013} and dual connectivity \cite{Rosa2016}. 
For each \ac{UE}, 
a \ac{CC} is defined as its \ac{PCell}~\cite{Pedersen2011}, 
which acts as the anchor \ac{CC}, 
and is thus used for basic functionalities, 
including mobility support and \ac{RLF} monitoring. 
Additionally configured \acp{CC} are denoted as \acp{SCell}, 
and they can be added, changed, or removed to optimise the \ac{BS} performance.

When modelling the power consumption of a system using \ac{CA}, 
it is necessary to take into account how the power consumption scales with the number of active \acp{CC}, ${N_{\mbox{cc}}}$. 
In this line, 
the work in~\cite{Yu2015} has presented the following power consumption model for \ac{CA}:
\begin{equation*}
   P_{\mbox{\scriptsize{BS}}}=\sum\limits_{j=1}^{\mbox{\scriptsize{N}}_{\mbox{\scriptsize{cc}}}}  \left (P_{{\mbox{\scriptsize{TX}}}_j}+{B}_j P_{\mbox{\scriptsize{CP}}_j}^{\mbox{\scriptsize{CA}}}\right ) +P_{\mbox{\scriptsize{CP}}}^{\mbox{\scriptsize{CAi}}},
    \label{eq:linPCCA}
\end{equation*}
where 
$P_{{\mbox{\scriptsize{TX}}}_j}=\frac{P_{\mbox{\scriptsize{out}}_j}}{\eta_{\mbox{\scriptsize{PA}}}}$, $B_j$, and $P_{\mbox{\scriptsize{CP}}_j}^{\mbox{\scriptsize{CA}}}$
are the effective transmit power used by \ac{CC} $j$, the bandwidth of \ac{CC} $j$, and the variable circuit power consumption,
which scales linearly with both the number of active \acp{CC}, ${N_{\mbox{cc}}}$, and their bandwidth, $B_j$, 
respectively,
while as in the general model,
$P_{\mbox{\scriptsize{CP}}}^{\mbox{\scriptsize{CAi}}}$ is the load independent circuit power consumption of the \ac{CA} system,
i.e., of the hardware components shared by the distinct \acp{CC}. 

Embracing the complexity of \ac{CA},
the variables, $P_{\mbox{\scriptsize{CP}}}^{\mbox{\scriptsize{CA}}}$ and $P_{\mbox{\scriptsize{CP}}}^{\mbox{\scriptsize{CAi}}}$, may take different values,
depending on the specific \ac{CA} implementation. 
For instance,
contiguous \ac{CA} can be realized with a single \ac{FFT} and a single \ac{RF} transceiver module for all \acp{CC}, 
while non-contiguous \ac{CA} usually requires multiple of them~\cite{Pedersen2011}.
In the worst case, 
each \ac{CC} would require a fully dedicated hardware, 
and thus the load independent circuit power consumption of the \ac{CA} system, ${P}_{\mbox{\scriptsize{CP}}}^{\mbox{\scriptsize{CAi}}}$, would scale linearly with the number of active \acp{CC}, ${{N}_{\mbox{cc}}}$~\cite{Li2020}.

\subsubsection{\ac{mMIMO} Power Consumption Model}
\label{sec:PC_RAN_mMIMO}

With regard to \ac{mMIMO}, 
and similarly as for the \ac{CA} case,
the linear power consumption model in eq.~\eqref{eq:linPC} has also been extended to take into consideration the large number of antenna elements and the new architecture of a \ac{mMIMO} \ac{BS}.
In this line, 
the work in~\cite{Tombaz2015} proposed the following power consumption model for \ac{mMIMO}:
\begin{equation}
    {P}_{\mbox{\scriptsize{BS}}}=
    {P}_{\mbox{\scriptsize{TX}}}+{N}_{\mbox{\scriptsize{TRX}}}^{\mbox{\scriptsize{A}}}{P}_{\mbox{\scriptsize{CP}}}^{\mbox{\scriptsize{A}}}+{P}_{\mbox{\scriptsize{CP}}}^{\mbox{\scriptsize{li}}},
 \label{eq:linPCMM}
\end{equation}
where 
${N}_{\mbox{\scriptsize{TRX}}}^{\mbox{\scriptsize{A}}}$ is the number of \ac{RF} \ac{mMIMO} transceiver modules,
which do not need to be necessarily equal and may be smaller than the number of antenna elements, 
${P}_{\mbox{\scriptsize{CP}}}^{\mbox{\scriptsize{A}}}$ is the power consumption of the \ac{RF} and digital processing needed to support each \ac{RF} \ac{mMIMO} transceiver module,
and ${P}_{\mbox{\scriptsize{CP}}}^{\mbox{\scriptsize{li}}}$ is the load-independent circuit power consumption.

Importantly,
it should be noted that ${N}_{\mbox{\scriptsize{TRX}}}^{\mbox{\scriptsize{A}}}$ depends on the type of beamforming architecture implemented at the \ac{BS}~\cite{DeDomenico2017}. 
Digital beamforming provides high flexibility, 
but requires a large number of \ac{RF} \ac{mMIMO} transceiver modules. 
In contrast, 
analog beamforming decreases the power consumption by significantly reducing the number of \ac{RF} \ac{mMIMO} transceiver modules at the cost of a lower spatial resolution capability of the beamforming and a larger latency to select the proper beam weights/configuration. 
Hybrid beamforming combines the advantages of the two architectures.

The linear \ac{mMIMO} power consumption model in eq.~\eqref{eq:linPCMM} provides a simple description of the relation between the number of \ac{RF} \ac{mMIMO} transceivers and the power consumption in a \ac{mMIMO} system. 
However, more advanced works have highlighted that it is of paramount importance to also take into account the impact of multi-user scheduling. 
Specifically, 
the research in~\cite{bjornson2015optimal} has described the steps to derive a more complete model for  \ac{mMIMO} \acp{BS}, 
which accounts for both downlink and uplink communications, 
under the assumption of \ac{ZF} processing\footnote{
Models for other specific precoding and combining schemes are discussed in~\cite{Bjornson2017}.}. 

\begin{figure}[t]
    \centering
    \includegraphics[width=9cm]{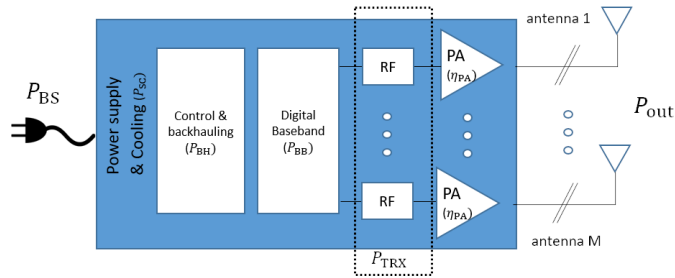}
    \caption{Power consumption block diagram of a \ac{mMIMO} \ac{BS}.}
    \label{fig:mMIMO_block}
\end{figure}
As described in Fig.~\ref{fig:mMIMO_block}, 
the main blocks of a \ac{mMIMO} \ac{BS}, 
from a power consumption perspective, 
are the \ac{PA}, the analog front-end, the digital baseband, the control and network
backhaul platform, and the power system (which includes also the cooling) \cite{Debaillie2015}.
Accordingly, 
the authors of~\cite{Bjornson2017} has described the \ac{mMIMO} \ac{BS} power consumption as the sum of the effective transmit power, ${P}_{\mbox{\scriptsize{TX}}} = \frac{{P}_{\mbox{\scriptsize{out}}}}{\eta_{\mbox{\scriptsize{PA}}}}$, and the circuit power, ${P}_{\mbox{\scriptsize{CP}}}$, as follows:
\begin{equation}
    {P}_{\mbox{\scriptsize{BS}}}={P}_{\mbox{\scriptsize{TX}}}+{P}_{\mbox{\scriptsize{CP}}},
\end{equation}
where
${P}_{\mbox{\scriptsize{CP}}}$ accounts for the power contributions due to the analog front-end, ${P}_{\mbox{\scriptsize{TRX}}}$, the digital baseband, ${P}_{\mbox{\scriptsize{BB}}}$, the control and network
backhaul platform, ${P}_{\mbox{\scriptsize{BH}}}$, and the power system, ${P}_{\mbox{\scriptsize{SC}}}$, and it can computed as:
\begin{equation}
    {P}_{\mbox{\scriptsize{CP}}}={P}_{\mbox{\scriptsize{TRX}}}+{P}_{\mbox{\scriptsize{BB}}}+{P}_{\mbox{\scriptsize{BH}}}+{P}_{\mbox{\scriptsize{SC}}}={P}_{\mbox{\scriptsize{CP}}}^{\mbox{\scriptsize{li}}}+{P}_{\mbox{\scriptsize{CP}}}^{\mbox{\scriptsize{ld}}}.
    \label{eq:linCPmMIMO}
\end{equation}
For clarity,
we will denote by ${P}_{\mbox{\scriptsize{CP}}}^{\mbox{\scriptsize{li}}}$ the sum of the fixed and load-independent power consumption required for the site power supply, control signalling, backhaul, signal processors, and \ac{RF} transceivers, 
and by ${P}_{\mbox{\scriptsize{CP}}}^{\mbox{\scriptsize{ld}}}$ the term that includes the variable load-dependent part of the circuit power consumption, ${P}_{\mbox{\scriptsize{CP}}}$. 
In addition, it is worth to highlight that the functions that are the main contributors to the power consumption of the digital processing unit are the channel estimation process, the channel coding and decoding units, and the beamforming processing.

Accordingly, 
the \ac{mMIMO} \ac{BS} power consumption model in eq.~\eqref{eq:linCPmMIMO} can be expressed as follows~\cite{Hossain2018}~\cite{Bjornson2014}:
\begin{equation}
    {P}_{\mbox{\scriptsize{BS}}}=\frac{{K}\cdot{P}_{\mbox{\scriptsize{UE}}}}{\eta_{\mbox{\scriptsize{PA}}}}+{P}_{\mbox{\scriptsize{CP}}}^{\mbox{\scriptsize{li}}}+\mathcal{C}_{\mbox{\scriptsize{3}}}{K}^3+\mathcal{D}_{\mbox{\scriptsize{0}}}{M}+\mathcal{D}_{\mbox{\scriptsize{1}}}{M}\cdot {K}+\mathcal{D}_{\mbox{\scriptsize{2}}}{M}\cdot {K}^2+ \mathcal{A}{K}\cdot{R}_{\mbox{\scriptsize{UE}}},
    \label{eq:mMIMOPC}
\end{equation}
where 
${K}$ is the number of simultaneously multiplexed \acp{UE} at the \ac{BS}, 
${P}_{\mbox{\scriptsize{UE}}}$ is the downlink output power per \ac{UE} 
(i.e., ${P}_{\mbox{\scriptsize{out}}}={K}\cdot{P}_{\mbox{\scriptsize{UE}}}$), 
$\mathcal{C}_{\mbox{\scriptsize{3}}}$ is the part of the beamforming processing, ${P}_{\mbox{\scriptsize{SP}}}$,
which scales linearly with ${K}^3$,  
$\mathcal{D}_{\mbox{\scriptsize{0}}}$ is the power consumed by the transceiver module attached to each antenna, 
$\mathcal{D}_{\mbox{\scriptsize{1}}}$ is the part of the beamforming processing, ${P}_{\mbox{\scriptsize{SP}}}$,
which scales linearly with ${M}\cdot {K}$,
$\mathcal{D}_{\mbox{\scriptsize{2}}}$ is the sum of the contributions of the channel estimation process, ${P}_{\mbox{\scriptsize{CE}}}$, and the beamforming processing, ${P}_{\mbox{\scriptsize{SP}}}$,
which scale linearly with ${M}\cdot {K}^2$,
${R}_{\mbox{\scriptsize{UE}}}$ is the \ac{UE} throughput, 
and $\mathcal{A}$ is the aggregated power consumption per bit of information required by the coding/decoding operations and by the load-dependent part of the backhaul. 
Table~\ref{tab:PC parameters} describes typical values of the parameters in eq.~\eqref{eq:mMIMOPC}.

\begin{table}[!ht]
    \caption{Typical values for \ac{mMIMO} power consumption model parameters~\cite{bjornson2015optimal}~\cite{Hossain2018}~\cite{Bjornson2014}.}
    \label{tab:PC parameters}
    \small
    \centering
    \begin{tabular}{|c|c||c|c|} 
    \hline
    Parameter  &  Value & Parameter & Value \\
    \hline 
    $\eta_{\mbox{\scriptsize{PA}}}$ & 0.39 & ${P}_{\mbox{\scriptsize{CP}}}^{\mbox{\scriptsize{li}}}$ &20 [W] \\
    \hline 
    $\mathcal{C}_{\mbox{\scriptsize{3}}}$ & 10$^{-7}$ [W] & $\mathcal{D}_{\mbox{\scriptsize{0}}}$ &1 [W] \\
    \hline 
    $\mathcal{D}_{\mbox{\scriptsize{1}}}$ & 3$\cdot$10$^{-3}$ [W] & $\mathcal{D}_{\mbox{\scriptsize{2}}}$ &9.4$\cdot$10$^{-7}$ [W] \\
    \hline 
    $\mathcal{A}$ & 1.15 [W/Gbps] &  & \\
    \hline 
    \end{tabular}
\end{table}

To conclude this section, 
Table \ref{tab:DRAN} summarizes the main contributions of the presented works on the power consumption models for \ac{DRAN} architectures.

\begin{table}[!ht]
\caption{Summary of Power Consumption Models for DRAN.}
\label{tab:DRAN}
\scriptsize
\centering
  \begin{tabular}{|c||c|c|} 
  \hline
  Paper & Model &Main parameters \\
  \hline
  \cite{Auer2011} & non-mMIMO BSs & Number of RF transceivers, load-dependent power  \\ & &  consumption and static power consumption  \\ \hline
  ~\cite{Yu2015} & non-mMIMO BSs with CA & Number of CCs, transmit power per CC, bandwidth per\\ & &  CC, variable circuit power, and load independent circuit power \\ \hline
  \cite{Tombaz2015} & mMIMO BSs & Number of RF transceivers, transmit power consumption, power \\ & & consumption of the RF and digital processing per transceiver, and static power consumption \\ \hline
  ~\cite{Hossain2018}~\cite{Bjornson2014} & mMIMO BSs& Number of multiplexed UEs, number of antennas, transmit \\ & & power consumption, beamforming power consumption \\ \hline
  \end{tabular}
\end{table}

\subsection{Power Consumption Model for Centralised \ac{RAN}}
\label{sec:PC_CRAN}
Considering a more sophisticated \ac{RAN} architecture,
the work in~\cite{Fiorani2016} has studied the power consumption modelling of a \ac{CRAN} considering different functional splits between the \ac{gNB}-\acp{DU} and the \ac{gNB}-\ac{CU}.
In terms of energy efficiency, 
centralisation enables three main benefits with respect to a decentralised architecture: 
stacking gain, pooling gain, and cooling gain~\cite{Fiorani2016}. 
The stacking gain refers to the capability of deploying less processing units (in the central node) to serve the same amount of cells in a given area,
while the pooling gain refers to the capability of using a limited amount of centralised resources to operate a large amount of cells, 
by exploiting the load variations in the network. 
The cooling gain appears due to the reduced amount of energy required to cool the cell site and the more advanced cooling solutions that can be implemented at the central node~\cite{Rowell2020}.

\begin{figure}[t]
    \centering
    \includegraphics[width=12cm]{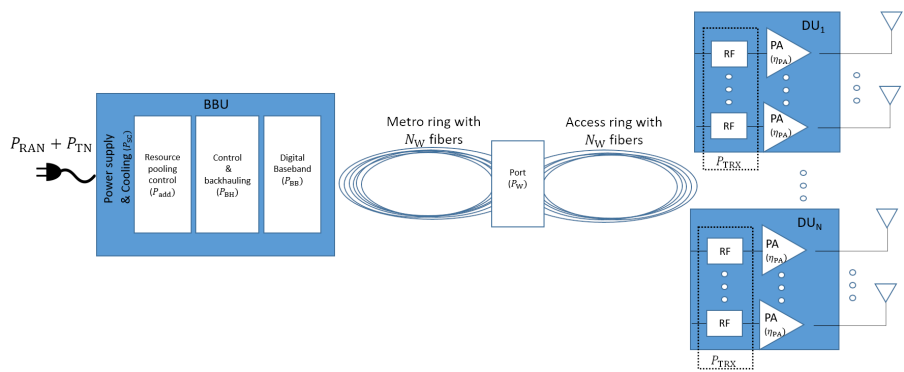}
    \caption{Power consumption block diagram of a CRAN system with optical transport network.}
    \label{fig:CRAN_P_block}
\end{figure}

In case of a CRAN, 
the network power consumption shall be modelled by taking into account the contribution of both the access network, ${P}_{\mbox{\scriptsize{RAN}}}$, including all CUs and DUs, and the transport network, ${P}_{\mbox{\scriptsize{TN}}}$, (see Fig. \ref{fig:CRAN_P_block}).

In this centralized architecture, 
the aggregated power consumption at the \ac{RAN}, ${P}_{\mbox{\scriptsize{RAN}}}$, can be computed as~\cite{Fiorani2016}:
\begin{equation}
    \sum_i {P}_{\mbox{\scriptsize{BS}}_{i}}=\sum_i {P}_{\mbox{\scriptsize{DU}}_{i}}\left(1-\frac{{P}_{\mbox{\scriptsize{co}}_{i}}+{P}_{\mbox{\scriptsize{NF}}_{i}}}{100}\right)+{P}_{\mbox{\scriptsize{BBU}}},
    \label{eq:linCRANPC}
\end{equation}
where ${P}_{\mbox{\scriptsize{DU}}_{i}}$ is the power consumption of the site where the $i$-th \ac{DU} is deployed,
which can be computed as, e.g., eq.~\eqref{eq:mMIMOPC}, 
${P}_{\mbox{\scriptsize{BBU}}}$ is the power consumption of the \ac{BBU} host where the \ac{CU} is located\footnote{
This model can be easily generalised to the case where a larger network with multiple \acp{CU} is considered.},
${P}_{\mbox{\scriptsize{NF}}_{i}}$ is the fraction of power consumption that corresponds to the \acp{NF} moved to the \ac{CU}, 
and ${P}_{\mbox{\scriptsize{co}}_{i}}$ is the fraction of power consumption that corresponds to the cooling related to the $i$-th \ac{DU}.
Importantly, 
note that the higher the number of \acp{NF} moved to the \ac{CU},
the higher the values of both parameters, ${P}_{\mbox{\scriptsize{co}}_{i}}$ and ${P}_{\mbox{\scriptsize{NF}}_{i}}$.

Moreover,
the power consumption of the \ac{BBU} host, ${P}_{\mbox{\scriptsize{BBU}}}$, can be computed as:
\begin{equation}
    {P}_{\mbox{\scriptsize{BBU}}}= \sum_i {P}_{\mbox{\scriptsize{DU}}_{i}}\left ( \frac{1/{G}_{\mbox{\scriptsize{co}}}}{100}+\frac{{P}_{\mbox{\scriptsize{NF}}_{i}}/{N}_{\mbox{\scriptsize{BS}}}}{100} \left \lceil\frac{{N}_{\mbox{\scriptsize{BS}}}}{{G}_{\mbox{\scriptsize{st}}}{G}_{\mbox{\scriptsize{po}}}} \right\rceil {P}_{\mbox{\scriptsize{add}}}\right),
    \label{eq:BBUPC}
\end{equation}
where ${G}_{\mbox{\scriptsize{st}}}$, ${G}_{\mbox{\scriptsize{po}}}$, ${G}_{\mbox{\scriptsize{co}}}$, and ${P}_{\mbox{\scriptsize{add}}}$ 
are the stacking gain, the pooling gain, the cooling gain, and the additional power consumption in the \ac{BBU} needed to enable resource pooling, 
respectively. 
The work in \cite{Fiorani2016} has provided a preliminary analysis to estimate the values of these parameters.
The authors in \cite{Checko2016} have used tele-traffic theory and simulations to evaluate the resource pooling gains at the \ac{BBU} and the fronthaul for different functional splits, 
and then investigated how to optimize the cost for \ac{BBU} pool, fronthaul capacity, and radio resource utilization. 
More recently, 
the authors in \cite{Shehata2018} have further extended this work, 
and provided an analysis of the power consumption of the different \ac{RAN} functional splits, 
considering the multiplexing gain arising in each split option.

It is important to note that eq.~\eqref{eq:BBUPC} assumes that the \ac{CU} is deployed on a dedicated hardware platform,
which has a similar architecture to the one used for a single \ac{DU} but with larger computational capacity. 
In the case of a \ac{VRAN},
the goal is to move the processing to \acp{GPP}, 
which are capable to provide real-time processing to maintain the timing in the \ac{RAN} protocols, 
while equipped with efficient and elastic resources
---CPU, memory, and networking---
to perform intensive digital processing. 
This approach promises further improved energy savings, 
which depend on the specifically used architecture~\cite{Nikaein2017}. 
One option is to use dedicated hardware (e.g., system on chip) for managing the layer 1 functions on dedicated hardware, 
while higher-layer functions are implemented in a software-based architecture.
This solution may enable, however, limited additional energy savings with respect to \ac{CRAN}. 
To reduce the power consumption and increase flexibility, 
an alternative option is to deploy only the most computationally-intensive functions on dedicated hardware, 
such as turbo decoding and encryption/decryption. 
As an extreme option, 
in the full \ac{GPP} architecture, 
all the \ac{RAN} functions are implemented in a virtual environment,
which may enable large power savings at the expense, however, of performance if the \acp{GPP} and infrastructure around cannot cope with the workload.  
Considering the \ac{GPP} architecture, 
the work in~\cite{SIGWELE20171} proposes a power consumption model for virtualized \acp{BBU} in a cloud node,
which takes into account the impact of the cooling system, the workload dispatcher switch, and the \acp{GPP} as follows:
\begin{equation*}
    {P}_{\mbox{\scriptsize{VBBU}}}={P}_{\mbox{\scriptsize{Vco}}}+{P}_{\mbox{\scriptsize{Dis}}}+\sum_i {P}_{\mbox{\scriptsize{GPP}}_{i}},
    \label{eq:cloudPC}
\end{equation*}
where ${P}_{\mbox{\scriptsize{Vco}}}$, ${P}_{\mbox{\scriptsize{Dis}}}$ and ${P}_{\mbox{\scriptsize{GPP}}_{i}}$ 
are the power consumption contributions in the virtualized \ac{BBU} due to the cooling system, the switch, and the $i$-th \ac{GPP}, 
respectively. 

To characterize the \ac{GPP} power consumption, 
a linear power consumption  model can be used \cite{3GPPTTR32.972}~\cite{Younis2018},
i.e.,
\begin{equation*}
    {P}_{\mbox{\scriptsize{GPP}}}={P}_{\mbox{\scriptsize{GPP}}_0}+\Delta_{\mbox{\scriptsize{GPP}}}{P}_{\mbox{\scriptsize{GPP}}_{\mbox{\scriptsize{m}}}} \rho_{\mbox{\scriptsize{GPP}}},
    \label{eq:GPPPC}
\end{equation*}
where ${P}_{\mbox{\scriptsize{GPP}}_0}$ and ${P}_{\mbox{\scriptsize{GPP}}_{\mbox{\scriptsize{m}}}}$ 
are the power consumption of the \ac{GPP} when it is in idle mode and at maximum load, 
respectively, 
and $\Delta_{\mbox{\scriptsize{GPP}}}$ and $\rho_{\mbox{\scriptsize{GPP}}}$ are the slope of the power consumption model,
which is related to the specific \ac{GPP} architecture, 
and the load of the \ac{GPP}, respectively,
where the latter depends on the \ac{GPP} computational capacity and the computational resources required by the hosted \acp{VNF}.

To model the \ac{GPP} computational load,
the work in~\cite{Younis2018} has used an experimental platform to infer the relationship between the downlink throughput and
the percentage of \ac{CPU} usage at the \ac{BBU}.
Instead,
the authors in~\cite{Sabella2014}\cite{Werthmann2013} have proposed an analytical characterisation of the \ac{GPP} computational load, 
which jointly depends on the functional split, the number of used transceiver modules/antennas, the bandwidth, the rate and the number of spatial \ac{MIMO}-layers used at the virtualized \ac{BS}. 
Importantly, 
the authors in~\cite{Chaudhary2019} have further analytically investigated the computational complexity of lower layer functions in \ac{3GPP} \ac{NR},
and this research has shown that, today, due to the form factor and power consumption of \ac{GPP}, 
dedicated hardware is the only feasible option for deploying full \ac{3GPP} \ac{NR} capabilities.

Complementing the above,
the authors in~\cite{Khatibi2018} have also recently developed an empirical model to describe the computational requirements of the \ac{RAN} \acp{NF},
focusing on the \ac{PHY} layer, 
which includes the most computationally expensive functionalities. 
Their results highlight that \acp{NF} can be classified according to their complexity into three classes:
\begin{itemize}
    \item
\ac{PHY} \acp{NF}, 
whose computational complexity only depends on the system configuration, 
and does not change with time 
(e.g., \ac{FFT}), 
    \item
\ac{PHY} \acp{NF}, 
whose computational complexity depends on both throughput requirements and channel quality 
(e.g., encoding/decoding), and 
    \item
higher layer \acp{NF}, 
whose computational complexity only depends on throughput requirements 
(e.g., packet scheduling).
\end{itemize}

Finally, 
to connect each \ac{DU} to the \ac{CU}, 
a fronthaul link, 
whose capacity fits the functional split throughput and latency requirements is needed.
Therefore, 
the total power consumption of the fronthaul is a function of the transport network technology, its topology, the capacity required by each \ac{BS}, and the number of \acp{BS} deployed in the \ac{RAN}.
For instance, 
when the fronthaul is based on optical dense wavelength division multiplexing with a ring architecture and optical switching, 
the power consumption of the transport network, ${P}_{\mbox{\scriptsize{TN}}}$, is due to all the fiber links connecting the \acp{DU} and the associated \acp{CU}.
The contribution of a link connecting the $i$\emph{-th} \ac{DU} with its \ac{CU}, ${P}_{\mbox{\scriptsize{FH}}_i}$, can be modelled as~\cite{Fiorani2016}:
\begin{equation*}
    {P}_{\mbox{\scriptsize{FH}}_i}=\left \lceil\frac{{C}_{\mbox{\scriptsize{FS}}_i}}{{R}_{\mbox{\scriptsize{tr}}}} \right\rceil  \left( 2\cdot{P}_{\mbox{\scriptsize{tr}}}+\frac{{P}_{\mbox{\scriptsize{W}}}}{{N}_{\mbox{\scriptsize{W}}}} \right),
    \label{eq:BHPC}
\end{equation*}
where ${C}_{\mbox{\scriptsize{FS}}_i}$ is the transport capacity required by the functional split at the $i$-th \ac{gNB}, 
${P}_{\mbox{\scriptsize{W}}}$ is the power consumption of a port used to interconnect access and metro rings in the transport network, 
${N}_{\mbox{\scriptsize{W}}}$ is the number of wavelengths per fiber, 
and ${R}_{\mbox{\scriptsize{tr}}}$ and ${P}_{\mbox{\scriptsize{tr}}}$ are the rate and the associate power consumption of the transport network nodes, 
respectively.
It is important to highlight that a centralized architecture leads to multiplexing gains not only at the \ac{BBU} side, 
but also in the communication network. 
Specifically, the work in \cite{Wang2017} has proposed an analytical model to derive an upper bound of the fronthaul statistical multiplexing gain, 
and shown that even pooling a moderate number of \acp{BS} can result in notable radio resource savings.

\begin{table}[!ht]
\caption{Summary of Power Consumption Models for CRAN.}
\label{Tab:CRAN}
\scriptsize
\centering
  \begin{tabular}{|c||c|c|} 
  \hline
  Paper & Model &Main parameters \\
  \hline
  \cite{Fiorani2016} & CRAN architecture & BBU power consumption, DU power consumption,  \\ & &  and stacking, pooling, and cooling gains  \\ \hline
  ~\cite{SIGWELE20171} & CRAN architecture with GPPs & Power consumption due to the cooling system, the switch, and GPPs \\ \hline
  \cite{3GPPTTR32.972}~\cite{Younis2018} & GPP power consumption & GPP power consumption in idle and maximum load, \\ & & load-dependent slope of the power model \\ \hline
  ~\cite{Sabella2014}{\cite{Werthmann2013}} & GPP power consumption& Functional split, number of antennas, bandwidth,\\ & & rate,  MIMO-layers \\ \hline
    ~\cite{Fiorani2016} & FH power consumption& Transport network capacity, number of wavelengths per fiber, \\ & & bandwidth, rate and power of the transport network nodes \\ \hline
  \end{tabular}
\end{table}

\begin{figure}
\centering
    \subfloat[]
    {\includegraphics[width = 7cm,height = 7cm]{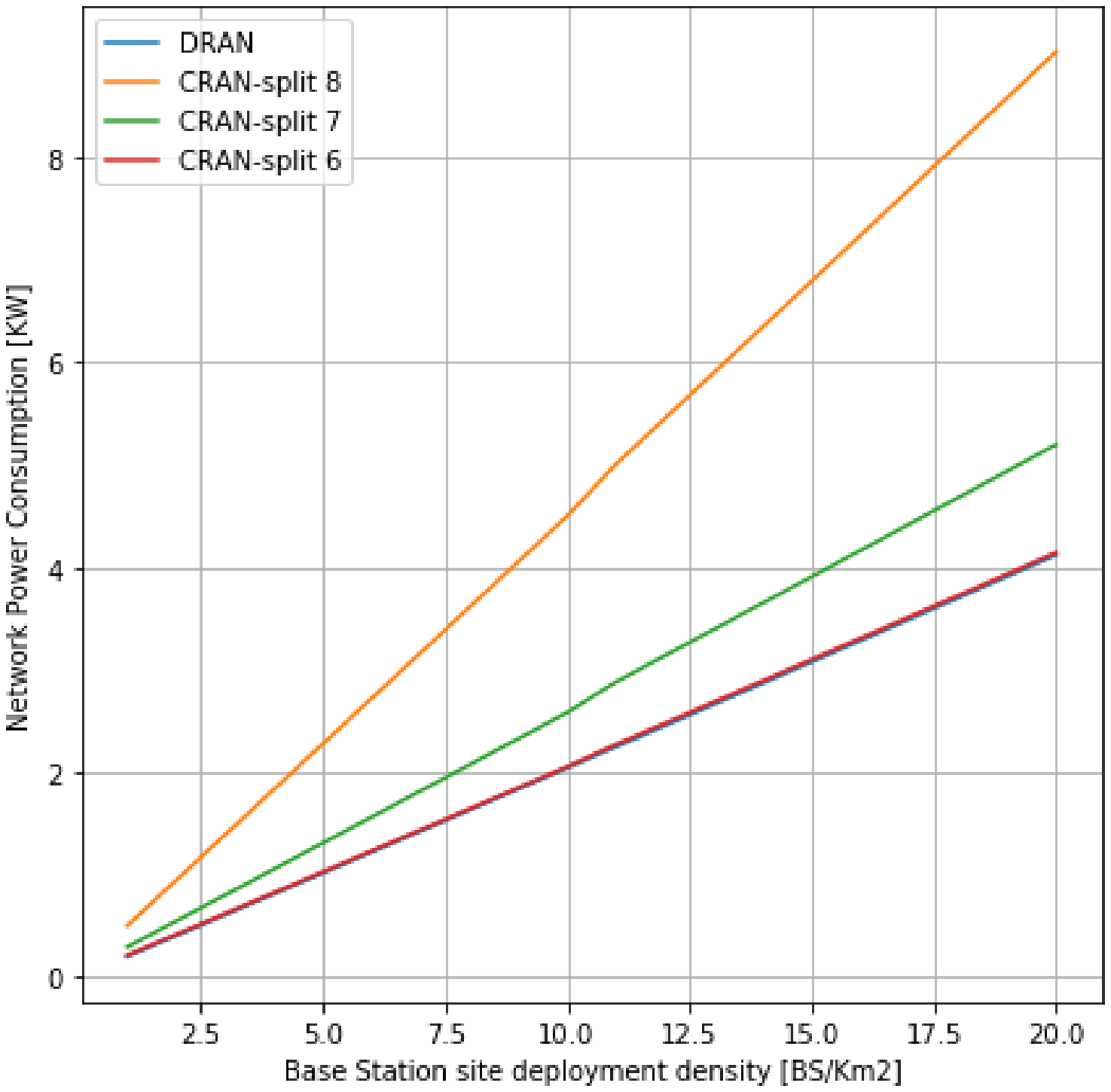}
    \label{fig:CRAN_PC}}
    \subfloat[]
    {\includegraphics[width = 7cm,height = 7cm]{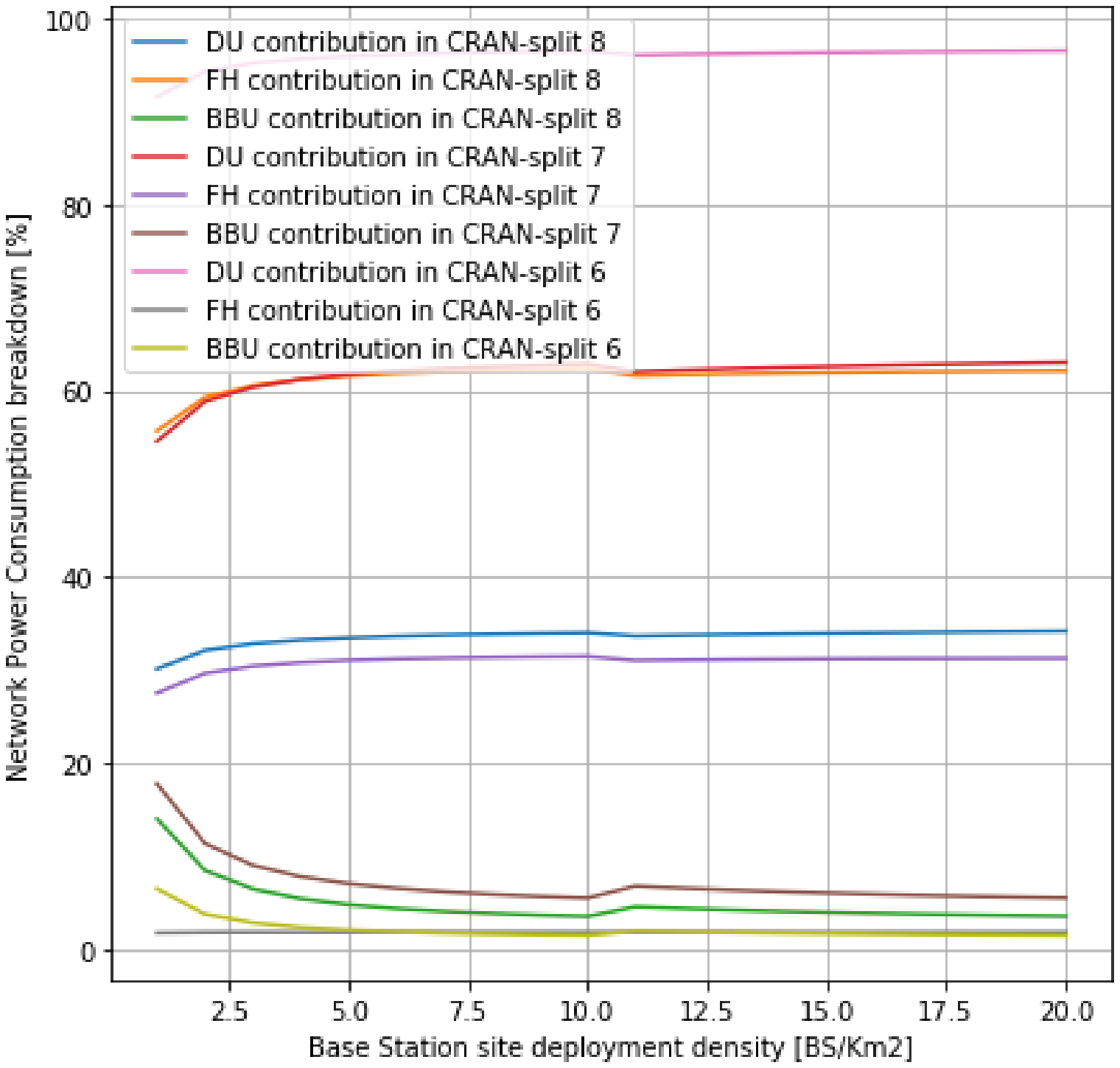}
    \label{fig:CRAN_PC_break}}
    \caption{Network power consumption (a) and its breakdown (b) as a function of the \ac{BS} deployment density for classic \ac{DRAN} and different \ac{CRAN} architectural options.}
\end{figure}

\bigskip
Table \ref{Tab:CRAN} provides a summary of the main contributions presented in this section, which discuss the power consumption models for \ac{CRAN} architecture.
To provide a general view,
and highlight the impact of the transport network on the overall system power consumption, 
we show in Fig.~\ref{fig:CRAN_PC} the network power consumption with respect to the \ac{BS} deployment density for a classic \ac{DRAN} and different \ac{CRAN} architectural options, 
using the parameters indicated in Table \ref{tab:PC parameters2}. 
Moreover, Fig.~\ref{fig:CRAN_PC_break} shows the impact of each contributor,
i.e., \acp{DU}, fronthaul nodes, and \ac{BBU}, 
with respect to the aggregated \ac{CRAN} power consumption. 
Fig.~\ref{fig:CRAN_PC} highlights that only a \ac{CRAN} with functional split option~6 (see Fig.~\ref{fig:split}) has similar power consumption to the classic \ac{DRAN}.  
\ac{CRAN} architectures with lower functional splits provide larger centralisation gains, 
but lead to higher power consumptions. 
In addition, 
Fig. \ref{fig:CRAN_PC_break} shows that the contribution from the transport network cannot be neglected in the overall power consumption characterisation,
particularly when some architectures are adopted. 
Specifically, 
while for split option~6, 
it only amounts for around 2\,$\%$ of the network power consumption, 
in split option~7 and~8, 
it contributes for around 30\,$\%$ and 60\,$\%$ of the network power consumption, 
respectively.

\begin{table}[!ht]
    \caption{Typical values for \ac{CRAN} power consumption model parameters ~\cite{Fiorani2016}~\cite{Fiorani2016ICC}.}
    \label{tab:PC parameters2}
    \small
    \centering
    \begin{tabular}{|c|c||c|c||c|c|} 
    \hline
    \ac{CRAN} split 8  &  Value & \ac{CRAN} split 7   & Value& \ac{CRAN} split 6   & Value \\
    \hline 
    ${P}_{\mbox{\scriptsize{co}}}$  & 10  & ${P}_{\mbox{\scriptsize{co}}}$ & 10 &   ${P}_{\mbox{\scriptsize{co}}}$ & 0 \\
    \hline 
    ${P}_{\mbox{\scriptsize{NF}}}$  & 15  & ${P}_{\mbox{\scriptsize{NF}}}$ & 10.5 &   ${P}_{\mbox{\scriptsize{NF}}}$ & 3 \\
    \hline
    ${G}_{\mbox{\scriptsize{po}}}$ & 5  & ${G}_{\mbox{\scriptsize{po}}}$ &5 & ${G}_{\mbox{\scriptsize{po}}}$  & 5\\
    \hline
    ${G}_{\mbox{\scriptsize{st}}}$ & 2  & ${G}_{\mbox{\scriptsize{st}}}$ & 2 & ${G}_{\mbox{\scriptsize{st}}}$  & 2\\
    \hline
    ${P}_{\mbox{\scriptsize{add}}}$ & 2 & ${P}_{\mbox{\scriptsize{add}}}$ & 2 &${P}_{\mbox{\scriptsize{add}}}$  &2 \\
    \hline
    ${G}_{\mbox{\scriptsize{co}}}$ & 0.2  & ${G}_{\mbox{\scriptsize{co}}}$ & 0.2 &${G}_{\mbox{\scriptsize{co}}}$  & 1 \\
    \hline
    ${C}_{\mbox{\scriptsize{FS}}}$ & 3044 Gbps & ${C}_{\mbox{\scriptsize{FS}}}$  & 882 Gbps  & ${C}_{\mbox{\scriptsize{FS}}}$  & 10 Gbps  \\
    \hline 
    ${R}_{\mbox{\scriptsize{tr}}}$ & 100 Gbps & ${R}_{\mbox{\scriptsize{tr}}}$  & 100 Gbps & ${R}_{\mbox{\scriptsize{tr}}}$  & 10 Gbps \\
    \hline 
    ${N}_{\mbox{\scriptsize{W}}}$ & 40  & ${N}_{\mbox{\scriptsize{W}}}$ & 40 & ${N}_{\mbox{\scriptsize{W}}}$  & 80\\
    \hline 
    ${P}_{\mbox{\scriptsize{W}}}$ & 2.2 W  & ${P}_{\mbox{\scriptsize{W}}}$ & 2.2 W &${P}_{\mbox{\scriptsize{W}}}$  & 2.2 W\\
    \hline
    ${P}_{\mbox{\scriptsize{tr}}}$ & 4.5 W & ${P}_{\mbox{\scriptsize{tr}}}$  & 4.5 W  & ${P}_{\mbox{\scriptsize{tr}}}$  & 2 W \\
    \hline 
    \end{tabular}
\end{table}

\bigskip

To conclude this section,
let us also highlight that the research and industry communities have spent notable efforts to characterize the \ac{RAN} power consumption
---whether distributed or centralised---
while considering the most relevant architectures. 
However, this is still an open research field.
In particular, 
we argue that due to the large ecosystem of equipment vendors and their different implementation of solutions,
there is a need for further data and experimentally driven research to validate the proposed models and find their appropriate parameters on real \ac{3GPP} \ac{NR} equipment.
In addition, 
further research is also needed to characterise \ac{CA} and \ac{mMIMO} \ac{BS} power consumption in the presence of multiple configurations and more complex technologies, 
such as \ac{CoMP}, 
and the characterisation of the power consumption of sites where multiple \acp{RAT} co-exist.

%% file: sections/Metrics.tex
To evaluate the impact of energy efficiency mechanisms on energy savings, 
energy efficiency metrics are as important as the used power consumption models. 
These metrics must be comprehensive, reliable and widely accepted to allow comparisons.
In addition, they have to capture both the energy consumed by the system under investigation as well as the performance measured at network level 
(such as coverage, capacity, and delay).

To achieve these goals, 
the \ac{ETSI} Environmental Engineering technical committee, the \ac{ITU}-T Study Group 5 and the \ac{3GPP} Technical Specification Group \ac{RAN} have specified metrics to assess mobile network energy efficiency under different operating conditions. 
The contribution of these \acp{SDO} has mainly focused on global system performance,
while considering different demographic areas, load scenarios, and radio access technologies.
In contrast, 
the academic research community has contributed to this effort by proposing energy efficiency link-specific metrics, 
which enable a more tailored energy efficiency network optimisation~\cite{Zappone2014}.

\bigskip 

In the following, 
we will first overview the contributions from different \acp{SDO}, 
considering both interference- and noise-limited scenarios,
and then touch on energy-delay related metrics.
Subsequently,
we also describe alternative metrics proposed by the academic research community to drive energy efficiency optimisation problems, 
indicating the merits and the drawbacks of each one of them. 

\subsection{Energy Efficiency Metrics for Interference-Limited Networks}
\label{sec:DVEE}

As one of the main targets of \ac{5G} networks to enhance energy efficiency is to adapt the system capacity 
---and the associated power consumption--- 
to the network load, 
load-aware metrics are key for the next generation of green communication networks. 
In this context, 
\ac{ETSI} has defined the mobile network data energy efficiency metric, $\mbox{EE}_{\mbox{\scriptsize{DV}}}$ $[\mbox{bit/J}]$~\cite{ETSIES203228},
which is the ratio between the data volume, $\mbox{DV}$, delivered in the network and the network energy consumption, $\mbox{EC}$, observed during the time period required to deliver such data,
i.e.,
\begin{equation}
   \mbox{EE}_{\mbox{\scriptsize{DV}}}=\frac{\mbox{DV}}{\mbox{EC}},
    \label{EE_data}
\end{equation}
where $\mbox{EC}$ should be computed by integrating eq.~\eqref{eq:RANPC} over an observation period that includes distinct load levels\footnote{
the \ac{3GPP} recommends that the performance should be evaluated considering at least 3 load levels~\cite{3GPPTR38.913}.}. 
Note that the data volume, $\mbox{DV}$, includes both downlink and uplink traffic for both circuit switched services and packet switched services,
and that this metric can be used to characterise a single or multiple \acp{BS},
operating in urban and/or dense-urban areas, 
in which the network experiences highly variable loads,
i.e. interference-limited scenarios. 
However, it is not appropriate for scenarios where the traffic load is low. 
Specifically, 
since the energy efficiency metric, $\mbox{EE}_{\mbox{\scriptsize{DV}}}$, is not weighted according to the global reference, 
a small energy saving in the low energy consumption region in a low load scenario may comparatively lead to apparently large energy efficiency gains,
while a large energy saving in the high energy consumption region may comparatively result into limited energy efficiency gains. 

To characterise the energy efficiency,
while considering distinct deployment scenarios,
e.g., dense-urban, urban, sub-urban, rural, or deep rural areas,
\ac{ETSI} has extended the previous metric to the total energy efficiency metric, $\mbox{EE}_{\mbox{\scriptsize{Total}}}$ $[\mbox{bit/J}]$~\cite{ETSIES203228},
which can be defined as the weighted sum of the energy efficiency in each deployment scenario,
i.e.,
\begin{equation*}
    \mbox{EE}_{\mbox{\scriptsize{Total}}}=\frac{\sum\limits_m\mbox{PoPP}_m \cdot \mbox{EE}_{\mbox{\scriptsize{DV}}_m}}{\sum\limits_m \mbox{PoPP}_m},
\end{equation*}
where 
$m$ is the index of the $m$-th scenario, 
$\mbox{EE}_{\mbox{\scriptsize{DV}}_m}$ is the energy efficiency of the $m$-th scenario,
and $\mbox{PoPP}_m$ is the weight (percentage) representing the typicality of the $m$-th scenario in the network under test.
This metric shall be used to characterise the energy efficiency performance of a large scale network, 
e.g., the network of an operator in a specific country.

To jointly consider deployment scenarios and traffic loads, 
the \ac{3GPP} has further complemented the work from \ac{ETSI}, 
introducing the network energy efficiency metric, $\mbox{EE}_{\mbox{\scriptsize{global}}}$ $[\mbox{bit/J}]$~\cite{3GPPTR38.913},
which can be defined as the sum of the energy efficiencies in multiple deployment scenarios and under different traffic loads,
i.e.,
\begin{equation*}
    \mbox{EE}_{\mbox{\scriptsize{global}}}=\sum\limits_m b_m \cdot \mbox{EE}_{\mbox{\scriptsize{DS}}_m}=\sum\limits_m  \sum\limits_l b_m \cdot a_l \cdot \mbox{EE}_{\mbox{\scriptsize{DV}}_{m,l}},
\end{equation*}
where
$\mbox{EE}_{\mbox{\scriptsize{DV}}_l}$ is the energy efficiency of the network for an observed deployment scenario, $m$, and traffic load level, $l$,
while $b_m$ and $a_l$ are the weights for the corresponding deployment scenario, $m$, and traffic load level, $l$, respectively. 
The \ac{3GPP} recommends to compute $b_m$ considering the proportion amongst the different deployment scenarios in terms of 
\emph{i)} power consumption, \emph{ii)} traffic load, or \emph{iii)} connection density ~\cite{3GPPTTR21.8663}.
With respect to $a_l$,
and giving an example, 
if 10$\%$, 30$\%$, and 50$\%$ traffic load levels are investigated, 
the corresponding weights based on a daily traffic model could be 6/24, 10/24 and 8/24, respectively~\cite{3GPPTTR21.8663}. 
$\mbox{EE}_{\mbox{\scriptsize{global}}}$ shall be used to characterise the energy efficiency performance of a large scale network, 
considering the multiple hours of the day, during which the spatial distribution of the load in the network may significantly change due to the daily human habits.

Although the above presented metrics provide a useful indication on how the energy consumption scales with the increase of the data rate requirements, 
it does not give any information about actual economic costs. 
To address this challenge, 
the work in~\cite{Yan2017} has introduced the economical energy efficiency metric, E$^3$ $[\mbox{bit/J}]$, 
which is defined as the ratio of effective system throughput to the associated energy consumption, 
weighted by a cost coefficient,
i.e.,
\begin{equation*}
    \mbox{E}^3=\frac{\sum\limits_{k \in \mathcal K} \alpha_k R_k}{\sum\limits_{n \in \mathcal N} P_{\mbox{\scriptsize T}n}+P_{\mbox{\scriptsize 0}n}C_n},
\end{equation*}
where $\mathcal K$  and $\mathcal N$ are the sets of \acp{UE} and serving \acp{BS}, respectively, 
$\mbox{R}_k$ is the effective throughput perceived by the $k$-th \ac{UE}, 
$\alpha_k$ is the priority weight related to the $k$-th \ac{UE}, 
$\mbox{P}_{\mbox{\scriptsize 0}n}$ and $\mbox{P}_{\mbox{\scriptsize T}n}$ are the static and the load-dependent power consumption, 
respectively, 
and $\mbox{C}_{n}$ is the cost coefficient of the $n$-th \ac{BS}. 
Note that the cost coefficient, $\mbox{C}_{n}$, is calculated as the ratio of the corresponding device cost to a predefined benchmark cost, 
such that the metric, E$^3$, has the same unit as the energy efficiency.
In scenarios where multiple access, fronthaul, computing, and other mechanisms coexist to provide efficient services to the end-users, 
this economical energy efficiency metric, E$^3$, is able to discriminate amongst solutions, 
not only in terms of provided network throughput and energy consumption,
but also in terms of additional (deployment and operational) costs. 
This allows the analysis of advanced network deployment and resource management schemes.

\subsection{Energy Efficiency Metrics for Noise-Limited Networks}
\label{sec:CoEE}

To complement the energy efficiency metrics presented in the previous section and deal with scenarios with sustained low data volumes, 
in particular in rural or in deep rural areas, 
\ac{ETSI} also introduced the mobile network coverage energy efficiency metric, $\mbox{EE}_{\mbox{\scriptsize{MN,CoA}}}$ $[\mbox{m$^2$/J}]$~\cite{ETSIES203228},
which is the ratio between the area covered by the network, $\mbox{CoA}$, and the network energy consumption observed during one year, $\mbox{EC}$,
i.e.,
\begin{equation}
    \mbox{EE}_{\mbox{\scriptsize{CoA}}}=\frac{\mbox{CoA}}{\mbox{EC}}.
    \label{EE_cov}
\end{equation}
This metric shall be used to characterise the energy efficiency performance in rural areas, 
where coverage is the main objective, 
or in the context of low-power wide-area networks.

In addition to the above,
the \ac{3GPP} has extended the coverage energy efficiency metric in eq.~\eqref{EE_cov} to consider a global metric,
the mobile network data energy efficiency metric, $\mbox{EE}_{\mbox{\scriptsize{global, CoA}}}$~\cite{3GPPTTR32.972}, 
which can be used to analyse the energy efficiency
over distinct areas, each one having distinct size or relevance for the operator,
i.e.,
\begin{equation*}
    \mbox{EE}_{\mbox{\scriptsize{global, CoA}}}=\sum\limits_i C_i \cdot \mbox{EE}_{\mbox{\scriptsize{CoA,i}}},
\end{equation*}
where 
$C_i$ 
is the weight for each such distinct area/deployment scenarios, 
which is computed by taking into account the area covered per deployment scenario and the relevance of the scenario itself to the total power consumption, 
e.g., the percentage of power consumption in dense urban areas to that in rural areas.
The energy computation in the actual area under coverage 
---and how it is affected by a network technology---, 
however, can be complex to estimate, 
and it may require the collection of a large number of \ac{UE} measurement reports.

In this context, 
\ac{ETSI} has defined a coverage quality factor, $\mbox{CoA}{\mbox{\scriptsize{Q}}}$ $[\%]$~\cite{ETSIES203228}, to estimate the quality of the coverage and measure the amount of connection failures due to coverage issues, load congestion or significant interference effects, 
i.e.,
\begin{equation*}
    \mbox{CoA}{\mbox{\scriptsize{Q}}}=(1-\mbox{FR}_{\mbox{\scriptsize{RRC}}})(1-\mbox{FR}_{\mbox{\scriptsize{RAB}}_{\mbox{\scriptsize{S}}}})(1-\mbox{FR}_{\mbox{\scriptsize{RAB}}_{\mbox{\scriptsize{R}}}}),
\end{equation*}
where 
$\mbox{FR}_{\mbox{\scriptsize{RRC}}}$, $\mbox{FR}_{\mbox{\scriptsize{RAB}}_{\mbox{\scriptsize{S}}}}$, and $\mbox{FR}_{\mbox{\scriptsize{RAB}}_{\mbox{\scriptsize{R}}}}$ are the radio resource control setup failure ratio, the radio access bearer setup failure ratio, and radio access bearer release failure ratio, 
respectively. 

\subsection{Energy Efficiency of E2E Network Slicing}
\label{sec:QoSEE}

Network slicing is a promising \ac{5G} technology to simultaneously support multiple services with diverse
characteristics and requirements in a single \ac{5G} network. 
The \ac{3GPP} is currently focusing on three main families of services, i.e., \ac{eMBB}, \ac{URLLC}, and \ac{mMTC}. 
In Release~17,
the \ac{3GPP} has worked towards an approach to evaluate the energy efficiencies for these families of slices~\cite{3GPPTS28.554}. 
Importantly, a slice is typically defined end-to-end, including \ac{RAN}, \ac{5GC}, and transport network.\footnote{
However, these metrics can be used also for evaluating the performance of a slice defined only over one of these (sub)networks, 
by considering only the corresponding measurements.}

For \ac{eMBB} slices, 
the \ac{3GPP} has specified the usage of the metric, $\mbox{EE}_{\mbox{\scriptsize{DV}}}$, which we have defined in eq.~\eqref{EE_data} \cite{3GPPTS28.554}. 
More interestingly, 
for \ac{URLLC} and \ac{mMTC} slices, 
data volume is not the main \ac{KPI}, 
and the \ac{3GPP} has introduced more appropriate metrics to characterize energy efficiency in these cases.

For \ac{URLLC} slices, 
the \ac{3GPP}, 
in line with \ac{ETSI} efforts~\cite{ETSIES203228}, 
has defined the metric, $\mbox{EE}_{\mbox{\scriptsize{URLLC, Lat}}}$ $[\mbox{s$^{-1}$/J}]$~\cite{3GPPTS28.554}, 
which is the inverse of the average end-to-end latency of the network slice divided by the energy consumption of the network slice, 
i.e.,
\begin{equation*}
    \mbox{EE}_{\mbox{\scriptsize{URLLC, Lat}}}=\frac{1}{T_{\mbox{\scriptsize{e2e}}}\cdot \mbox{EC}},
\end{equation*} 
where 
$\mbox{T}_{\mbox{\scriptsize{e2e}}}$ is the overall system end-to-end latency.

In cases where latency and throughput are both important \acp{KPI} of the \ac{URLLC} slice, 
or when the operator wants to evaluate the slide energy efficiency over different periods of time with distinct loads, 
the \ac{3GPP} has specified the metric, $\mbox{EE}_{\mbox{\scriptsize{URLLC,DV,Lat}}}$ $[\mbox{bit/s/J}]$~\cite{3GPPTS28.554}. 
This metric is defined as the slice data volume divided by the product between its average end-to-end latency and its energy consumption, 
i.e.,
\begin{equation*}
    \mbox{EE}_{\mbox{\scriptsize{URLLC,DV,Lat}}}=\frac{\mbox{DV}}{T_{\mbox{\scriptsize{e2e}}}\cdot \mbox{EC}}.
\end{equation*}

Finally, for \ac{mMTC} slices, 
where the system main target is to provide service to a large number of mobile devices, 
the \ac{3GPP} has specified the metric, $\mbox{EE}_{\mbox{\scriptsize{mMTC}}}$~\cite{3GPPTS28.554}, 
which is defined as the ratio of the number of \acp{UE}\footnote{
The \ac{3GPP} has currently defined two variants of this metric, 
where the first one considers the maximum number of registered \acp{UE}, 
while the second one considers the average number of active \acp{UE}.} 
in the slice divided to the associated energy consumption:
\begin{equation*}
    \mbox{EE}_{\mbox{\scriptsize{mMTC}}}=\frac{N_{\mbox{\scriptsize{UE}}}}{\mbox{EC}}.
\end{equation*}


\subsection{Link-aware Energy Efficiency Metrics}
\label{sec:LAEE}

The energy efficiency metrics presented in Sections \ref{sec:DVEE},  \ref{sec:CoEE}, and \ref{sec:QoSEE} should be used to evaluate the impact of network-wide solutions specifically designed to increase the system energy efficiency (as those presented in Sections \ref{sec:TimeD}, \ref{sec:CarrierD}, and~\ref{sec:AntennaD}), 
and to compare these solutions with the current baseline systems. 
In contrast, the metrics presented in this section are more appropriate to make \textit{greener} the classic RRM schemes, 
such as beamforming or scheduling, 
which mainly target spectral efficiency.
Hence, they should be used to optimize specific mechanisms rather than to evaluate a system.
The way in which RRM schemes are optimized in terms of energy efficiency is not discussed in this survey, 
as this is well reviewed in previous manuscripts (see for instance \cite{Zappone2014}).  

Specifically, 
to aid a finer grain performance optimization,
the research community has extended the network energy efficiency metric, $\mbox{EE}_{\mbox{\scriptsize{DV}}}$, presented in eq.~\eqref{EE_data} by taking into consideration the different requirements of its distinct links. 
In fact, 
since the energy efficiency metric, $\mbox{EE}_{\mbox{\scriptsize{DV}}}$, can be seen as the aggregated sum of the energy efficiencies of each network link, 
its optimisation through radio resource management may tend to favour the links that can provide the largest throughput, 
which may limit, e.g., the cell-edge \acp{UE} performance,
and thus the network fairness.

To address this issue,
and denoting by $\mathcal L$ the set of links in the network, 
the \ac{WSEE} metric, WSEE $[\mbox{bit/J}]$~\cite{Zappone2014}, is defined as the weighted mean of the different energy efficiencies measured at each link $i \in \mathcal L$,
i.e.,
\begin{equation}
    \mbox{WSEE}=\sum\limits_{i \in \mathcal L} w_i \cdot \mbox{EE}_{\mbox{\scriptsize{DV,i}}},
    \label{eq:WSEE}
\end{equation}
where 
$\mbox{EE}_{\mbox{\scriptsize{DV,i}}}$ is the link energy efficiency, 
defined as in eq.~\eqref{EE_data}, 
and $w_i$ is the weight of $i$-th link. 
Importantly,
note that the use of different weights for different links enables assigning them different priorities during the resource allocation to increase the system fairness. Accordingly, the weights are defined a-priori, for instance based on the data plan of the specific user.

To enable an even more fair resource allocation, 
researchers have also proposed the \ac{WPEE} metric, WPEE $[\mbox{bit/J}]$~\cite{Zappone2014}, 
which is defined as the exponentially weighted product of the different energy efficiencies measured at each link $i \in \mathcal L$,
i.e.,
\begin{equation*}
    \mbox{WPEE}=\prod\limits_{i \in \mathcal L} \left( \mbox{EE}_{\mbox{\scriptsize{DV,i}}}\right)^{w_i }.
\end{equation*}
In particular, 
the \ac{WPEE} metric maximisation ensures that no link experiences a zero throughput, 
and has been shown to converge to the Nash Bargaining solution~\cite{peters1992axiomatic}.
Following this approach, however,
it is not possible to improve the energy efficiency of the $i$-th link, $\mbox{EE}_{\mbox{\scriptsize{DV,i}}}$, without decreasing the energy efficiency of other link, e.g. the $j$-th link, $\mbox{EE}_{\mbox{\scriptsize{DV,j}}}$.

To achieve a better trade-off among overall system performance and fairness, 
a max-min fair resource allocation policy can be adopted. 
An energy efficiency resource allocation scheme that is max-min fair can be designed 
by maximising the \ac{WMEE} metric, WMEE $[\mbox{bit/J}]$,
i.e.,
\begin{equation*}
    \mbox{WMEE}=\min\limits_{\left\{i \in \mathcal L \right\}} \left(w_i \cdot \mbox{EE}_{\mbox{\scriptsize{DV,i}}}\right).
\end{equation*}
Note that when the \ac{WMEE} metric is optimised, 
the resource allocation achieves the same product, $w_i \cdot \mbox{EE}_{\mbox{\scriptsize{DV,i}}}$, for all links $i \in \mathcal L$. 
Thus, if the weights are set equal, 
this max-min fair resource allocation will provide the same energy efficiency for each link.

\begin{figure}[t]
    \centering
    \includegraphics[width=9cm]{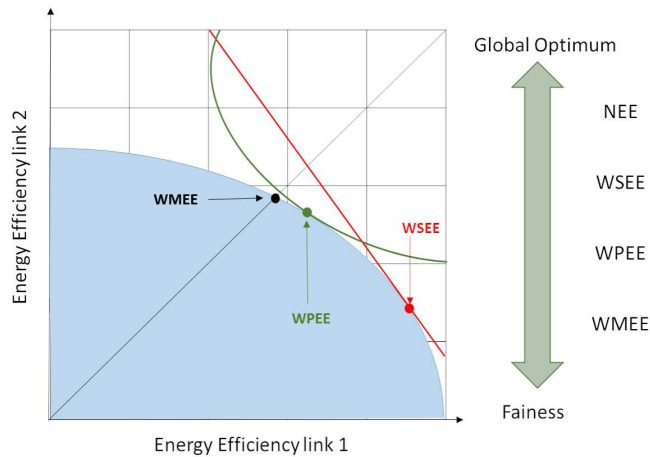}
    \caption{Operating regions of the energy efficient optimisation based on the \ac{NEE}, the \ac{WMEE}, the \ac{WPEE}, and the \ac{WSEE} metrics~\cite{Zappone2014},
    where \ac{NEE}$=\sum\limits_{i \in \mathcal L} \mbox{EE}_{\mbox{\scriptsize{DV,i}}}$.}
    \label{fig:Pareto}
\end{figure}

To facilitate the reader's understanding,
Fig.~\ref{fig:Pareto} provides a qualitative comparison of the performance achieved by a network with two links when optimising its energy efficiency using the \ac{WMEE}, the \ac{WSEE} and the \ac{WPEE} metrics with respect to the Pareto boundaries. 
Each of the three approaches leads to a solution that belongs to the energy efficient Pareto region.
However, the \ac{WSEE} and the \ac{WPEE}  metrics allocate more resources to the link~1, 
which is characterized by a better energy efficiency in order to get closer to the global optimum. 
In contrast, 
the \ac{WMEE} metric shares the resources between the two links such that they are characterized by the same energy efficiency. 

\bigskip

To conclude this section,
Table~\ref{tab: EE metrics} provides a summary of the most relevant energy efficiency metrics for \ac{5G} system optimisation,
where it is important highlighting that these metrics have been mainly designed for assessing \ac{4G} networks, 
focusing on \ac{eMBB} services, 
and considering only data rate, latency, and coverage requirement.
In contrast,
with respect to other \ac{5G} use cases,
e.g. \ac{mMTC} and \ac{URLLC},
there is a lack of specific and well understood metrics,
and further research is needed in this area. 
\ac{mMTC} applications call for energy efficiency metrics that takes into account the number of connection handled by the network in combination with the area covered by the network, 
e.g., an extension of the mobile network coverage energy efficiency metric, $\mbox{EE}_{\mbox{\scriptsize{MN,CoA}}}$.
Similarly, 
\ac{URLLC} applications require a metric able to capture the system reliability in combination with the end-to-end delay, 
e.g., an extension of the latency metric, $\mbox{EE}_{\mbox{\scriptsize{L}}}$. 

\begin{table}[!ht]
    \caption{Summary of EE metrics.}
    \label{tab: EE metrics}
    \scriptsize
    \centering
    \begin{tabular}{|c||c|c|c|c|} 
\hline
    Metric & Unit  &  Calculation & KPI& Application \\
    \hline 
    $\mbox{EE}_{\mbox{\tiny{DV}}}$ \cite{ETSIES203228} & $[\mbox{bit/J}]$ & $\frac{\mbox{DV}}{\mbox{EC}}$ & Load-aware &Evaluate network energy efficiency of eMBB systems,\\&&&& characterized by large load variations\\
    \hline 
    $\mbox{EE}_{\mbox{\tiny{Total}}}$ \cite{ETSIES203228} & $[\mbox{bit/J}]$ & $\frac{\sum\limits_m\mbox{PoPP}_m \cdot \mbox{EE}_{\mbox{\tiny{DV,m}}}}{\sum\limits_m \mbox{PoPP}_m}$ &Load-aware &Evaluate network energy efficiency of eMBB systems\\&&&& spanning across multiple areas\\
    \hline 
    $\mbox{EE}_{\mbox{\tiny{global}}}$ \cite{3GPPTR38.913} & $[\mbox{bit/J}]$ & $\sum\limits_m  \sum\limits_l b_m \cdot a_l \cdot \mbox{EE}_{\mbox{\tiny{DV}}_l}$ &Load-aware &Evaluate network energy efficiency of eMBB systems\\&&&&considering scenarios with distinct loads\\
    \hline 
    $    \mbox{EE}_{\mbox{\tiny{CoA}}}$ \cite{ETSIES203228} & $[\mbox{m$^2$/J}]$ & $\frac{\mbox{CoA}}{\mbox{EC}}$ &Coverage-aware &Evaluate network energy efficiency of rural areas\\
    \hline 
    $    \mbox{EE}_{\mbox{\tiny{global, CoA}}}$ \cite{3GPPTTR32.972} & $[\mbox{m$^2$/J}]$ & $\sum\limits_i C_i \cdot \mbox{EE}_{\mbox{\tiny{CoA,i}}}$ &Coverage-aware &Evaluate network energy efficiency of rural areas\\&&&& with distinct sizes or priorities\\
    \hline 
    E$^3$  \cite{Yan2017} & $[\mbox{bit/J}]$ & $\frac{\sum\limits_{k \in \mathcal K} \alpha_k R_k}{\sum\limits_{n \in \mathcal N} P_{Tn}+P_{0n}C_n}$ &Load \& cost-aware & Evaluate network energy efficiency with\\&&&& consideration on deployment costs\\
    \hline 
        $\mbox{EE}_{\mbox{\scriptsize{URLLC, Lat}}}$  \cite{3GPPTS28.554}  & $[\mbox{s$^{-1}$/J}]$ & $    \frac{1}{\mbox{T}_{\mbox{\scriptsize{e2e}}}\cdot \mbox{EC}}$ &Latency & Evaluate network energy efficiency for URLLC systems\\
    \hline 
        $\mbox{EE}_{\mbox{\scriptsize{URLLC,DV,Lat}}}$  \cite{3GPPTS28.554} & $[\mbox{bit/s/J}]$& $    \frac{\mbox{DV}}{\mbox{T}_{\mbox{\scriptsize{e2e}}}\cdot \mbox{EC}}$ &Latency \& load & Evaluate network energy efficiency for URLLC systems\\&&&& with consideration on distinct load scenarios\\
    \hline 
       $\mbox{EE}_{\mbox{\scriptsize{mMTC}}}$  \cite{3GPPTS28.554} &$[\mbox{UE/J}]$& $    \frac{\mbox{N}_{\mbox{\scriptsize{UE}}}}{\mbox{EC}}$ &Connectivity &Evaluate network energy efficiency of mMTC systems\\
    \hline 
    WSEE  \cite{Zappone2014} & $[\mbox{bit/J}]$ & $\sum\limits_{i \in \mathcal L} w_i \cdot \mbox{EE}_{\mbox{\tiny{DV,i}}}$ &Link \& Load-aware & Optimize energy efficient resource management\\&&&&considering link priorities\\
    \hline 
    WPEE  \cite{Zappone2014} & $[\mbox{bit/J}]$ & $\prod\limits_{i \in \mathcal L} \left( \mbox{EE}_{\mbox{\tiny{DV,i}}}\right)^{w_i }$ &Link \& Load-aware & Optimize energy efficient resource management\\&&&& preventing links with zero throughput\\
    \hline 
    WMEE  \cite{Zappone2014} & $[\mbox{bit/J}]$ & $\min\limits_{\left\{i \in \mathcal L \right\}} \left(w_i \cdot \mbox{EE}_{\mbox{\scriptsize{DV,i}}}\right)$ &Load-aware & Optimize energy efficient resource management\\&&&& with max-min fairness across radio links \\
    \hline 
    \end{tabular}
\end{table}

%% file: sections/mMIMO.tex
\ac{5G} networks are targeted at a 100 times higher energy efficiency with respect to previous generations
(see Fig.~\ref{fig:imtRequirements}), 
and \ac{mMIMO} has been identified as a key technology to reach such target. 
As discussed in previous sections,
by leveraging its extensive beamforming and spatial multiplexing capabilities, 
\ac{mMIMO} can significantly network capacity,
but also reduce the transmit power required at the \ac{BS} to achieve a targeted rate,
given a frequency band of operation and coverage area. 
However, running such larger number of antennas at the \ac{BS},
together with the more signal processing required to handle the larger capacity in a \ac{mMIMO} cell,
also increases the energy consumption of the \ac{BS}.

Multiple studies have set out to fundamentally understand the challenges of \ac{mMIMO} networks in general, 
and the above presented energy efficiency trade-off in particular. 
Mainly concentrating on providing insights for network design, 
a large body of research has focused on deriving theoretical bounds on the capacity and the energy consumption of \ac{mMIMO} systems, 
as well as the interplay between different network parameters. 
More practical research, 
on the other hand, 
has tackled \ac{mMIMO}-based network design, \ac{BS} deployment, as well as radio resource management and optimisation problems, 
while taking into account end-users' \ac{QoS} demands.
Extensive specification work has also be carry out in \ac{3GPP} \ac{NR} to accommodate \ac{mMIMO} capabilities in both the control- and user-planes.

One of the main challenges faced in \ac{mMIMO} is around the accuracy and the tractability of the models used to characterise performance at the network-level. 
\ac{mMIMO} capacity and rates, for instance, significantly degrade in the presence of channel correlation\footnote{
Spatial correlation means that there is a correlation between the received average signal gain and the angle of arrival of such signal.
Spatial correlation generally degrades the performance of multi-antenna systems,
as it decreases the number of independent channels that can be created by precoding.} 
and/or pilot contamination,
which is a function, among others, of the \ac{BS} and the \ac{UE} distributions, the scenario topology and the wireless channel, 
as well as of independent \ac{BS} scheduling decisions and the resulting interference.
However, this is hard to capture in tractable models.
Same issues also revolve around the accuracy and complexity of the power consumption models,
as discussed in Section~\ref{sec:PC_Models}. 

In light of these tractability issues, 
most theoretical findings on energy efficiency related to \ac{mMIMO} networks,
which are well grounded now, 
are limited to single-cell scenarios, 
where model simplifications can be more easily justified.
In contrast,
the estimation of energy efficiency metrics and trade-offs in multi-cell scenarios tends to be addressed through computer aided numerical evaluations. 

\ac{3GPP} \ac{NR} \ac{mMIMO} specification work and enhancements can be framed within the broader context of improved mobile broadband, 
and that such \ac{3GPP} work does not necessarily relate to energy efficiency \emph{per se},
but mostly to control signalling definition and related procedures. In constrast, 
in the following,
we concentrate on providing an overview of the fundamental understanding of energy efficiency in the context of \ac{mMIMO} systems,
in both single- and multi-cell scenarios,
and focus on green large-scale \ac{mMIMO} network deployment and optimization, where energy efficiency is defined as the ratio of the achievable data rate to the related power consumption [bit/Joule].\footnote{For more details on \ac{3GPP} \ac{NR} related \ac{mMIMO} specification,
the reader is referred to~\cite{8826541}, 
and references therein.} 


\subsection{Single-cell scenario}
\label{sec:mMIMO_single_cell}
\input{sections/mMIMO_single_cell}


\subsection{Multi-cell scenario}
\label{sec:mMIMO_multi_cell}
\input{sections/mMIMO_multi_cell}

%% file: sections/mMIMO_single_cell.tex
In this subsection, 
we focus
---and survey--- 
energy efficiency bounds and trade-offs when considering a single-cell \ac{mMIMO} scenario,
which are fundamentally well understood,
and already helping \acp{MNO} to design their networks.
Explicit closed-form expressions are provided, 
and the impact of different \ac{mMIMO} parameters into energy efficiency and power consumption are carefully analyzed.

\subsubsection{Bounds}
The pioneering work in~\cite{ngo2013energy,bjornson2014massive,hoydis2013massive} provided a first analysis of the energy efficiency in a single-cell \ac{mMIMO} system,
mostly based on the assumption of perfect \ac{CSI} being available at the \ac{BS}.
The research in~\cite{ngo2013energy} showed that the performance of a \ac{mMIMO} system with $M$ antennas at the \ac{BS} and a \ac{BS} transmit power, $P_\mathrm{TX}/M$, is equal to the performance of a \ac{SISO} system with a \ac{BS} transmit power, $P_\mathrm{TX}$, without any intra-cell interference.
This result indicated that, 
by using a large number, $M$, of \ac{BS} antennas 
the \ac{BS} transmit power, $P_\mathrm{TX}$, can be proportionally scaled down by a factor, $1/M$. 
This work also suggested that the spectral efficiency in a \ac{mMIMO} system can be increased by a factor, $K$, when serving $K$ \acp{UE} in the same time-frequency resource. 
The findings in~\cite{bjornson2014massive,hoydis2013massive} also resonated with these conclusions,
reporting that a power reduction proportional to $1/M$  can be achieved in \ac{TDD} systems\footnote{The power reduction is proportional to $1/\sqrt{M}$ in the case of imperfect \ac{CSI}.}, 
while maintaining non-zero rates, 
as the number, $M$, of antennas grows to infinity.
This was a promising result,
indicating that the energy efficiency of the system under study could be monotonically increased with the number, $M$, of antennas, 
without any trade-off.
However, it is important to mention that such conclusions were obtained with significant assumptions in the \ac{BS} power consumption model. 

Soon after, however,
further developments in this area of research showed that the energy efficiency bounds in \ac{mMIMO} systems are hidden behind more complex \ac{BS} power consumption models,
and that inaccuracies and/or oversimplifications, 
not reflecting the essence of \ac{mMIMO} hardware implementations, 
can lead to misleading practical insights. 
For example,
when not considering the circuit power consumption related to running a larger number of antennas,
as it was the case in~\cite{ngo2013energy,bjornson2014massive,hoydis2013massive}, 
one can be let to believe that an unbounded energy efficiency can be achieved by adding more and more antennas. 
Deploying more and more antennas, however, requires additional circuitry, 
incurring a larger power consumption,
which needs to be accurately captured by the analysis to draw the appropriate conclusions.
In this line,
the pioneering work in~\cite{bjornson2015optimal} stands out,
which considered a more sophisticated
---and realistic---
\ac{mMIMO} \ac{BS} power consumption model,
and in turn,
found significantly different conclusions to the previously stated ones.

In the following, 
we survey in detail this work,
which represent the most advanced 
---and settled---
understanding of energy efficiency in single-cell \ac{mMIMO} systems.

In more detail, 
the authors in~\cite{bjornson2015optimal},
using the more detailed \ac{BS} power consumption model already presented in eq.~\eqref{eq:linCPmMIMO},
analyzed for the first time in a systematic manner the total energy efficiency of the \ac{UL} and \ac{DL} in a single-cell \ac{mMIMO} network with respect to 
\begin{itemize}
    \item 
    the total transmit power, $P_\mathrm{TX}^\mathrm{tot}$, 
    accounting for the downlink and the uplink transmit powers, 
    \item
    the number, $K$, of simultaneously multiplexed \acp{UE}, and
    \item 
    the number, $M$, of antennas, 
    where $M\geq K+1$.
\end{itemize}
Remarkably,
closed-form expressions for the energy efficiency optimal operating point with respect to each of these parameters, 
when the others are fixed, 
were provided for the case where both \ac{ZF} processing and perfect \ac{CSI} knowledge are considered. 
In particular, 
the authors considered a \ac{TDD} system, 
in which the pilot signaling occupies $\tau^{(\mathrm{ul})}K$ and $\tau^{(\mathrm{dl})}K$ symbols in the uplink and downlink, 
respectively, 
where the inequality, $\tau^{(\mathrm{dl})},\tau^{(\mathrm{ul})}\geq 1$, must be true to enable orthogonal pilot sequences among \acp{UE},
and with such system model,
they provided a formulation of the gross downlink rate, 
expressed in bit per second,
as follows:
\begin{equation}
    \bar{R}=B\log(1+\rho (M-K)),
\end{equation}
where $\rho$ is a design parameter proportional to the received \ac{SNR}.
Following this formulation, 
the total transmit power, $P_\mathrm{TX}^\mathrm{tot}$, required to serve each \ac{UE} with a gross rate, $\bar{R}$, was shown to be:
\begin{equation}
    P_\mathrm{TX}^\mathrm{tot} = \frac{B\sigma^2 \rho \mathcal{S}_{\mathbf{x}}}{\eta}K,
\end{equation}
where 
$\sigma^2$ is the channel noise power, 
$\mathcal{S}_\mathbf{x}$ is a term that accounts for the \ac{UE} distribution and propagation environment,
and $\eta$ is a term that accounts for the efficiency of the \acp{PA}.

From the above formulation,
it is important to restate that the value of the design parameter, $\rho$, is proportional to the \ac{UE} \ac{SNR}, 
which in turn,
is directly proportional to the total transmit power, $P_\mathrm{TX}^\mathrm{tot}$, when considering \ac{ZF} processing. 
Thus, finding the optimal total transmit power, $P_\mathrm{TX}^\mathrm{tot*}$, which maximizes the joint \ac{UL} and \ac{DL} energy efficiency involves deriving the optimal design parameter, $\rho^*$.
After some manipulations,
the authors of in~\cite{bjornson2015optimal} found a lower bound for the optimal design parameter, $\rho^*$, 

which shows that the optimal total transmit power, $P_\mathrm{TX}^\mathrm{tot*}$, increases with the load-independent circuit power consumption, $P_\mathrm{CP}^\mathrm{li}$, the power required by all the circuit components of each single-antenna \ac{UE}, $\mathcal{C}_1$, and the power consumed by the transceiver module attached to each antenna, 
$\mathcal{D}_0$.
This result also shows that the optimal strategy to improve the energy efficiency is to increase the total transmit power, $P_\mathrm{TX}^\mathrm{tot}$, with the number, $M$, of antennas, 
not in an arbitrary manner, 
but while considering the effect of the circuit power consumption, $P_\mathrm{CP}$.

These conclusions are in stark contrast with those in~\cite{ngo2013energy,bjornson2014massive,hoydis2013massive},
which as depicted earlier,
concluded instead that 
the \ac{BS} transmit power, $P_\mathrm{TX}$, 
---the downlink part of the total transmit power, $P_\mathrm{TX}^\mathrm{tot}$--- 
monotonically decreases with the increase of the number, $M$, of antennas,
while the system remains energy efficient.
This implies a linear relationship between the energy efficiency and the number, $M$, of antennas. 
This optimistic finding only applies, however, in the idealistic case where the circuit power consumption, $P_\mathrm{CP}$, plays a negligible role in the overall power consumption.  


\smallskip

Moreover,
the study on the optimal number, $K^*$, of multiplexed \acp{UE}, 
provided in~\cite{bjornson2015optimal}, 
highlights that, 
if the power consumption for the beamforming processing, $P_\mathrm{SP}$, and channel estimation, $P_\mathrm{CE}$, are negligible, 
then the optimal number, $K^*$, of multiplexed \acp{UE} can be approximated as follows:
\begin{equation}
    K^* \approx \flcl*{\left( \mu \sqrt{1+\frac{U}{\left(\tau^{(\mathrm{ul})}+\tau^{(\mathrm{dl})}\right)\mu}} -1 \right)},
    \label{eq:optK}
\end{equation}
where $U$ is the channel coherence time (in symbols), 
and 
\begin{equation}
    \mu = \frac{
    P_\mathrm{CP}^\mathrm{li} + \frac{B \sigma^2 \mathcal{S}_\mathbf{x}}{\eta} \rho K}{\mathcal{C}_1+\bar{\beta}
    \mathcal{D}_0},
\end{equation}
where $\bar{\beta}$ is the ratio of the number, $M$, of antennas to the number, $K$, of multiplexed \acp{UE}, 
(i.e., $\bar{\beta} = M/K$), 

This bound indicates that 
the optimal number, $K^*$, of multiplexed \acp{UE} decreases with the power required by all the circuit components of each single-antenna \ac{UE}, $\mathcal{C}_1$, and the power consumed by the transceiver module attached to each antenna, $\mathcal{D}_0$,
whereas it increases with 
the load-independent circuit power consumption, $P_\mathrm{CP}^\mathrm{li}$,
the total transmit power $P_\mathrm{TX}^\mathrm{tot}$ (proportional to $\rho$), the noise power, $\sigma^2$, and the parameter, $\mathcal{S}_\mathbf{x}$. 
Note that this last term, $\mathcal{S}_\mathbf{x}$, increases proportionally with the coverage radius of the cell, 
meaning that a larger number, $K$, of multiplexed \acp{UE} must be served when the coverage area increases in order to maximize the energy efficiency. 

\smallskip

In addition,
the study on the optimal number, $M^*$, of antennas, 
provided in~\cite{bjornson2015optimal}, 
also indicates that such number, $M^*$, is lower bounded by:
\begin{equation}
    M^* \geq \frac{\frac{B\sigma^2\mathcal{S}_\mathbf{x}}{\eta\mathcal{D}'}\rho + \frac{\mathcal{C}'}{\mathcal{D}'} + K -\frac{1}{\rho}}{\ln(\rho)+\ln{\left( \frac{B\sigma^2\mathcal{S}_\mathbf{x}}{\eta\mathcal{D}'}\rho + \frac{\mathcal{C}'}{\mathcal{D}'} + K - \frac{1}{\rho}\right) - 1}} - \frac{1}{\rho}.
    \label{eq:lbM}
\end{equation}

This bound indicates that 
the optimal number, $M^*$, of antennas increases with the load-independent circuit power consumption, $P_\mathrm{CP}^\mathrm{li}$, and the power required by all the circuit components of each single-antenna \ac{UE}, $\mathcal{C}_1$, 
whereas it decreases with 
the power consumed by the transceiver module attached to each antenna, $\mathcal{D}_0$.

Importantly,
when the design parameter, $\rho$, becomes large,
the lower bound given in eq.~\eqref{eq:lbM} can be approximated as:
\begin{equation}
    M^* \approx \frac{B\sigma^2\mathcal{S}_x}{2\eta\mathcal{D}'}\frac{\rho}{\ln(\rho)},
\end{equation}
which shows that there is an almost linear scaling law of the optimal number, $M^*$, of antennas with respect to the design parameter, $\rho$, in its high value regime, 
and thus with respect to the total transmit power, $P_\mathrm{TX}^\mathrm{tot}$.

It should also be noted the linear dependence of the optimal number, $M^*$, of antennas with the \ac{UE} distribution and propagation environment, 
captured by the variable, $\mathcal{S}_\mathbf{x}$, 
which implies that a larger optimal number, $M^*$, of antennas is needed, 
as the size of the coverage area increases, 
since the variable, $\mathcal{S}_\mathbf{x}$, increases with the cell radius.

\smallskip

\begin{figure}
    \centering
    \includegraphics[scale=0.8]{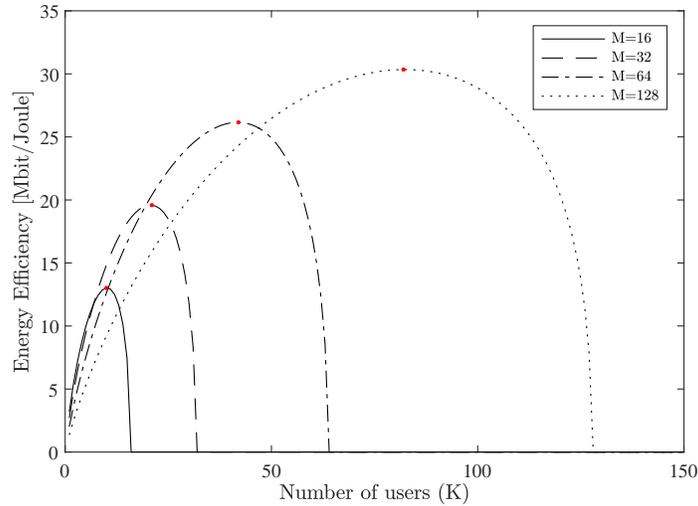}
    \caption{Energy efficiency with respect to the number, $K$, of multiplexed \acp{UE} for a given number, $M$, of antennas. 
    The maximum energy efficiency values are highlighted in red.}
    \label{fig:bounds_EE_K}
\end{figure}

For the sake of clarity,
in Fig.~\ref{fig:bounds_EE_K},
we illustrate the achievable energy efficiency for four different values of the number, $M$, of antennas, 
when varying the number, $K$, of multiplexed \acp{UE}. 
Note that this energy efficiency curves are characterized by a concave shape in all the considered antenna configurations, 
and that the maximum achievable values are highlighted in red. 
This plot highlights that the optimal number, $K^*$, of multiplexed \acp{UE} increases sub-linearly with the number, $M$, of antennas.

\begin{figure}
    \centering
    \includegraphics[scale=0.8]{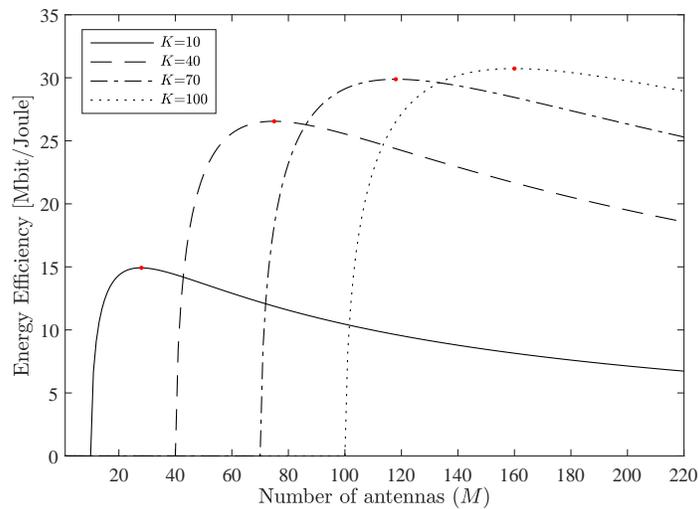}
    \caption{Energy efficiency with respect to the number, $M$, of antennas for a given number, $K$ of multiplexed \acp{UE}. 
    The maximum energy efficiency values are highlighted in red.}
    \label{fig:bounds_EE_M}
\end{figure}

In a similar way, 
in Fig.~\ref{fig:bounds_EE_M},
we illustrate the achievable energy efficiency for different values of the number, $K$, of multiplexed \acp{UE}, 
when varying the number, $M$, of antennas. 
The energy efficiency curves have, 
in this case, 
a quasi-concave shape, 
and the maximum achievable values are also highlighted in red. 
These results confirm that, 
for a given value of the number, $K$, of multiplexed \acp{UE}, 
augmenting the number, $M$, of antennas increases the energy efficiency up to a maximum energy efficiency value,
where the \ac{UE} rate gain due to the further increasing the number, $M$, of antennas is not sufficient anymore to counterbalance the cost incurred by their associated power power consumption.

Finally,
it should be noted from Fig.~\ref{fig:bounds_EE_K} and Fig.~\ref{fig:bounds_EE_M} that deploying hundreds of antennas to serve a large number of \acp{UE} is the optimal solution from an energy efficiency perspective,
confirming the energy efficiency-enabler role of \ac{mMIMO}.
These results are well establish and at the forefront of the state-of-the-art.  

\subsubsection{Trade-offs}

The spectral efficiency, 
defined as the system throughput per unit of bandwidth [bit/s/Hz],
has been historically adopted as the key optimisation metric to optimise mobile network deployments. 
As energy efficiency has become a new important \ac{KPI} to the operation of \ac{5G} networks, 
the relation between these two metrics is thus of fundamental importance. 
How much spectral efficiency is traded to realize an energy efficiency gain? 

Early works in the literature, 
not accounting for the circuit power consumption, $P_\mathrm{CP}$, in the \ac{BS} power consumption model, 
concluded that a 10x spectral efficiency improvement could be achieved for a given \ac{BS} transmit power, $P_\mathrm{TX}$, 
when adopting hundreds of antennas at the \ac{BS}~\cite{Marzetta2010}.  
In this idealized scenario,
since the energy efficiency grows proportionally to the number, $M$, of antennas, 
the corresponding energy-spectral efficiency curve is pushed outwards, 
meaning that both metrics can be simultaneously maximized.

As shown earlier,
however,
the circuit power consumption, $P_\mathrm{CP}$, linearly grows with the number, $M$, of antennas, 
and when this number is large, 
it dominates the \ac{BS} overall power consumption, 
implying that the energy efficiency cannot be enhanced, 
unless the efficiency of the \ac{BS} hardware is improved accordingly~\cite{bjornson2015optimal}.
In this way, 
the circuit power consumption, $P_\mathrm{CP}$, breaks the monotonic relation between the energy and spectral efficiency, 
and makes these two metrics not consistent and conflicting with each other. 
For this reason, 
their trade-off must be carefully analyzed, 
and proper network optimisation techniques need to be developed to strike the right balance between these two metrics.

Studies on such energy and spectral trade-off are often carried out based on optimisation problems,
aiming, e.g., at maximizing the energy efficiency given a spectral efficiency requirement.
However, 
in some more sophisticated approaches, 
the balance between the energy and spectral efficiency is achieved by maximizing the \ac{RE} [bit/Joule]~\cite{tang2014resource},
which is defined as the weighted sum of the energy and spectral efficiency, 
and the weights assigned to the two terms allows to explore the trade-off.
It is important to highlight, however, that all such frameworks do not generally provide explicit equations for the energy-spectral efficiency trade-off.
Instead, 
they mostly build on the top of results obtained from tailored optimisation algorithms aimed at solving such problem,
which usually turns out to be intricate~\cite{hao2016energy,huang2018spectral,you2020spectral,liu2016energy}.

In this line,
the fundamental trade-off between the energy and spectral efficiency has been carefully studied in~\cite{liu2016energy}, 
when considering a single-cell \ac{mMIMO} system with linear precoding, 
i.e. \ac{ZF} and \ac{MRT}, 
and transmit antenna selection. 
In particular, 
the \ac{BS} transmit power, $P_\mathrm{TX}$, and the number, $M_a<M$, of active antennas are jointly considered as a resource to balance the energy and spectral efficiency,
and the adopted \ac{BS} power consumption model is equivalent to the one introduced in eq.~\eqref{eq:linCPmMIMO}.
Note, however, that this research did not take into account the impact of the number, $K$, of  multiplexed \acp{UE}, 
which may have a significant influence to the trade-off. 
Nonetheless, 
this study has importantly shown that different numbers, $M_a$, of active antennas lead to different energy efficiency-spectral efficiency curves. 
In more details, 
for a given number, $M_a$, of active antennas,
the energy efficiency is a quasi-concave function with respect to the spectral efficiency, 
which confirms the existence of a clear trade-off between these two metrics. 
Their numerical results in this work have also shown that, 
in the low \ac{SNR} region,
which corresponds to a low spectral efficiency, 
\ac{MRT} achieves a higher energy efficiency than \ac{ZF} due to its lower complexity.
In contrast, 
in the high spectral efficiency regime, 
\ac{ZF} outperforms \ac{MRT} in terms of spectral efficiency, 
for a given energy efficiency, 
owing to its ability of canceling intra-cell interference.

As discussed earlier,
most theoretical \ac{mMIMO} works in literature assume spatially uncorrelated channels and perfect \ac{CSI} knowledge due to tractability reasons,
as in the aforementioned study. 
However, \ac{CSI} acquisition is challenging, 
especially when dealing with the large antenna arrays in the \ac{mMIMO} case. 
As explained in Section~\ref{sec:energyEfficiencyMetrics:mMIMO},
in \ac{TDD} systems, 
the acquisition of downlink \ac{CSI} can be facilitated via uplink training by taking advantage of the channel reciprocity. 
However, even in this case, 
the \ac{CSI} may still be inaccurate due to practical hardware limitations, 
such as calibration errors in the transceivers~\cite{choi2014downlink}, 
and in high mobility scenarios, 
it can quickly become outdated.

Tackling this challenge,
the authors in~\cite{you2020spectral,huang2018spectral} have explored this problem,
and provided analyses of the energy and spectral efficiency trade-off,
considering statistical \ac{CSI} knowledge,
instead of an instantaneous one.
This type of feedback,
at the expense of a lower network capacity,
has the advantage of being stable during longer time periods and to be more easily obtainable by the \ac{BS} through long-term feedback or covariance extrapolation.

\begin{figure}
    \centering
    \includegraphics[scale=0.8]{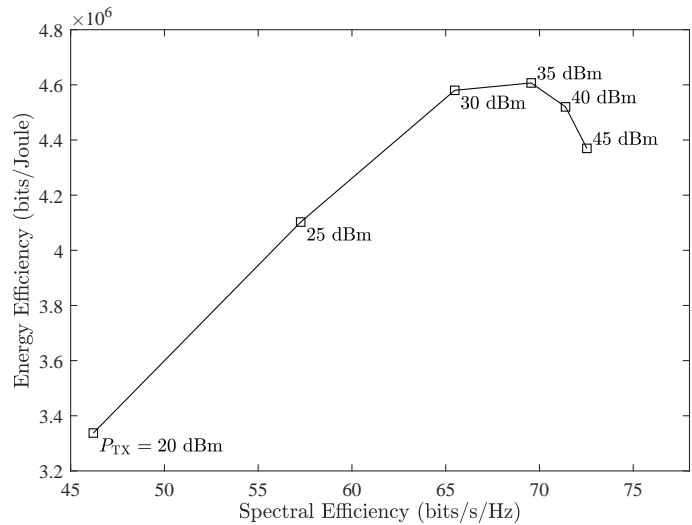}
    \caption{Trade-off between energy and spectral efficiency for different values of the \ac{BS} transmit power, $P_\mathrm{TX}$}
    \label{fig:single_tradeoff}
\end{figure}

Particularly, 
the work in~\cite{you2020spectral} has analysed the single-cell \ac{mMIMO} downlink case, 
where 
one \ac{BS} with $M$ antennas simultaneously transmits signals to $K$ \acp{UE}, 
the channel spatial correlation is captured using a jointly correlated Rayleigh fading model~\cite{gao2009statistical}, 
statistical \ac{CSI} is available at the transmitter,
and the considered \ac{BS} power consumption model accounts for 
the \ac{BS} transmit power, $P_\mathrm{TX}$, and the static circuit power consumption, $P_\mathrm{CP}$.
The energy-spectral efficiency trade-off has been investigated by maximizing the \ac{RE} metric~\cite{tang2014resource}, 
which strikes for a energy-spectral efficiency balance. 

Fig.~\ref{fig:single_tradeoff} shows the derived trade-off when considering different values of the \ac{BS} transmit power, $P_\mathrm{TX}$. 
Note that the energy-spectral efficiency curve is characterized by a concave shape. 
In particular, 
the energy and spectral efficiency can be jointly augmented by increasing the \ac{BS} transmit power, $P_\mathrm{TX}$, until reaching, in this case, the optimal energy-spectral efficiency point with a \ac{BS} transmit power, $P_\mathrm{TX}=35$~dBm.
After reaching such optimal point, 
which corresponds to the maximum achievable values of both energy and spectral efficiency,
the \ac{BS} transmit power, $P_\mathrm{TX}$, becomes the dominating consuming factor. 
Thus, increasing the \ac{BS} transmit power, $P_\mathrm{TX}$, allows to increase the spectral efficiency only at the expense of a reduced energy efficiency.

\bigskip

Finally, 
before concluding this section,
a summary of the main works on single-cell energy efficiency bound and trade-offs is presented in Table \ref{tab:single-cell energy efficiency bound}.

\begin{table}[!ht]
\caption{Summary of single-cell energy efficiency bound and trade-offs in the literature}
\label{tab:single-cell energy efficiency bound}
\scriptsize
\centering
  \begin{tabular}{|c||c|c|c|c|c|} 
  \hline
  Paper & Type & KPI & Parameters  \\
  \hline
  \cite{bjornson2018energy} & bound & EE & bandwidth, \ac{BS} transmit power \\ \hline
  \cite{bjornson2015optimal} & bound & EE & transmitting antennas, multiplexed \acp{UE}, \ac{BS} transmit power \\ \hline
  \cite{liu2016energy} & trade-off & EE-SE & transmitting antennas, \ac{BS} transmit power  \\ \hline
  \cite{you2020spectral} & trade-off & RE & \ac{BS} transmit power \\  \hline
  \cite{huang2018spectral} & trade-off & RE & \ac{BS} transmit power  \\
  \hline
  \end{tabular}
\end{table}

\bigskip

As concluding remark,
it should also be noted that,
to enhance the energy efficiency of \ac{mMIMO} systems,
large research efforts have also being spent on the understanding and optimization of both practical precoding and \ac{UE} scheduling techniques,
while considering single-cell \ac{mMIMO} scenarios.
The optimal precoding to achieve optimal energy efficiency under ideal and known channel conditions has been derived in~\cite{Belmega2011}.
Optimal energy efficient precoding schemes, 
while considering imperfect \ac{CSI} at the transmitter, 
have been studied in~\cite{Varma2013}.
\ac{UE} scheduling algorithms that take channel orthogonality into account,
avoiding to schedule two \emph{nearby} \acp{UE} in the same time-frequency resource,
have been proposed in~\cite{Liu2012,Mao2012,Zhang2013,Li2018}.
Other more experimental approaches considered to increase energy efficiency have been modulation diversity~\cite{Lee2010}, cognitive radios~\cite{He2011}, and spatial modulation~\cite{Renzo2014}. 

%% file: sections/mMIMO_multi_cell.tex
In this subsection, 
we focus
---and survey--- 
the latest most representative developments on the understating of the energy efficiency in large-scale multi-cell \ac{mMIMO} networks,
introducing the most relevant tools used to carry the analyses out and differentiating among the uplink and downlink case.

\begin{figure}
    \centering
    \includegraphics[scale=0.8]{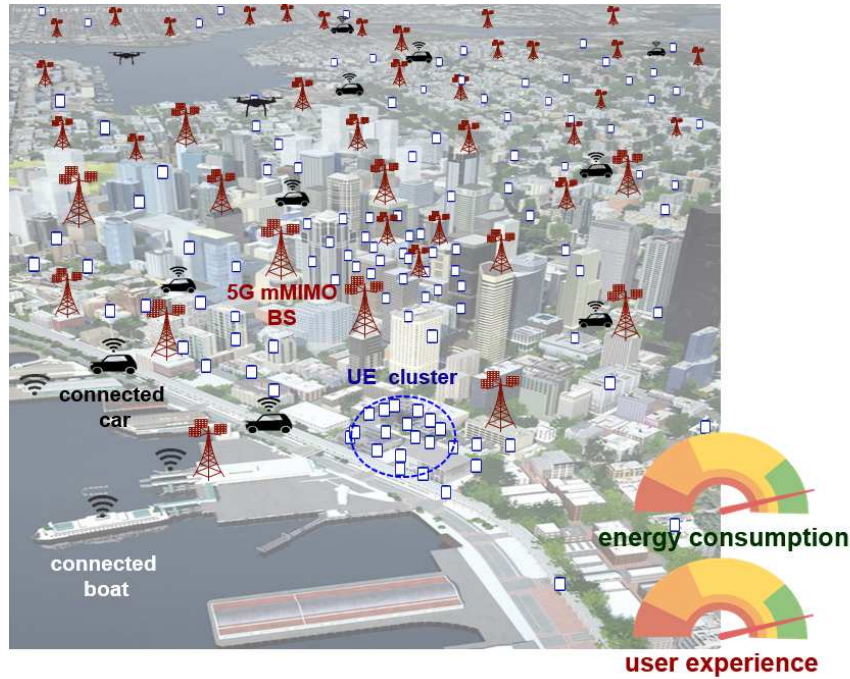}
    \caption{Example of a large-scale multi-cell \ac{mMIMO} network.}
    \label{fig:mMIMOlargescale}
\end{figure}

In contrast to the single-cell \ac{mMIMO} scenario,
it should be noted, however, that the multi-cell \ac{mMIMO} one is not so well theoretically understood as of today,
due to its much higher complexity.
Indeed,
large-scale networks are in general hard to model and analyse 
(see Fig.~\ref{fig:mMIMOlargescale}),
due to their
\begin{itemize}
    \item complex topology,
    \item evolving \ac{UE} distributions,
    \item dynamic traffic demands,
    \item fluctuating wireless channels,
    \item sophisticated protocols and algorithms, and
    \item large number of parameters to tune. 
\end{itemize}
This poses a challenge on their theoretical comprehension.
In fact,
there are no available explicit ---and general--- closed-form expressions that describe the energy efficiency bounds and trade-offs of a multi-cell \ac{mMIMO} network in a holistic manner.
Instead, 
the research in this area is scattered and focused on particular aspects of the system,
which together with some of the assumptions taken, 
have helped to increase the tractability of the problem, 
and derive some initial insights on the energy efficiency problem.
Further research on the topic is needed.


\subsubsection{Available Mathematical Tools}

Given that the comprehension of the power consumption of a \ac{mMIMO} \ac{BS} has reached some degree of maturity
(see Sections~\ref{sec:PC_Models} and~\ref{sec:mMIMO_single_cell}), 
one of the main difficulties to unlock a fundamental understanding of the energy efficiency in multi-cell \ac{mMIMO} networks is the derivation of its capacity.
Stochastic geometry~\cite{Book_Haenggi2012},
the state-of-the-art tool for the theoretical analysis of multi-cell networks,
particularly small cell ones~\cite{Jeff2011,Dhillon2012,Bai2015,Ming2016},
has been recently started to be used to address this issue.

Embracing the randomness of today's deployments,
the work in~\cite{Bai2016} has
analysed the uplink \ac{mMIMO} performance,  
considering a large-scale multi-cell network and the pilot contamination problem. 
This paper has presented a \ac{mMIMO} network capacity scaling law as a function of the number of antennas and multiplexed \acp{UE} per \ac{mMIMO} \ac{BS}, 
and investigated the performance gains attainable through a practical fractional uplink pilot reuse\footnote{
Fractional uplink pilot reuse refers to a class of schemes where cell-centre \acp{UE} of every cell reuse the available set of uplink pilots for \ac{CSI} estimation, 
while cell-edge ones can only reuse a subset of them, 
with this subset being orthogonal in neighboring or more coupled cells.},
used to mitigate pilot contamination.
It is important to note, however, that significant assumptions were taken for tractability reasons.
The study assumed that there was a large number of \acp{UE} in each \ac{BS} at all times, 
and thus that all \acp{BS} in the network were active and that all uplink pilots available were in use in all \acp{BS}.
This makes it difficult to study off-peak hours where energy savings are more likely to be harvested. 
Moreover, a pure \ac{NLoS} path loss model with Rayleigh multi-path was considered, 
which cannot capture the important effect of channel correlation on \ac{mMIMO} performance,
Unfortunately, 
even though these important assumptions and simplifications were made,
the resulting expressions are still not tractable, 
requiring few folds of integrals,
making difficult to infer relationships among system parameters. 
This is particularly true when capacity expressions need to be coupled with complex \ac{BS} power consumption models,
as those presented in Section~\ref{sec:PC_Models},
to derive the energy efficiency.

Using similar stochastic geometry tools,
the authors in~\cite{Zhang2017} studied the downlink \ac{mMIMO} performance,
while considering a heterogeneous network comprised of \ac{mMIMO} macrocells and single-antenna small cells,
together with the effect of \ac{LoS} and \ac{NLoS} transmissions,  pilot contamination, and cross-tier interference 
(i.e., interference from macro to small cells and vice versa). 
However, 
similar as in the previous case,
a fully loaded network is considered with all uplink pilots at use, 
which significantly affects the pilot contamination,
and does not allow to analyse low and medium traffic load scenarios.
In addition, important assumptions also revolve around the use of Rayleigh fading for the \ac{LoS} transmissions. 
Some of these issues have been recently addressed by the research in~\cite{8632721},
where pilot allocation schemes are considered with the objective of reducing the pilot contamination.
However, the expressions are still too complex to infer parameter relationships without a numerical evaluation. 

\subsubsection{Uplink \ac{mMIMO} network deployment perspectives}

Aware of such complexities,
the authors in~\cite{Bjornson2016} proposed a tractable stochastic geometry-based analysis of the energy efficiency of a \ac{TDD} multi-cell \ac{mMIMO} network subject to end-users' \ac{QoS} demands, 
while using a simplified but yet complete system model.
In more detail,
the authors targeted at maximizing the uplink energy efficiency of such multi-cell \ac{mMIMO} network, 
while ensuring a minimum uplink spectral efficiency to the average \acp{UE},
and used this formulation to obtain some practical insights into the \ac{mMIMO} energy efficiency problem. 
The adopted \ac{HPPP}-based system model\footnote{
It should be also noted that this general system model allows to compare \ac{mMIMO} setups, with few \acp{BS} and many antennas per \ac{BS}, to small cell ones, with many \acp{BS} and few antennas per \acp{BS},
thus providing guidance on the design of future green wireless networks, 
from the uplink perspective.}
accounted for
\begin{itemize}
    \item 
    the \ac{BS} density, $\lambda$,
    \item
	the number, $M$, of antennas per \ac{BS},
	\item
    the number, $K$, of multiplexed \acp{UE} per \ac{TTI} at each \ac{BS}, 
    \item
    an idealised uplink fractional power control,
    \item
    imperfect channel estimation through pilot contamination\footnote{
    Rayleigh fading was assumed,
    and thus the effect of spatial correlation was ignored in this analysis.},
    \item 
    hardware impairments, 
    modelled as a reduction of the signal power by a factor, $1-\epsilon^2$,
    \item
    \ac{MRC} received filters at the \ac{BS},
    \item
    single-antenna \acp{UE},
    and 
    \item
    an uplink pilot reuse scheme.
\end{itemize}
Importantly,
the complexity of calculating the average spectral efficiency of the network as the sum of the spectral efficiencies of all \acp{UE} through this framework was acknowledged,
indicating the need for heavy numerical evaluations of integrals.
To address this issue,
and obtain explicit expressions,
the authors focused on the performance of the average \ac{UE} instead,
and derived the following tractable 
---but yet tight--- 
lower bound for its uplink \ac{SINR}:
\small
\begin{equation}
    {\rm SINR_{UL}} = \frac{M(1-\epsilon^2)^2}
    {\big( K + \frac{\sigma^2}{\zeta}\big) \big( 1+\frac{2}{\beta(\alpha-2)}+\frac{\sigma^2}{\zeta}\big)
    + \frac{2K}{\alpha-2}\big(1 + \frac{\sigma^2}{\zeta}\big)
    + \frac{K}{\beta}\big(\frac{4}{(\alpha-2)^2} + \frac{1}{\alpha-1} \big)
    + M (1-\epsilon^2)\big( \frac{1}{\beta(\alpha-1)} + \epsilon^2\big)},
\end{equation}
\normalsize
where
$\alpha$ is the path loss exponent,
$\beta$ is the pilot reuse factor,
$\zeta$ is the path loss compensation power control coefficient,  
and $\sigma^2$ is the noise power.

Leveraging this expression, 
the authors formulated the uplink average \ac{UE} spectral efficiency and the resulting uplink \ac{ASE}, 
and with that, 
they derived the uplink \ac{APC} of the multi-cell \ac{mMIMO} network, 
using a linear version of the \ac{BS} power consumption model presented in~\eqref{eq:mMIMOPC},
i.e., 
\begin{equation}
    {\rm APC_{UL}} = \lambda \Bigg( \bigg( 1-\frac{\beta K -1 }{S}\bigg) \frac{\zeta \omega}{\eta} \frac{\Gamma(\frac{\alpha}{2}+1)}{(\pi\lambda)^(\frac{\alpha}{2})} K 
    + P_{\rm CP}^{\rm li} + {\cal C}_1K + {\cal D}_0M + {\cal D}_1MK \Bigg) 
    + {\cal A}\cdot{\rm ASE},
    \label{eq:mMIMOPC2}
\end{equation}
where
$S$ is coherent block length,
which is related to the channel coherence time, $U$, in eq.~\eqref{eq:optK},
$\eta$ is the \ac{PA} power efficiency, 
$\Gamma()$ is the Gamma function,
and ${\rm ASE}$ is the area spectral efficiency.

With these formulations, 
the authors defined an optimisation problem to find the most uplink energy efficient network deployment, 
while provisioning the average \ac{UE} with a minimum \ac{SINR}.  
Through some mathematical manipulations detailed in~\cite{Bjornson2016},
the authors derived  
\begin{itemize}
    \item 
    the uplink spectral efficiency feasibility region,
    \item
    the optimal uplink pilot reuse factor, $\beta^*$,
    \item
    the optimal \ac{BS} density, $\lambda^*$,
    \item
    the optimal number, $K^*$, of multiplexed \acp{UE} per \ac{TTI} at each \ac{BS},
    subject to a given ratio, $\frac{M}{K}$, of the number, $M$,  of antennas to the number, $K$, of multiplexed \acp{UE},
    and
    \item
    the optimal number, $M^*$, of antennas per \ac{BS},
    subject to a given number, $K$, of multiplexed \acp{UE},
\end{itemize}
and demonstrated that reducing the cell size 
---increasing the \ac{BS} density--- 
is beneficial for energy efficiency, 
but that such positive effect saturates when the circuit power consumption dominates over the transmission power. 
Their results also showed that adding more antennas in a controlled manner to the \ac{BS} to bring it to the \ac{mMIMO} regime also enhances the energy efficiency\footnote{
These conclusion are inline with that presented in the analysis of the single-cell \ac{mMIMO} case in the previous section,
indicating that the optimal strategy to improve the energy efficiency is to increase the total transmit power, $P_\mathrm{TX}^\mathrm{tot}$, with the number, $M$, of antennas, 
not in an arbitrary manner, 
but while considering the circuit power consumption, $P_\mathrm{CP}$.}.
In more detail,
their numerical examples on a typical scenario resulted in the maximum energy efficiency when having 91 antennas and 10 \acp{UE} multiplexed per \ac{BS}, 
which resembles a \ac{mMIMO} 
---and not a small cell---
setup. 
It should be noted that the energy efficiency gains, 
in this case, 
mostly came from the intra-cell interference suppression provided by \ac{mMIMO}, 
and by sharing the circuit power costs among the various multiplexed \acp{UE} in the same time-frequency resource. 
Moreover, the analysis showed that a large pilot reuse factor can be used to protect the network against inter-cell interference, 
and that it can be tailored to guarantee a certain average \ac{UE} spectral efficiency.

\bigskip 

\begin{figure}
   \centering
   \subfloat[\ac{AEE} and \ac{ASE} trade-off w.r.t. the number of antennas, $M$.]
   {\includegraphics[scale=0.6]{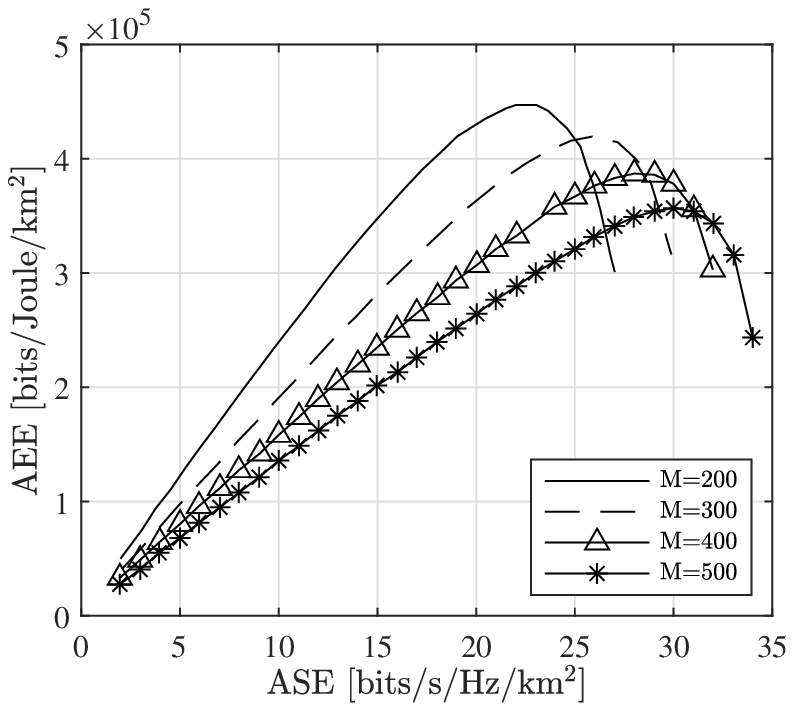}
   \label{fig:multicellA}}
   \vspace{0.5cm}
   \subfloat[\ac{AEE} and \ac{ASE} trade-off w.r.t. the number of multiplexed \acp{UE}, $K$.]
   {\includegraphics[scale=0.6]{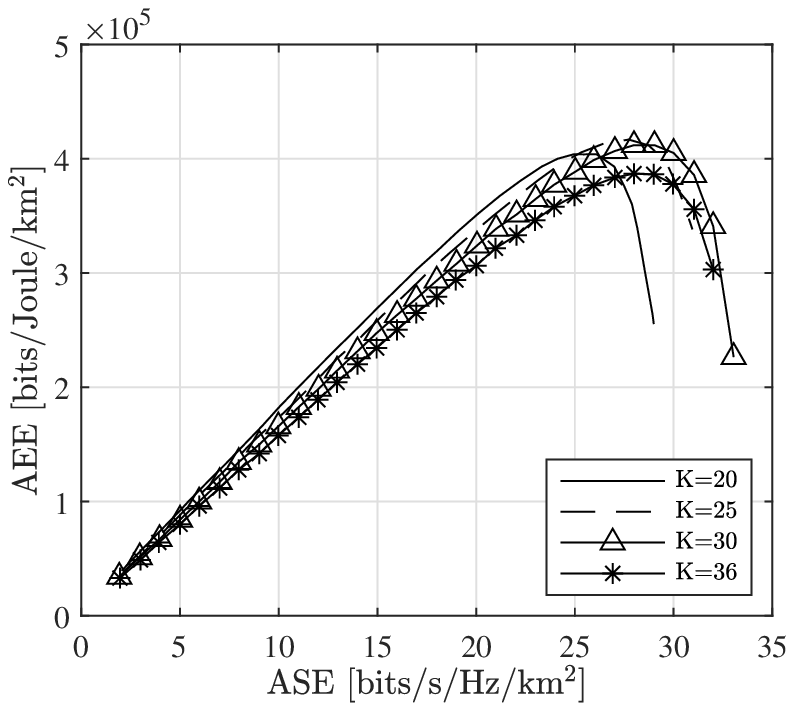}
   \label{fig:multicellB}}
   \vspace{0.5cm}
   \subfloat[\ac{AEE} and \ac{ASE} trade-off w.r.t.the pilot-to-data power ratio, $\beta$.]
   {\includegraphics[scale=0.6]{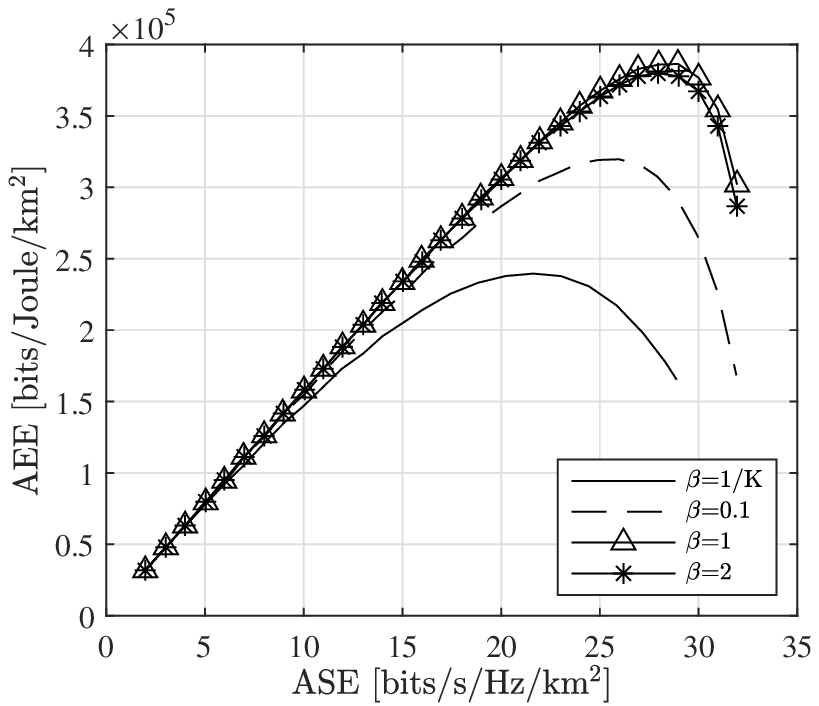}
   \label{fig:multicellC}}
   \caption{Multi-cell \ac{AEE} and \ac{ASE} trade-offs}
\end{figure}

Using a different modelling tool than stochastic geometry, 
based on numerical evaluations,
but also focusing on the uplink energy efficiency,
the authors in~\cite{Xin2016} provided an analysis of the \ac{AEE} and the \ac{ASE} trade-off for different system parameters,
such as the pilot reuse ratio and number of antennas and multiplexed \acp{UE} per \ac{BS}. 
Importantly, 
even if this work does not focus on minimising power consumption, 
but maximising the \ac{AEE},
it reaches similar general conclusions than those of~\cite{Bjornson2016}.
The multi-cell \ac{mMIMO} network always performs better in terms of \ac{ASE} with an increasing number of antennas. 
However, adding more antennas or multiplexed \acp{UE} might not achieve the optimal \ac{AEE}.
The reason is that the power consumed by the transceiver module attached to each antenna and the related signal detection and processing have a non-negligible effect on the total \ac{BS} power consumption.
In more detail,
their results have shown that,
when considering smaller \ac{ASE} targets,
a lower number of  antennas (see Fig.~\ref{fig:multicellA}) and multiplexed \acp{UE} (see Fig.~\ref{fig:multicellB}) at the \ac{BS} suffices to achieve the optimal \ac{AEE}.
However,
more antennas and multiplexed \acp{UE} are required to satisfy higher \ac{ASE} requirements,
which makes the network less energy efficient,
resulting in a smaller \ac{AEE}, 
as soon as the transmit power dominates the circuit and processing power consumption. 
Finally,
in this work,
the authors have also derived the pilot-to-data power ratio that maximizes the \ac{AEE},
and studied how such parameters affects the \ac{AEE} and the \ac{ASE} trade-off (see Fig.~\ref{fig:multicellC}).
Moreover, they show that in relevant scenarios, 
the optimal number of multiplexed \acp{UE} for maximizing the \ac{AEE} is much smaller than half of the coherent clock length, 
as it is for maximizing the \ac{ASE}~\cite{Marzetta2010}.

\subsubsection{Downlink \ac{mMIMO} network deployment perspectives}

Using a similar stochastic geometry approach and \ac{BS} power consumption model as in~\cite{Bjornson2016}, 
the authors in~\cite{Bjornson2015} extended the previous work to the downlink case,
aiming at optimising the downlink energy efficiency of a \ac{TDD} multi-cell \ac{mMIMO} network,
while providing a minimum spectral efficiency to the average \ac{UE}.
It is important to note that perfect \ac{CSI} was assumed in this case due to the complexity of modelling in the same framework both the channel estimation phase in the uplink and the data transmission phase in the downlink\footnote{
Such downlink dependency on the uplink is the main reason why the theoretical downlink energy efficiency of \ac{mMIMO} networks has been less rigorous studied.}.
As a result,
the effect of pilot contamination was neglected,
decreasing the accuracy of the model.
\ac{ZF} precoders at the \ac{BS} and single-antenna \acp{UE} were considered.
Importantly, 
it should be noted that embracing the same methodology as in~\cite{Bjornson2016}, 
the authors also focused their analysis on the performance of the average \ac{UE},
and derived the following tractable lower bound for its downlink \ac{SINR}:
\begin{equation}
    {\rm SINR_{DL}} = \frac{(1-\epsilon^2)(M-K)}
    {\frac{2K}{(\alpha-2)} 
    + \epsilon^2(M-K) 
    + \frac{\Gamma(\alpha/2+1)}{(\pi\lambda^2)} \frac{\omega \sigma^2}{\rho}},
\end{equation}
where 
$\rho$ now represents the downlink transmit power allocated by the \ac{BS} to the average \ac{UE}.  

With such expression, 
and using the corresponding downlink \ac{APC} model,
shown in the following for the sake of clarity:
\begin{equation}
     {\rm APC_{DL}} = \lambda \bigg(\frac{K\rho}{\eta} + P_{\rm CP}^{\rm li} + {\cal C}_1K + {\cal D}_0M + {\cal D}_1MK \bigg) + {\cal A}\cdot{\rm ASE},
\end{equation}
the authors derived, 
following similar steps as in the uplink counterpart work,
the following optimal operation points:
\begin{itemize}
    \item
    the optimal downlink transmit power per \ac{UE}, $\rho^*$
    \item 
    the optimal \ac{BS} density, $\lambda^*$,
    \item
    the optimal number, $K^*$, of multiplexed \acp{UE} per \ac{TTI} at each \ac{BS}, 
    subject to a given ratio, $\frac{M}{K}$, of the number, $M$,  of antennas to the number, $K$, of multiplexed \acp{UE},
    and
    \item
    the optimal number, $M^*$, of antennas per \ac{BS},
    subject to a given number, $K$, of multiplexed \acp{UE}.
\end{itemize}

The analysis of the obtained closed-form expressions showed that the same conclusions obtained from the uplink analysis apply to the downlink one.
The optimal energy efficiency is achieved by a \ac{mMIMO}-like deployment, 
where in their particular example, 
the optimum number of antennas and multiplexed \acp{UE} per BS are 193 and 21, 
respectively. 
Note that the results for the downlink case in this section and those for the uplink one in the previous section differ due to the different \ac{UE} requirements and the different power consumption of a \ac{BS} and a \ac{UE} while transmitting.

\bigskip

Finally, 
before concluding this section,
a summary of the main works on multi-cell energy efficiency modelling, bound derivation, optimization and trade-offs is presented in Table \ref{tab:multi-cell energy efficiency bound}.

\begin{table}[!ht]
\caption{Summary of multi-cell energy efficiency bound and trade-offs in the literature}
\label{tab:multi-cell energy efficiency bound}
\scriptsize
\centering
  \begin{tabular}{|c||c|c|c|c|c|} 
  \hline
  Paper & Type & KPI & Model features  \\
  \hline
  \cite{Bai2016} & uplink/stochastic geometry (SG) & coverage probability/ASE & macrocells, pilot contamination \\ \hline
  \cite{Zhang2017} & downlink/SG  & coverage probability/ASE & HetNet, pilot contamination, LoS/NLoS \\ \hline
  \cite{Bjornson2016} & uplink/SG and optimization & APC/EE & macrocells, pilot contamination, hardware impairments  \\ \hline
  \cite{Xin2016} & downlink/analytical model/trade-off & AEE-ASE &  macrocells, pilot contamination, pilot reuse \\  \hline
  \cite{Bjornson2015} & downlink/SG and optimization & APC/EE & macrocells, hardware impairments   \\
  \hline
  \end{tabular}
\end{table}

\bigskip 

As a concluding remark,
it is also important to highlight that, 
the presented uplink and downlink results in this section advocate for \ac{mMIMO} deployments 
---and their dimensioning optimisation---,
according to end-users' \ac{QoS} demands,
as an important tool to increase energy efficiency in \ac{5G} networks.
Using an inadequate number of \ac{BS} or antennas per \ac{BS} to meet a given end-users' \ac{QoS} demands can result in highlight suboptimal energy efficiency performances. 
This is an important area of research were more work is needed. 

Moreover, it should me mentioned that,
despite the lack of work on a holistic \ac{mMIMO} deployment dimensioning and operation optimization,
there is a large body of work around \ac{mMIMO} power control optimization for energy efficiency maximization in multi-cell \ac{mMIMO} networks.
In this line, 
the following research stands out. 
In~\cite{Zappone2014}, 
systematic approaches to solve energy efficiency maximization problems are extensively discussed. 
In this regard, 
the framework presented in~\cite{Zappone2016} has provided network- and \ac{UE}-centric downlink power control algorithms,
where minimum rate constraints are imposed and the \ac{SINR} takes a general form,
able to deal with complex \ac{mMIMO} systems/configurations.
Centralised algorithms are also developed,
which are guaranteed to converge, 
with affordable computational complexity, 
to a Karush–Kuhn–Tucker point of the considered non-convex optimisation problem.
Building on such framework,
the work in~\cite{Zappone2017} has proposed a framework to compute suboptimal power control strategies with even more affordable complexity. 
This is achieved by jointly using fractional programming and sequential optimisation. 
Numerical evidence has shown that such sequential fractional programming framework achieves global optimality in several practical communication scenarios.

%% file: sections/ON_OFF_symbol.tex
As motivated in Section~\ref{sec:energyEfficiencyEnabler},
future wireless communication systems require efficient hardware and mechanisms that enable the adaptation of the network functionalities and parameters to the load variations in an on-line manner in order to avoid excessive energy consumption,
while ensuring the end-users' \ac{QoS} demands. 
These enabling technologies can be classified, for example, by observing the time-scale and the domain in which they operate, 
e.g., time, frequency, or spatial (antenna) domains. 

In this section,
we concentrate on time-domain mechanisms,
also known as symbol shutdown.
These schemes refer to those solutions that are targeted at harvesting energy savings by considering the short-term traffic load variations in a cell.
They operate in the time scale of hundreds of microseconds to hundreds of milliseconds, 
taking advantage of the \ac{3GPP} \ac{NR} lean carrier design, 
to rapidly (de)activate hardware components of the \ac{BS} according to the presence or absence of traffic. 
Note that such short-term traffic load variations are a function of several factors, 
such as the number of active \acp{UE}, the traffic types, and the interference,
and that to realise such energy savings,
fast reacting \ac{BS} hardware and radio resource management mechanisms, 
which operate in the \ac{3GPP} \ac{NR} \ac{OFDM} or slot time scale, 
are thus required to dynamically minimum energy consumption,
while avoiding any \ac{UE} performance degradation.
Given that such radio resource management mechanisms may operate at an \ac{OFDM} symbol time scale,
and that they need to guarantee cell service continuity, 
it is important to note that they must be designed and operated embracing the \ac{3GPP} \ac{NR} signalling framework provided by the lean carrier design, 
to make sure they do not violate any control-plane protocol.

In the following, 
we overview the \ac{3GPP} \ac{NR}  lean carrier design capabilities, 
and present state of the art symbol shutdown mechanisms and frameworks for their optimization. 

\begin{figure}
    \centering
    \includegraphics[scale=0.5]{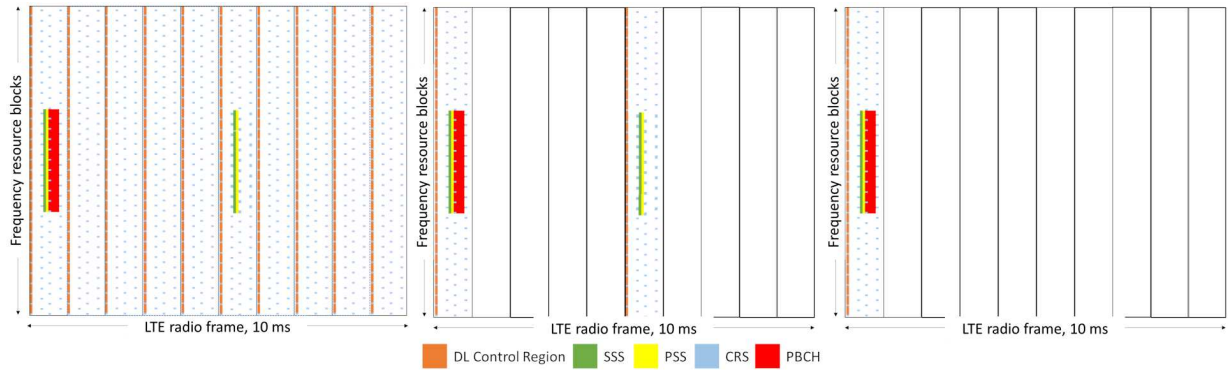}
    \caption{LTE frame for (left) unicast cell; (middle) Release~14 \ac{MBSFN}-dedicated cell; (right) Release~14 \ac{FeMBMS}/unicast-mixed cell.}
    \label{fig:LTE_frame}
\end{figure}

In \ac{3GPP} \ac{LTE}, 
cell \ac{DTX} can be used for deactivating \ac{BS} hardware components,
such as the \ac{PA}, 
when transmissions are absent in a given frame~\cite{Frenger2011}. 
The benefits of cell \ac{DTX} are, however, limited by the control signalling required by \ac{3GPP} \ac{LTE} to drive \ac{UE} cell camping procedures and others,
even in the absence of traffic. 
In more detail,
the \ac{3GPP} \ac{LTE} frame lasts 10~ms, 
and it is composed by 10 sub-frames, 
each one including 14 \ac{OFDM} symbols with a duration of 71.4~\mbox{$\mathrm{\mu}$}s. 
In the unicast mode, 
\acp{CRS} are transmitted in every sub-frame, 
\acp{PSS} and \acp{SSS} are transmitted every 5~ms, 
and \acp{BCH} are repeated in every first sub-frame of a frame 
(see the left side of Fig.~\ref{fig:LTE_frame}).
Therefore, the \ac{BS} is only able to sleep for a few \ac{OFDM} symbols, 
and needs to wake-up one \ac{OFDM} symbol before the transmission of any control signal. 
In this case, 
cell \ac{DTX} can achieve at most a 33$\%$ sleep ratio at zero load~\cite{Frenger2019}.

To allow longer sleep periods in \ac{3GPP} \ac{LTE}, 
the use of the multicast-broadcast single-frequency network (\ac{MBSFN}) frame was proposed, 
which is a technology introduced to enable mobile television broadcasting,
characterized by the need of less frequent signalling. 
To standardise such technology,
the \ac{3GPP} evaluated in~\cite{3GPPTR37.910} both the sleep ratio (at symbol level and sub-frame level) and the sleep duration, 
when using 
\begin{itemize}
    \item 
    \ac{3GPP} \ac{LTE} Release~14 \ac{FeMBMS}/unicast-mixed cells, 
    in which two out of ten sub-frames, 
    i.e. sub-frames 0 and 5, 
    contain unicast control signaling
    (see the central plot in Fig.~\ref{fig:LTE_frame}),
    and 
    \item
    \ac{3GPP} \ac{LTE} Release~14 \ac{MBSFN}-dedicated cells, 
    where only one non-\ac{MBSFN} sub-frame with standard unicast signaling is transmitted every 40~ms (see the right side of Fig.~\ref{fig:LTE_frame}).
\end{itemize}
The results from the \ac{3GPP} studies showed that, 
in the \ac{FeMBMS}/unicast-mixed mode, 
a \ac{BS} could stay in energy saving mode for up to 4~ms, 
which leads to an 80$\%$ sleep ratio. 
Importantly, 
with the \ac{MBSFN}-dedicated mode, 
a \ac{BS} could sleep even further up to 39~ms, 
which results into a 93.75$\%$ sleep ratio.

\begin{figure}
    \centering
    \includegraphics[scale=0.7]{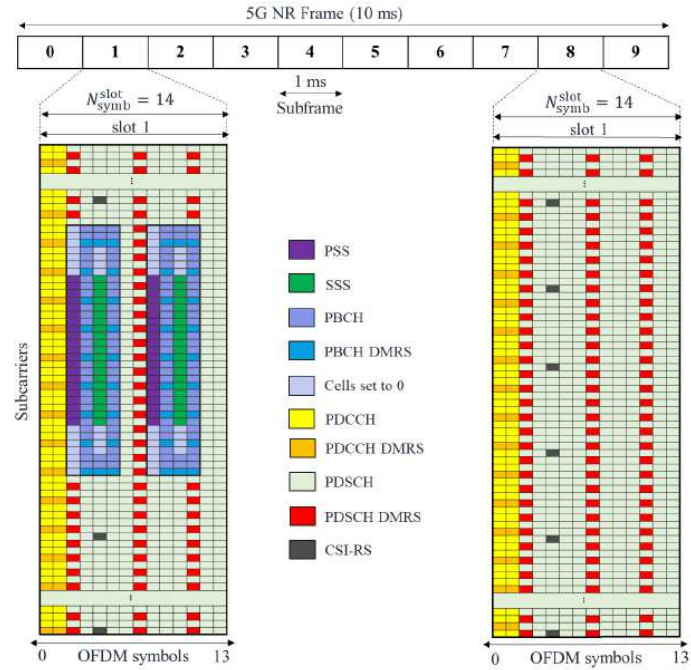}
    \caption{\ac{NR} frame structure example \cite{Fuentes2020}.}
    \label{fig:NR_frame}
\end{figure}

As discussed in Section~\ref{sec:energyEfficiencyMetrics:leanCarrier}, 
in contrast to \ac{3GPP} \ac{LTE}, 
\ac{3GPP} \ac{NR} is characterized by a user/data-specific signalling instead of a cell-specific one
---the lean carrier 
(see Fig.~\ref{fig:NR_frame}). 
Specifically:  
\begin{itemize}
    \item 
    The \ac{CRS} is not used anymore in \ac{3GPP} \ac{NR}, 
    and a \ac{SS} burst set, 
    including one or multiple \acp{SSB}
    ---each one of them in turn comprised of \ac{PSS}, \ac{SSS}, and \ac{BCH}---
    is transmitted to support \ac{UE} cell (re)selection and handover procedures with a larger periodicity, 
    i.e., 5, 10, 20, 40, 80, and 160\,ms~\cite{Book_Dahlman2018}. 
    \item 
    Since the \ac{CRS} is not used,
    \ac{CSI} acquisition procedures have also been redesigned,
    and on demand \acp{CSI-RS} are reused 
    ---and further extended---
    to provide support for beam and mobility management as a complement to the \ac{SSB}~\cite{Book_Dahlman2018}.
    \item
    The minimum required \ac{SI} broadcast in \ac{3GPP} \ac{NR} has also been reduced with respect to \ac{3GPP} \ac{LTE},
    and the part not strictly necessary for network entry is transmitted on-demand now~\cite{Yang2019}.
    \item
    Moreover, 
    in \ac{3GPP} \ac{NR}, 
    a single-antenna port can be used to transmit the mandatory control signals, 
    while, 
    in \ac{3GPP} \ac{LTE}, 
    all the antenna ports are used to transmit such mandatory control signals~\cite{Frenger2019}.
\end{itemize}

This lean carrier design enables both larger sleep ratios and longer sleeping duration, 
whose particular values depend on the specific system numerology, 
i.e., subcarrier spacing, the number of \acp{SSB} per \ac{SS} burst set, and the \ac{SS} burst set periodicity.
For example,  
when considering a subcarrier spacing of 15~kHz and two \ac{SS} blocks per \ac{SS} burst set,
a \ac{3GPP} \ac{NR} cell can stay in sleep mode up to 19\,ms, 
which leads to a sleeping ratio of 95\% for a \ac{SS} burst set periodicity of 20~ms. 
Importantly,
when the \ac{SS} burst set periodicity is maximised to 160\,ms, 
a  \ac{3GPP} \ac{NR} cell  can stay in sleep mode up to 159~ms,  
which leads to a sleeping ratio of 99.38\%~\cite{3GPPTR37.910}.

As introduced earlier,
the longer sleeping duration enabled by the lean carrier design also allows for deeper sleeps,
where more \ac{BS} hardware components can be switched off. 
In this line,
multiple \ac{BS} sleep states have been defined, 
where each sleep mode is associated to a given sleeping duration ~\cite{Debaillie2015}~\cite{Salem2019}.
As a rule of thumb,
all the \ac{BS} hardware components that can enter and exit a sleep mode fast enough with respect to the sleep mode duration can be easily deactivated in such period, 
while the components having a longer latency need to remain active.
In more detail, 
and to give an example according to \ac{3GPP} discussions~\cite{Salem2019}: 
\begin{itemize}
    \item 
    In sleep mode~1 (i.e., cell \ac{DTX}~\cite{Frenger2011}), 
    characterized by a duration (deactivation plus reactivation time) of 71\,\mbox{$\mathrm{\mu}$}s, 
    the \ac{PA} and some components of the digital \ac{BBU} and the analog \ac{FE} can be deactivated.
    \item 
    In sleep mode~2, 
    which has a minimum duration of 1\,ms, 
    additional components of the analog \ac{FE} can be switched off.
    \item 
    In sleep mode~3, 
    which has a minimum duration of 10\,ms, 
    the \ac{BS} can additionally deactivate all the digital \ac{BBU} processing, and almost all the analog \ac{FE} 
    (except the clock generator).
    \item 
    Finally, 
    in sleep mode~4, 
    which has a minimum duration of 1\,s, 
    only the wake-up functionalities are maintained.
\end{itemize}
Interested readers should note that the complete list of \ac{BS} elements that can be switched off for each sleep mode can be found in~\cite{salem:tel-02500618}.

\begin{figure}
    \centering
    \includegraphics[scale=0.5]{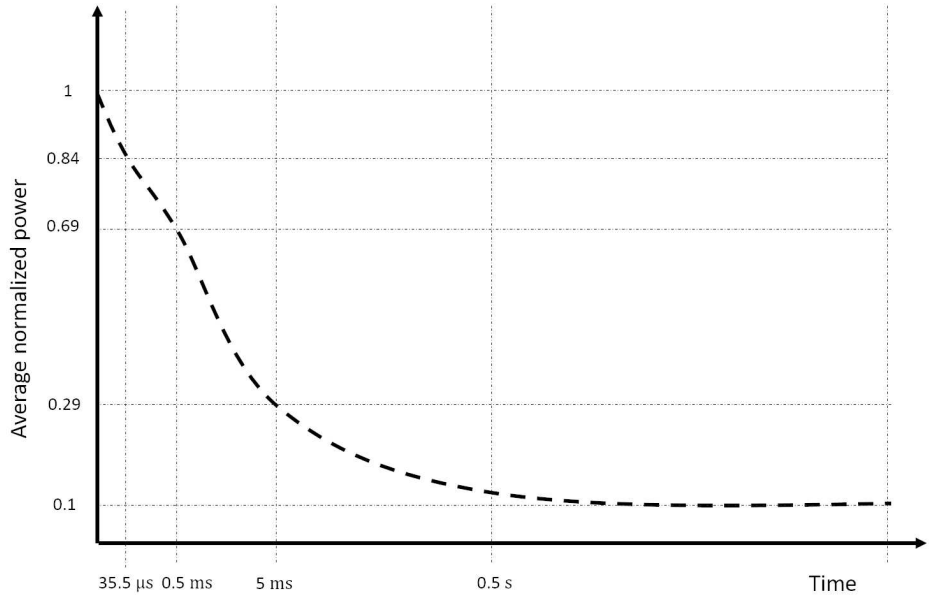}
    \caption{Power consumption trend as the \ac{BS} enters in successive sleep modes~\cite{Debaillie2015}. 
    Activation time is assumed to be equal to the half of the minimum sleep duration~\cite{Salem2019}.}
    \label{fig:ASM}
\end{figure}

In periods with absence of traffic,
a \ac{BS} can go through the presented sleep modes subsequently to reduce its energy consumption.
This process is often referred as \ac{ASM} 
(see Fig.~\ref{fig:ASM}). 
When the cell load rises and \ac{UE} traffic appears, 
such \ac{UE} traffic is buffered, 
and the \ac{BS} has to immediately switch on its functionalities, 
and serve the required data to satisfy the end-users' \ac{QoS} demands. 
However,
it should be noted that,
since hardware activation and deactivation times are not negligible, 
and their lengths increase with the number of involved \ac{BS} hardware components~\cite{Salem2019}, 
\acp{ASM} may increase the \ac{UE} perceived latency, 
and this can accordingly affect the overall \ac{UE} performance~\cite{Tano2019}. 
Therefore, 
there is a need for optimising the `path' through the different sleep modes, 
and proactively activate the \ac{BS} hardware components before traffic arrives in order to limit performance losses.

To assess the positive impact of different \acp{ASM} and their optimization on energy efficiency, 
a new \ac{BS} power consumption model was proposed in~\cite{Tombaz2016},
i.e.,
\begin{equation}
\mbox{P}_{\mbox{\scriptsize{BS}}}= 
\begin{array}{l}
\begin{cases}
\mbox{P}_{\mbox{\scriptsize{TX}}}+\mbox{P}_{\mbox{\scriptsize{CP}}},  &\mbox{if } \mbox{P}_{\mbox{\scriptsize{TX}}}>0, \\
  \delta_{\mbox{\scriptsize{s}}}\mbox{P}_{\mbox{\scriptsize{CP}}}^{\mbox{\scriptsize{li}}}, & \mbox{if } \mbox{P}_{\mbox{\scriptsize{TX}}}=0 \mbox{ and sleep mode is active}, 
  \end{cases} \label{eq:mMIMOPC_DTX}
    \end{array} 
\end{equation}
where $\mbox{P}_{\mbox{\scriptsize{TX}}}$,  $\mbox{P}_{\mbox{\scriptsize{CP}}}$, and $\mbox{P}_{\mbox{\scriptsize{CP}}}^{\mbox{\scriptsize{li}}}$ are the transmit power, the circuit power, and the load independent part of the circuit power consumption, 
respectively, 
and $\delta_{\mbox{\scriptsize{s}}}$ is the fraction of the load-independent circuit power consumption, $\mbox{P}_{\mbox{\scriptsize{CP}}}^{\mbox{\scriptsize{li}}}$, required by the cell in sleep mode. 
In this case,
the authors assume that the fraction, $\delta_{\mbox{\scriptsize{s}}}$, is equal to 0.84, 0.69, and 0.29 for sleep mode~1, sleep mode~2, and sleep mode~3, 
respectively, 
while for sleep mode~4, 
the fraction, $\delta_{\mbox{\scriptsize{s}}}$, can be lower than 0.1. 
It is important to note that current \ac{3GPP} \ac{NR} implementation cannot realise this last mode of operation,
as it requires 1\,s of continuous sleeping period,
and \ac{3GPP} \ac{NR} can only do up to 160\,ms. 
 
Embracing this framework,
the work in \cite{Tano2019} has proposed the use of dedicated timers to control when to deactivate components and go into deeper sleep modes.
The authors have highlighted that these timers should depend on the type of traffic carried out by each single \ac{BS},
and to make a more flexible usage of the sleep modes, 
they have designed a \ac{RL} model,
based on Q-Learning, 
to optimise the duration of each sleep state.
Energy savings and experienced delay are balanced using this technique,
using as enabler the average packet inter-arrival time.
Importantly,
their results have shown that it is possible to implement sleep modes and achieve significant energy savings,
even with stringent delay constraints, 
for low to medium traffic load scenarios.
In more detail,
up to 80$\%$ energy savings can be obtained when replacing \ac{3GPP} \ac{LTE} with \ac{3GPP} \ac{NR} technology and using the proposed sleep modes.
Nevertheless, 
it should be noted that this type of scheme relies on a continuous exchange of network signalling,
which may impact, 
e.g., the performance of cell (re)selection processes~\cite{3GPPRP-190326}.
More work is thus needed in this direction to understand how sleep modes impact the network load distribution, the resulting inter-cell interference, the related \ac{RRM} procedures,
and finally the end-users' \ac{QoS}.

To finalise this section,
let us highlight that to further enhance the performance of sleep modes,
avoiding the large access delay for \acp{UE} due to the time incurred by some \ac{BS} hardware components to power up,
the latest proposals in \ac{3GPP} \ac{NR} Release~18 suggest the use of some potential assisting information sent from the \ac{UE} to the \ac{BS} to setup \ac{ASM} and transmission parameters and maximize the power saving without greatly degrading the end-users' \ac{QoS}.
Such information may include
---but it is not limited to---
latency requirements. 
Multi-carrier operation schemes where one carrier can transmit the control signalling 
---or a lighter version of it--- 
of another sleeping carrier due to sleep mode operation have also be proposed to avoid \ac{UE} performance degradation due to transients~\cite{RWS-210447}.

Table \ref{tab:time-domain} provides a summary of the main contributions of the time-domain energy saving schemes discussed in this section.
\begin{table}[!ht]
\caption{Summary of main time-domain energy saving solutions.}
\label{tab:time-domain}
\scriptsize
\centering
  \begin{tabular}{|c||c|c|} 
  \hline
  Paper & Focus &Main contribution/idea \\
  \hline
  ~\cite{Frenger2011} & Cell DTX & Enable cell shutdown if there \\ & &  is no data in a given frame  \\ \hline
  ~\cite{3GPPTR37.910} & Compare EE in NR and LTE & Evaluate sleep ratio and sleep duration in NR and LTE  \\ \hline
  ~\cite{Debaillie2015}~\cite{Salem2019} & ASM & Leveraging lean carrier design, \\ & & introduce deeper sleep modes \\ \hline
  ~\cite{Tombaz2016} & Sleep mode EE evaluation& Introduce a power model for sleep modes \\ \hline
  ~\cite{Tano2019} & Study and improve sleep modes& Introduce timers to control sleep modes \\ \hline
  \end{tabular}
\end{table}

%% file: sections/ON_OFF_carrier.tex
\begin{figure}
    \centering
    \includegraphics[scale=0.5]{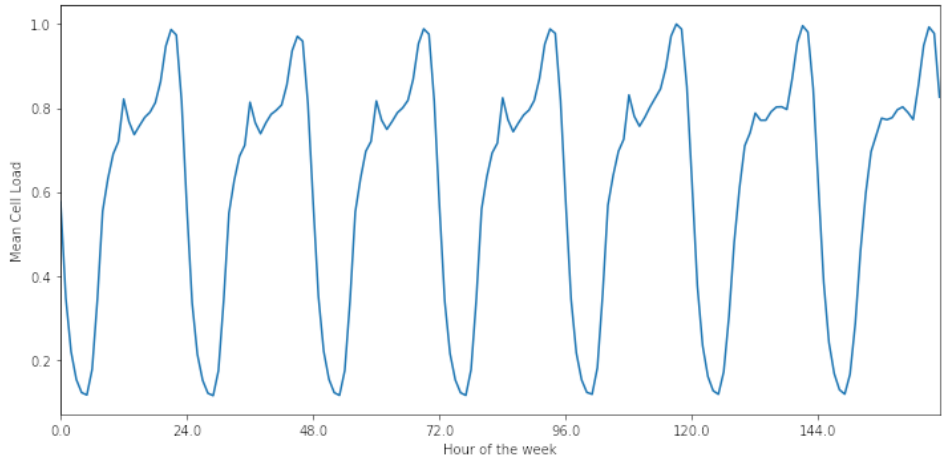}
    \caption{Normalized weekly load for a typical cell in a dense urban scenario.}
    \label{fig:cell_load_prof}
\end{figure}

In this section,
we concentrate on carrier-domain mechanisms,
also known as carrier shutdown.
These schemes refer to those solutions that take advantage of deep dormancy to deactivate  hardware components of the \ac{BS} for long periods of time,
e.g. time scales of minutes, even hours. 
In more detail,
we focus on the mechanisms used to harvest energy savings by considering the long-term traffic load variations of a cell.
To give an example,
in most of the cells, 
the traffic daily profile shows a regular trend, 
with low load periods early in the morning, medium loads during the work-hours and high data rate in the late evening 
(see Fig.~\ref{fig:cell_load_prof}). 
Weekends may be characterized by lower traffic demands with respect to workdays.
As a result,
since mobile networks are sized to satisfy peak time traffic, 
energy may be wasted during low and medium load periods,
and thus energy saving mechanisms able to adapt the network configuration to this long-term traffic load variations by fully (de)activating cells are necessary.
Note that these schemes allow deeper sleeps than the symbol shutdown ones. 

It is important to note that,
since these solutions operate at large time scales, 
they are not usually designed for providing cell service continuity, 
but for switching off carriers.
As a result, 
carrier-domain mechanisms do not need to be designed and integrated with the \ac{3GPP} \ac{NR} signalling framework provided by the lean carrier design,
and can be operated over the top.
However, \ac{3GPP} \ac{NR} specification work was still required to coordinate carrier (de)activation among \acp{BS}. 

According to the degree of coordination among \acp{BS},
and the nature of such \acp{BS},
three main approaches to carrier-domain mechanisms have been considered by the \ac{3GPP},
i.e. intra-\ac{BS}, inter-\ac{BS}, and inter-\ac{RAT} energy saving mechanism~\cite{3GPPTR36.927},
which are further discussed in the following.

\subsection{Intra-\ac{BS} energy saving mechanisms}

In the first case,
intra-\ac{BS} energy saving,
a \ac{BS} may activate energy saving mechanisms to locally adapt its capacity to traffic requirements with the main aim of reducing \ac{RF} amplifier power consumption. 
At this level, 
the tasks of each \ac{BS} is to independently control its number of active cells.

In this line, 
%
to improve the system energy efficiency, 
besides switching off one or more sectors of a \ac{BS},
which may be risky in terms of coverage, 
it is practically more important and feasible to dynamically control the number of active \acp{CC}, 
when a multi-carrier system or \ac{CA} is deployed.
As discussed in Section~\ref{sec:PC_RAN_CA},
\ac{CA} is a \ac{3GPP} flagship technology introduced in \ac{3GPP} \ac{LTE-A},
which allows a \ac{BS} to simultaneously operate on different bands. 
In \ac{3GPP} \ac{NR} Release~15, 
the  dormant state was introduced in \ac{CA}, 
such that (de)activation delay for \acp{SCell} could be reduced,
and the set of active \acp{CC} could be rapidly adapted to match \acp{UE} requirements~\cite{Li2020}. 
In addition, 
the concept of \ac{BWP} was also introduced in \ac{3GPP} \ac{NR} Release~15~\cite{Kim2020}. 
With this mechanism, 
when the cell load is low, 
the \ac{BS} can configure only a part of a given \ac{CC} for actual transmission/reception,
which is referred to as a \ac{BWP}\footnote{
For each \ac{CC}, 
at most four \acp{BWP} can be defined for the downlink and the uplink communications.}. 
Importantly,
control and data signalling only occur within this part of the spectrum, 
enabling thus a reduced power consumption at both the \ac{BS} and \ac{UE} sides,
as they need to handle/monitor smaller bandwidths. 
\ac{BWP} can be (de)activated by a timer, \ac{DCI}, and/or \ac{RRC} signalling, 
which can enable faster bandwidth adaptation with respect to the original \ac{CA} framework.
In this context, 
the work in~\cite{Yu2015} stands out,
which investigated the optimal transmit power and \ac{CA} configuration to optimise the \ac{WSEE} metric 
(see eq.~\eqref{eq:WSEE} in Section~\ref{sec:PC_Models}), 
while satisfying the downlink and the uplink \acp{UE} requirements. 
Important energy savings were reported. 

It should be noted, however, that these intra-\ac{BS} energy saving mechanisms may have unwanted side effects on the overall network performance,
as hinted earlier, 
as they only take a \ac{BS}-centric viewpoint. 
For example,
while reducing its energy consumption,
a \ac{BS} may significantly impact the network layout (i.e., coverage) as well as the load and interference distributions across the network~\cite{Feng2017}. 
The turning off of such \ac{BS} may leave some \ac{UE} without coverage or proper service, 
and increase the traffic loads of multiple neighbouring \acp{BS}.
If these \acp{BS} operate on the same spectrum band,
\ac{BS} (de)activation will also impact the inter-cell interference pattern, 
which will also have important consequences on the rank and the \ac{MCS} selection, 
and thus on packet success rate.
To provide a proper network-wide energy consumption optimisation, 
while ensuring end-users' \ac{QoS}, 
it is thus necessary to take these network effects into account, 
when adjusting a \ac{BS} configuration. 

Table \ref{tab:intraBS} presents the main contribution on intra-\ac{BS} energy saving mechanism discussed in this section.
\begin{table}[!ht]
\caption{Summary of main intra-BS energy saving mechanisms.}
\label{tab:intraBS}
\scriptsize
\centering
  \begin{tabular}{|c||c|c|} 
  \hline
  Paper & Focus &Main contribution/idea \\
  \hline
  ~\cite{Li2020} & CA dormant state & Enable fast bandwidth adaptation \\ & &  in CA framework  \\ \hline
    ~\cite{Kim2020} & BWP concept& When the cell load is low,\\ & & only a part of a given CC is used \\ \hline
  ~\cite{Yu2015} & CA optimization & Study optimal transmit power\\ & &  and CA configuration  \\ \hline
  \cite{Feng2017} & BS ON-OFF switching & Discuss challenges in \\ & & Intra-BS energy saving \\ \hline
  \end{tabular}
\end{table}

\subsection{Inter-\ac{BS} energy saving mechanisms}

Inter-\ac{BS} mechanisms are suited to address the above mentioned challenge,
operating over multiple neighbouring \acp{BS}, 
in a centralised or a distributed manner, 
to jointly optimise the use of radio resources and provide energy savings without significantly affecting the end-users' \ac{QoS}. 
This resource management problem, however, is typically challenging, 
as it involves multiple network functionalities and complex mathematical formulations.

Taking a global perspective, 
without too many practical constraints, 
the authors of~\cite{Feng2017TWC} investigated how to optimise the network energy efficiency by jointly managing long-term \ac{BS} activation, \ac{UE} association, and power control in a \ac{HetNet} with \ac{mMIMO} capabilities. 
They modelled this problem using mixed-integer programming, 
and to address complexity,
proposed a sub-optimal centralised scheme, 
where \emph{i)} the integer variables  (i.e., the cell (de)activation and the \ac{UE} association variables) are relaxed, 
and \emph{ii)} the \ac{BS} activation, \ac{UE} association, and power control problems are iteratively solved. 
Since their centralised solution was still characterized by a large complexity, 
they also proposed a distributed solution based on game theory, 
which provided lower performance in terms of energy savings and \ac{UE} performance, 
but it is proven to converge to the Nash equilibrium.

From a more practical view point,
the \ac{3GPP} \ac{NR} has defined in Release~15 and beyond functionalities to support the (de)activation of \ac{3GPP} \ac{NR} capacity booster cells,
which are overlaid on a larger \ac{3GPP} \ac{NR} coverage cell~\cite{3GPPTR37.816,3GPPTS38.300}.
This solution builds upon the possibility for the capacity booster cell to autonomously decide to shutdown.
This switch off decision is typically based on cell load information,
and can also be taken by the \ac{OAM}, 
if needed be.
The capacity booster cell may initiate handover actions in order to off-load the cell being switched off, 
and may indicate the reason for handover with an appropriate cause value to support the target node in taking subsequent actions.
Dual connectivity may also be used to support the offloading procedure. 
During the shutdown time period of capacity booster cell,
idle mode \acp{UE} are prevented from camping on this cell or handover to it.
All neighbouring cells are informed by the coverage cell owning such capacity booster cell about its switch-off actions over the X$_{n}$ interface. 
Moreover, the coverage cell owning such capacity booster cell can request an inter-\ac{BS} cell re-activation over the X$_{n}$ interface to wake up the capacity booster cell,
if the former cell overloads. 
The switch on decision may also be taken by the \ac{OAM}.
All neighbouring cells are informed by the coverage cell owning such capacity booster cell about the re-activation by an indication on the X$_{n}$ interface.

In addition to their impact on network performance, 
i.e. coverage, rate, latency,
energy inter-\ac{BS} saving mechanisms can have a detrimental effects on the \ac{BS} hardware life-time, e.g. due to ping-pong effects. 
Specifically, 
deep and frequent transients between different status lead to large temperature gradients in the involved hardware components, 
which increase their failure rates,  
thus augmenting the maintenance costs to fix or replace the \ac{BS}. 
The impact due to these long-term energy saving mechanisms can be measured by the acceleration factor, 
which is defined as the ratio between the failure rate observed by applying such mechanisms over time and the one experience when keeping the \ac{BS} always active.
In this line, 
the work in~\cite{Chiaraviglio2017} has modelled hardware failure rate due to cell (de)activation, 
and proposed a heuristic to control the statuses of the \acp{BS} of a network, 
which minimises the acceleration factor growth over time, 
while satisfying the end-users' \ac{QoS} demands. 
In contrast to baseline solutions, 
which maximize the energy savings at the cost of increasing the \ac{BS} failure rate over time, 
the proposed approach achieves around a 30$\%$ of power savings in a \ac{3GPP} \ac{LTE} scenario,
while keeping the acceleration factor close to one,
which can be translated in further energy savings as \acp{BS} are maintained or replaced less often.

To further avoid excessively frequent status changes of the \acp{BS} and associated network performance losses, 
the \ac{BS} control policy in charge of (de)activation decisions could consider the load distribution and the manner in which it varies in time and space. 
To this end, 
load can either be characterized statistically or using data-driven approaches.
Embracing this concept, 
the authors of~\cite{Celebi2019} have considered a dense \ac{HetNet}, 
where small cells and \acp{UE} are randomly deployed, 
following a \ac{HPPP} distribution, 
and packets arrive to the transmission buffers according to an exponential distribution. 
Using this model, 
they have characterized the probability density function of the cell load using a gamma distribution,
and used this information to elaborate multiple \ac{BS} (de)activation strategies,
comparing them in terms of complexity, blocking rate probability, throughput, and energy efficiency.

Following the same line of thinking, 
the authors of~\cite{Saxena2016} have proposed a stochastic game, 
where distinct \ac{BS} instances in a \ac{CRAN} platform take advantage of spatial correlation, 
and jointly estimate the network traffic, 
exploiting past observations. 
The \ac{CRAN} then uses these estimates to decide the status of the \acp{RRH} in the network, distribute the \acp{UE} among the cells, control the \acp{RRH} transmit power, and setup cooperative transmission schemes to guarantee that coverage requirements are satisfied. 
The authors also demonstrated, 
through a \ac{CRAN} experimental platform, 
that the proposed solution using traffic estimations leads to large energy savings with respect to a dynamic \ac{BS} switching solution~\cite{Oh2013}, 
which is not aware of the traffic evolution.

\begin{figure}[!t]
    \centering
    \includegraphics[scale=0.5]{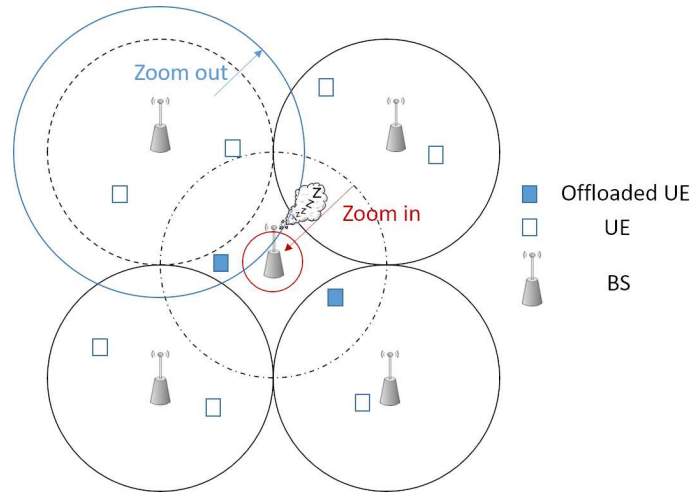}
    \caption{Cell zooming procedure \cite{Jiang2018}.}
    \label{fig:cell_zooming}
\end{figure}

As a different approach to inter-\ac{BS} energy saving,
cell zooming has also become a popular mechanism,
which is not based on carrier shutdown,
but consists in reducing the cell coverage of lightly loaded cells, 
while simultaneously increasing the area covered by neighbouring compensating ones~\cite{Niu2010} 
(see Fig.~\ref{fig:cell_zooming}). 
When using this mechanism,
topology changes should be smoothly implemented to limit service outage~\cite{Conte2011}. 
To face this challenge, 
the work in~\cite{Jiang2018} has proposed a data-driven approach to optimise the cell zooming mechanism. 
Specifically, 
this framework has represented the network through a graph, 
and used a \ac{BS} connectivity metric related to user-level data to construct the graph adjacency matrix. 
The authors have run a Markov process on the graph to identify network communities on which implementing the cell zooming. 
This process is realized through two steps. 
The first step consists on using a polynomial model to predict the expected community traffic load in the next hour,
and the second step operates the cell zooming to identify the \acp{BS} to deactivate and how to distribute the load among the active \acp{BS}. 
With respect to baseline solutions based on the knowledge of instantaneous traffic~\cite{Niu2010,Oh2013}, 
the authors have highlighted that the proposed solution leads to 20$\%$ energy saving gains at the cost of increasing the blocking rate only by 0.1$\%$. 
They have also shown that the achieved blocking rate is greatly affected by the large prediction noise, 
which results in inaccurate forecasts.

Table \ref{tab:interBS} summarizes the main contributions of the inter-\ac{BS} energy saving schemes discussed in this section.

\begin{table}[!ht]
\caption{Summary of main inter-BS energy saving mechanisms.}
\label{tab:interBS}
\scriptsize
\centering
  \begin{tabular}{|c||c|c|} 
  \hline
  Paper & Focus &Main contribution/idea \\
  \hline
  ~\cite{Feng2017TWC} & Network coordination in HetNets & Joint control of BS activation,\\ & & UE association, and power control \\ \hline
    ~\cite{3GPPTR37.816} & Network coordination in 5G NR& Defines message to support the\\ & &  (de)activation of capacity booster cells\\ \hline
  ~\cite{Chiaraviglio2017} & Impact of BS switching on HW failure &Propose a heuristic that limits\\ & & the HW failure rate with BS (de)activation\\ \hline
  ~\cite{Niu2010}\cite{Conte2011}\cite{Jiang2018} & Cell zooming & Control network layout to prevent \\ & & coverage losses due to energy saving schemes \\ \hline
  \end{tabular}
\end{table}


\subsection{Inter-\ac{RAT} energy saving mechanisms}

It should also be noted that,
in the early stage of \ac{5G} deployments, 
\ac{3GPP} \ac{NR} \acp{BS} are not uniformly distributed across a city area, 
and thus there is a need for a tight inter-working between the \ac{3GPP} \ac{NR} network and the underlying \ac{3GPP} \ac{LTE} one. 
As previously mentioned, 
\ac{3GPP} \ac{NR} \acp{BS} are characterized by a larger power consumption with respect to \ac{3GPP}  \ac{LTE} ones, 
as they integrate more complex hardware to operate on a wider bandwidth and use a larger number of antennas and transceiver modules (due to \ac{mMIMO}). 
Therefore, 
the wireless community is also currently developing inter-\ac{RAT} energy saving solutions to switch off \ac{3GPP} \ac{NR} booster capacity cells, 
when their traffic demand is low~\cite{3GPPTS28.310}. 

From a general and conceptual perspective, 
many of the inter-\ac{BS} energy saving optimization strategies presented earlier apply to this inter-\ac{RAT},
provided that such frameworks can capture the different characteristics of \ac{3GPP} \ac{NR} and \ac{LTE} \acp{BS}. 
Thus, we do not provide further details in this space. 

From a more practical view point,
however,
it is important to highlight that the \ac{3GPP} \ac{NR} Release~16 has done specification work in this field.  
Similarly as described in the inter-\ac{BS} energy saving case, 
the \ac{3GPP} \ac{NR} capacity booster cell can autonomously decide to switch off based on its own load in the inter-\ac{RAT} energy saving one.
The new enhancements, however, allow now to declare a \ac{3GPP} \ac{LTE} cell as the coverage cell of the \ac{3GPP} \ac{NR} capacity booster cell, 
and thus such \ac{3GPP} \ac{LTE} coverage cell can request an inter-system cell re-activation over the S1/NG interface based on its own cell load information or neighbour cell load information.
Further details can be found in~\cite{3GPPTR37.816}~\cite{R3-206885}.

%% file: sections/ON_OFF_channel.tex
While the \ac{mMIMO} frameworks presented in Section~\ref{sec:MMIMO} provided general important insights on the deployment and optimization of energy-efficient \ac{mMIMO} networks in both the uplink and the downlink,
it should be noted that,
once the network is deployed and running,
different approaches can be used to minimise energy consumption. 
For instance, 
when the traffic load is low, 
the energy consumption of a \ac{mMIMO} system may be reduced by using only a subset of the available \ac{BS} antennas and/or transceiver modules\footnote{These two terms, antennas and transceivers, although they represent different concepts, abusing notation, they are interchangeably used in this section.},
according to traffic requirements and avoiding resource waste.
This type of approaches are referred to as antenna selection or channel shutdown~\cite{Prasad2017,Asaad2018},
and are the focus of this section.

A number of frameworks have investigated channel shutdown subject to end-users' \ac{QoS} demands on short time scales on the basis of multi-path fast fading variations,
i.e., activating those \ac{BS} antennas with favorable channel conditions at each 
---or a small number of---
\acp{TTI}~\cite{Gao2013,Xu2013,Le2016,Makki2017,Lee2017,Arash2017}.
However, in wideband systems such as \ac{5G} with many subcarriers per carrier, 
it is unlikely that a \ac{BS} antenna is simultaneously not selected on all such subcarriers.
Moreover, 
antenna selection based on multi-path fast fading also requires all \ac{BS} antennas to be activated at least for channel estimation, 
thus limiting their sleeping time. 

Taking a more practical approach,
the authors in~\cite{Senel2019} investigated from a generic view point the antenna selection problem in the downlink considering larger time scales, basing the decision-making in large-scale fading parameters.
This work is targeted at finding both the optimal number of \ac{BS} antennas and their transmit powers to minimise the downlink power consumption of a \ac{mMIMO} network,
while meeting end-users' \ac{QoS} demands in the form of a minimum \ac{SINR} per \ac{UE}. 
Both single-cell and multi-cell scenarios were analysed,
where the system model in the latter accounted for multiple \acp{BS}, a number of antennas and simultaneously multiplexed \ac{UE} per \ac{BS}, and imperfect channel estimation through pilot contamination.
It should be noted, however, that the authors adopted a basic \ac{BS} power consumption model, 
which depends on the \ac{PA} efficiency, 
and is only linear with the number of \ac{BS} antennas.
Signal processing power consumption, 
for example, 
as a function of the number of simultaneously multiplexed \acp{UE} in each \ac{TTI} is not considered.
For the single-cell case,
the authors derived the optimal number of \ac{BS} antennas and their transmit powers in closed-form.
Importantly, 
these expressions prove that,
only when the circuit power consumption per \ac{BS} antenna is small,
the minimum \ac{BS} power consumption can be attained by activating all the \ac{BS} antennas. 
Otherwise, 
the \ac{BS} can save energy by deactivating a subset of them.
For the multi-cell case, 
and contrary to the single-cell one,
since pilot contamination was considered, 
a coherent interference term appeared in the \ac{SINR} formulation,
which scales with the number of \ac{BS} antennas in the pilot-sharing cells, 
thus limiting the achievable \acp{SINR} of \acp{UE}. 
As a consequence, 
the results indicated that increasing the number of \ac{BS} antennas still leads to lower transmit powers, 
but this is not necessarily the optimal to minimise the \ac{BS} power consumption, 
as there is a cost associated with using such \ac{BS} antennas. 
Unfortunately, 
the authors concluded that it is hard to obtain closed-form expressions for the optimal number of \ac{BS} antennas and their transmit powers for this more complex multi-cell case.
However, 
they showed that the joint optimisation problem can be relaxed as a geometric programming problem that can be solved efficiently,
and suggested that their algorithm can be used to optimally (de)activate \ac{BS} antennas to deliver the requested traffic with minimal power consumption.
However, a detailed system-level analysis is missing in this paper.  

The work in~\cite{Hossain2018} complements the above study investigating how to solve the antenna selection problem,
while considering daily load variations in multi-cell \ac{mMIMO} networks. 
To do so,
the authors computed the distribution of the active \acp{UE} in a cell for different network loads using queuing theory. 
They then modelled the distributed daily energy efficiency maximization problem,
using the active number of \ac{BS} antennas in each cell as a variable, 
and solved it through a game theory framework.
As a result,
a best response algorithm was proposed, 
where each \ac{BS} iteratively selects the strategy that produces its most favorable outcome given other \ac{BS} strategies. 
This \textit{selfish} approach did not achieved the global optimum,
but lead to a Nash equilibrium without the need for coordination across cells,
thus being adequate for an intra-\ac{BS} operation without any network coordination.
Noting that this work did not considered specific \ac{UE} rate requirements in the optimization,
results showed that,
at very low loads, 
the proposed adaptive antenna selection scheme could achieve significant energy efficiency gains as high as 250$\%$, 
when compared with a baseline system that does not adapt the number of antennas to the traffic profile,
at the expense of around a 50\% reduction in the average \ac{UE} data rate.

It should be considered, however, that even if most practical antenna selection schemes deployed in current \acp{BS} aim at turning off \ac{BS} antennas for a long period of time to achieve deeper sleeps and more power savings.
This may lead in some cases to a large activation delays,
and thus to a degraded \ac{UE} experience. 
For example, if parts of the transceiver chains are turned off semi-statically, 
the rate of cell-edge \acp{UE} could be significantly reduced,
and latency may be become unacceptable for some \ac{URLLC} services.

Considering both rate and latency requirements, 
the work in~\cite{Pramudito2020} has also recently investigated the power consumption minimisation problem in multi-cell \ac{mMIMO} networks at shorter time scales,
by implementing cell \ac{DTX} (see Section \ref{sec:TimeD}) in conjunction with precoding and antenna selection.
Note that differently than the references presented earlier in this section,
this work does not base its decision-making on the basis of multi-path fast fading variations. 
The authors, instead, have considered a multiple frame optimisation window, 
and proposed a strategy to select the right precoding technique for each transmission frame such that the total transmission time and latency are minimised.
Moreover, this work have proposed a technique to trade the \ac{UE} latency for additional energy savings, 
by reducing the number of active \ac{BS} antennas used in each frame. 
Numerical results have highlighted that the proposed adaptive antenna selection scheme provides large energy efficiency gains in lightly loaded scenarios without impacting the end-users' \ac{QoS}.
In more detail, 
for rate and latency requirements between 1 and 6\,Mbps per \ac{UE} and 0.001 and 5 duty cycles per frame, respectively,
the authors claim that the proposed technique can provide energy efficiency improvements between 125\,\% and 1124\,\% in the suburban scenario, 
and between 196\,\% and 952\,\% in the rural scenario,
without compromising \ac{QoS}.

To support this type of schemes,
and avoid the mentioned performance losses due to transients,
the latest proposals in \ac{3GPP} \ac{NR} Release~18 suggest the development of new specification enhancements to support a fine-grained (de)activation of \ac{BS} antennas, 
e.g., in the unit of an \ac{OFDM} symbol or a slot. 
In practice, 
considering the buffer sizes of active \acp{UE}, their data rate requirements and the expected transmission capacity, 
the number of active \ac{BS} antennas can be adjusted to match the end-users' \ac{QoS} demands and maximize the potential power savings without incurring large performance loss~\cite{RWS-210447}.

Table \ref{tab:antenna-domain} summarizes the main contribution presented in this section on antenna-domain energy saving schemes.

\begin{table}[!ht]
\caption{Summary of main antenna-domain energy saving mechanisms.}
\label{tab:antenna-domain}
\scriptsize
\centering
  \begin{tabular}{|c||c|c|} 
  \hline
  Paper & Focus &Main contribution/idea \\
  \hline
  ~\cite{Senel2019} & Antenna selection problem & Optimize the BS active antennas and transmit powers,\\ & & based on slow fading\\ \hline
    ~\cite{Hossain2018} & Antenna selection in multi-cell mMIMO& Use game theory to adjust the active antenna number\\ & &  to the network daily load variations\\ \hline
  ~\cite{Pramudito2020} & Cell DTX+precoding and antenna selection &Propose a multi-stage optimization scheme\\ &in multi-cell mMIMO & that trades-off user latency and energy efficiency\\ \hline
  \end{tabular}
\end{table}

%% file: sections/ML.tex
As previously discussed in Section~\ref{sec:energyEfficiencyMetrics:AI}, 
energy efficiency optimisation highly depends on the accuracy of the embraced models,
and unfortunately, 
many of the current models are rigid,
mostly the theoretical ones,
unable to adapt to specific channel characteristics, enabling technologies, or environment changes.
This may yield a considerable theory to practice gap.
Instead, 
data-driven optimisation may be able to close this gap,
learning the practical state of the network and inferring optimum network operation policies by means of \ac{AI} to enhance the energy efficiency of mobile networks~\cite{Cao2018}. 
In this line,
recently, 
several works in the literature are aiming at improving energy efficiency by exploiting state-of-the-art \ac{ML} algorithms. 

\begin{figure}
    \centering
    \includegraphics[scale=0.2]{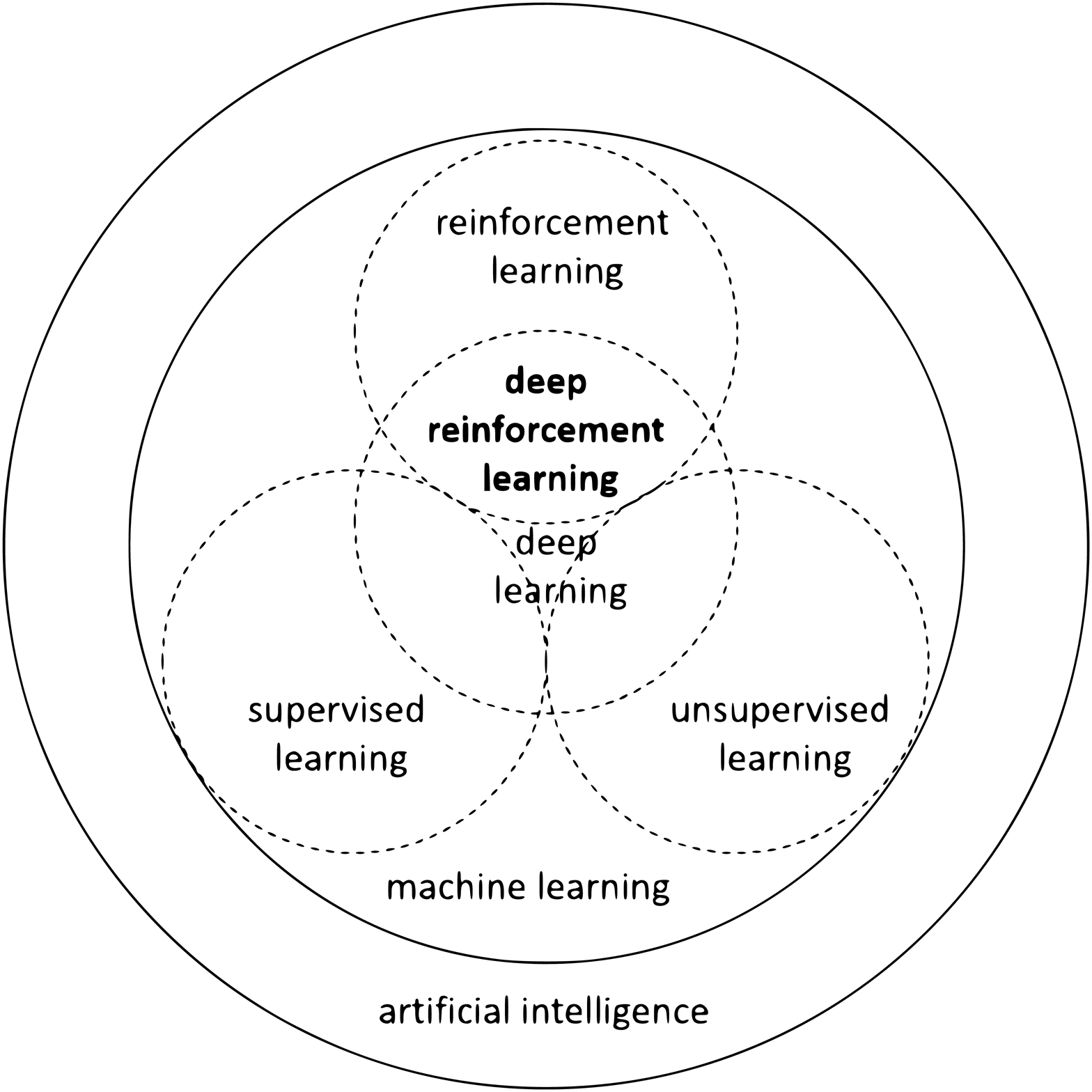}
    \caption{Relationships among deep reinforcement learning, deep learning, reinforcement learning, supervised learning, unsupervised learning, machine learning, and \ac{AI}.}
    \label{fig:DRL_overview}
\end{figure}

In Fig.~\ref{fig:DRL_overview}, 
we illustrate the relationships between the main concepts developed in the framework of \ac{AI} and \ac{ML}~\cite{Li2018DRL}. 
Specifically, 
today, 
within the \ac{AI} world, 
\ac{ML} comprises the family of algorithms which uses data to develop intelligent systems.
We can identify three main models: 
supervised learning, unsupervised learning, and \ac{RL}.

Supervised and unsupervised learning models have been investigated to characterise and forecast network traffic and predict \ac{5G} network behaviour,
leveraging rich data-sets. 
These models are, for example, becoming increasingly popular to define new solutions for \ac{PHY} layer functions~\cite{OShea2017}.
In contrast to supervised and unsupervised learning, 
the essence of \ac{RL} concerns learning to make online decisions through interactions with the environment to control network operations. Therefore, \ac{ML} models are key to enable an intelligent mobile network able to characterize its environment, predict system changes in time and space, and react accordingly in a real time manner. 

In the following, 
we review the literature related to the use of \ac{ML} techniques for traffic prediction and network optimization in green \ac{5G} networks.

\subsection{\ac{ML} for Traffic Prediction}
\label{sec:SL_LP}

\input{sections/ML_Pediction}

\subsection{\ac{ML} for 5G Energy Efficiency Optimisation}
\label{sec:RL_LP}

\input{sections/ML_RL}

%% file: sections/ML_Pediction.tex
One of the fundamental challenges along the path to enable full network adaptation to end-users' \ac{QoS} requirements is the accurate forecasting of the network traffic.
Such data forecasting can help driving energy efficient network decisions, 
e.g., carrier shutdown and others,
as it will be shown in the next section. 

The forecasting  of the network traffic presents, however, important challenges:
\begin{itemize}
\item 
End-users have different \ac{QoS} demands at different moments of the day and in different places. 
Therefore, 
traffic demands change in time and space, 
making the prediction task difficult.
\item 
The mobility of \acp{UE} introduces spatial dependencies among neighboring cells. 
Moreover, 
spatial dependencies can occur between distant cell towers, 
as efficient urban transportation systems easily enable \acp{UE} to travel across cities within half an hour.
\item 
The spatial distribution of \acp{UE} at the urban scale is further influenced by many factors, 
including land use, population, holidays, and various social activities. 
These further complicate the spatio-temporal dependencies among traffic in distinct cell towers.
\item 
The prediction time scale should match the decision periodicity of the energy saving mechanism.
For instance, 
when adjusting every hour the number of active carriers, 
the forecasting model should provide predictions of the cell load every hour. 
In contrast, 
if the mechanism works on a daily basis, 
a longer prediction, i.e., 24 hours, is required,
making the task more challenging.
\end{itemize}

The studies in the literature aiming at predicting the network traffic can be differentiated into two groups, 
according to the adopted methods, 
i.e., statistical-based and \ac{ML}-based approaches.

\subsubsection{Statistical-based methods}

Statistical-based methods rely on capturing the statistics of the network traffic. 
One of the most popular statistical approaches when predicting network traffic is \ac{ARIMA}~\cite{jagerman1997stochastic}, 
which originates from three models: 
the auto-regressive model, the moving average model, and their combination (ARMA). 
The predictions performed by this model are based on considering the lagged values of a given time-series, 
while accommodating for non-stationarity.  
The main limitation of \ac{ARIMA} is its inability to capture the seasonality 
---a time series with a repeating cycle---
of network traffic.
To overcome such limitation, 
an extension of this algorithm, 
named \ac{SARIMA}, 
has been proposed~\cite{hyndman2018forecasting}. 
\ac{SARIMA} adds three new hyper-parameters to specify the auto-regression, the moving average, and the differencing for the seasonal component of the series, 
as well as an additional parameter for the period of the seasonality.

Statistical methods like this, 
however, 
are not able to capture rapid traffic variations, 
since they rely on the mean value of the historical data. 
Moreover, they are mainly linear, 
and it has become clear that they cannot provide high accuracy when predicting network traffic, 
especially when considering complex network traffic behaviors observed in real scenarios~\cite{azari2019cellular}.

\subsubsection{\ac{ML}-based methods}

In contrast to statistical methods,
data-driven approaches based on \ac{ML} have been recently investigated as a solution for network traffic prediction, 
as they allow to model non-linearities, 
while taking advantage of the big amounts of data currently being collected by the \acp{BS}.

Traditional \ac{ML} algorithms such as \ac{KNN}~\cite{martinez2019methodology} and \ac{SVR}~\cite{wu2004travel} are able to model non-linear relationships.
However, they require well-tuned parameters to achieve accurate prediction results. 
Moreover, these methods are known to have short memory due to their limited parameter set and inefficient computing, 
which is detrimental for improving the prediction accuracy.

\paragraph{Recurrent Neural Networks}

\begin{figure}
    \centering
    \includegraphics[scale=0.8]{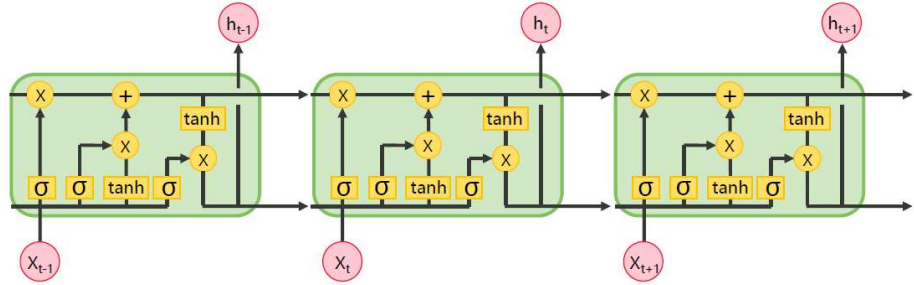}
    \caption{Example of \ac{RNN} composed of three stacked \ac{LSTM} units. The terms $x_t$ and $h_t$ are respectively the input and the output at time time $t$. Moreover, the sigmoid and hyperbolic tangent activation functions are represented by the $\sigma$ and $\tanh$ symbols, respectively.}
    \label{fig:LSTM}
\end{figure}

Research has then moved into \acp{RNN} to model more complicated nonlinear sequence patterns, 
which has provided promising results in many fields, 
such as speech recognition, image caption, and natural language processing.
In particular, 
\ac{LSTM} has been proposed as a solution to the problem of vanishing gradient in traditional \acp{RNN}~\cite{hochreiter1997long}. 
This neural network architecture allows to learn long-term dependencies from the time series provided in the input. 
A \ac{LSTM} unit is characterised by three gates, 
i.e., input gate, forget gate, and output gate. 
These gates control the unit operations by considering three inputs, 
i.e., the input vector, the memory of the previous time-step, and the output of the previous time-step. 
The non-linearity is modeled through a sigmoid unit and a hyperbolic tangent unit, 
which implement the respective functions. As an example, Figure~\ref{fig:LSTM} shows a neural network architecture composed of three stacked \ac{LSTM} units.

Based on such initial definition,
the authors in~\cite{zhang2020gated} improved the \ac{LSTM} state-of-the-art with an \ac{LSTM} architecture using an encoder-decoder model based on gated dilated causal convolution. 
In the encoder, 
the long-range memory capacity is enhanced by gated dilated convolutions without increasing the number of model parameters in order to learn a vector representation of the input sequence. 
Subsequently, 
different temporal-independent and temporal-dependent features, such as the daytime, holidays, weather, are fused with the representation vector.
This allows to provide to the model additional relevant information with respect to the network traffic time series. 
In the decoder, 
the model applies a \ac{RNN} with multiple \ac{LSTM} units to map the fused vector representation back to the variable-length target sequence. 

Attention mechanisms are also often used when adopting a \ac{LSTM} architectures to weight the importance of previous observations \cite{Vaswani2017}. 
However, experiments have highlighted that simply using lagged inputs 
(i.e., data points from one year ago, half a year ago, and a quarter before) 
allows reaching better prediction accuracy than using complex attention mechanisms due to the strong periodicity characterizing the network traffic time series. 

It should be noted, however, that the aforementioned \ac{LSTM} methods do not take into consideration the spatial dependencies between the traffic experienced by different \acp{BS},
although it has became apparent that capturing this information may be fundamental to provide accurate forecasting of network traffic~\cite{wang2019spatio}. 
Different extensions of the previous presented approaches have been proposed in this direction.

The authors in~\cite{hossain2020deep} have, for example, proposed a novel prediction model to forecast traffic congestion, such that the uplink to downlink resource ratio can be adjusted to improve networking efficiency. 
The proposed model is composed of a tree-based deep model, 
followed by an \ac{LSTM}. 
This tree-based model uses convolutional layers, 
which are useful to capture spatial information. 
Moreover, using a deep model allows to reduce the computational cost, 
because the convolution operations are performed in parallel in a tree-like structure.

In~\cite{wang2017spatiotemporal}, 
a hybrid deep learning model for spatio-temporal prediction is proposed instead to incorporate spatial correlation, 
in which the temporal dependence is captured by a \ac{LSTM}, 
whereas the spatial dependence is encapsulated by auto-encoders. 
Specifically, 
an auto-encoder is a neural network architecture used in unsupervised learning to represent a set of data,
while reducing its dimensionality~\cite{goodfellow2016deep}.
The auto-encoder learns to compress data from the input layer into a short code, 
i.e., the embedding, 
and then decompresses that code into a data structure that closely matches the original data 
i.e., the output  layer. 
With this architecture, 
the auto-encoders are used to model historical information from the neighboring \acp{BS}, 
and capture spatial dependencies.
The proposed model is shown to offer an average \ac{MAE} improvement of 31\% and 40\% with respect to \ac{SVR} and \ac{ARIMA}, respectively.

Even though the aforementioned attempts to capture spatial information,
\ac{LSTM} is not fundamentally adequate for it.  
In particular,
the gates that characterize this model are usually fully connected, 
and as a consequence, 
the number of parameters is large, 
requiring high memory and computation time for training the model. 
This model is thus highly complex and frequently turns overfitted.

\paragraph{Convolutional Neural Networks}

An evolution of \ac{LSTM}, 
named \ac{ConvLSTM} has been proposed to solve this problem, 
by replacing the inner dense connections with convolution operations~\cite{shi2015convolutional}. 
This architecture significantly reduces the number of parameters, 
and enhances the ability of capturing spatio-temporal information. 
Indeed, \acp{CNN} are widely adopted now to deal with image classification problems, 
and capture spatial information.
Similarly,
when considering the network traffic prediction case,
network traffic data is treated as images,
where the geographical space is modeled by a matrix, 
and the traffic distribution in different areas of the city is described by the elements of such matrix.

As a good example,
the authors in~\cite{zhang2018citywide} provide a traffic prediction architecture, 
in which spatio-temporal dependencies are  captured by utilizing densely connected \acp{CNN}. 
In a similar way, 
the authors in~\cite{zhang2018long} have  proposed a prediction algorithm, 
which can model both temporal and long-distance spatial dependencies. 
The proposed model follows an encoder-decoder paradigm, 
where a stack of \ac{ConvLSTM} and \ac{CNN} elements are combined. Numerical results show that such model can achieve up to 61\% lower \ac{NRMSE} as compared to \ac{ARIMA} and other statistical methods, while requiring up to 600 times shorter ground truth measurement durations.

A main limitation of this type of approaches, however, is that they only work with regular grid-based region partitions, 
which are not practical for cellular networks, 
and limits the prediction performance.

\paragraph{Graph Neural Networks}

To overcome this limitation, 
an architecture based on \ac{GNN} has recently been proposed to model the network traffic spatio-temporal dependencies using a graph representation.
In particular, 
given a direct graph, 
each \ac{BS} is modeled as a vertex, 
and each edge defines the spatial relations between adjacent \acp{BS}.

\begin{figure}
    \centering
    \includegraphics[scale=0.8]{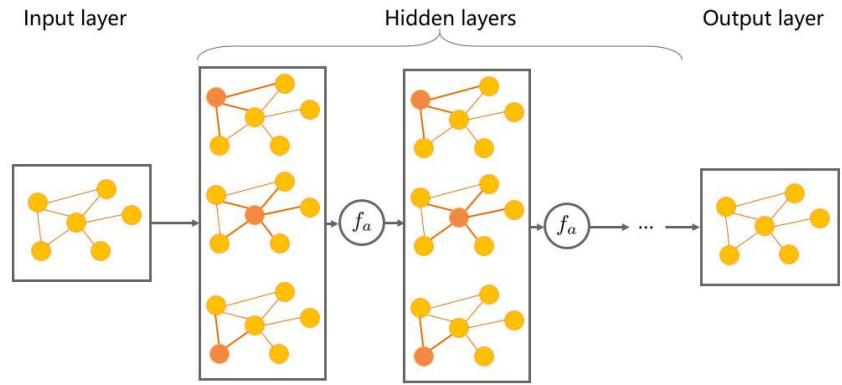}
    \caption{Example of a \ac{GNN} architecture. The activation function unit is indicated by the $f_a$ symbol.}
    \label{fig:GCN}
\end{figure}

The authors in~\cite{wang2019spatio} have adopted such \ac{GNN}-based architecture, 
and decomposed the total data traffic volume into in-tower traffic and inter-tower traffic, 
which corresponds, respectively, to the traffic serviced to the \acp{UE} residing within the coverage of a \ac{BS} and the traffic serviced to the \acp{UE} moving among areas covered by different \acp{BS}.
In the proposed architecture, 
each edge has a weight that depends on the total data traffic moving from the corresponding \acp{BS} vertexes. 
Importantly, 
it should be noted that complete directed graphs can contain a huge amount of edges, 
which hinder the efficient learning of model parameters. 
Therefore, low weights are treated as noise, 
and the corresponding edges are pruned by defining a threshold that allows to balance prediction accuracy and computing efficiency.
The presented numerical results have shown that this architecture achieves 16\% lower \ac{MAE} compared to a \ac{LSTM} model. Moreover, the performance analysis have shown that the traffic mobility induced by the roaming of humans plays a large role in the prediction accuracy, 
and showed that a combination of in-tower and inter-tower traffic patterns can be applied for network or social event forecasting.

\paragraph{External inputs: Point of Interests}

While the aforementioned research has mainly focused on the network traffic itself, 
it is also well understood that external factors, 
such as the \ac{POI} distribution, 
may influence the demand of network traffic.  
In particular, 
the analysis provided in~\cite{xu2017understanding} reveals that the dynamic urban network traffic usage exhibits five basic time domain patterns, 
which are correlated to the city functional zones, 
e.g., residential, transport, office.

In this line,
the work in~\cite{zhang2019deep} has targeted network traffic prediction by exploiting cross-domain data. Specifically,
three types of data sets are considered, 
i.e., \acp{BS} location, \acp{POI} distribution, and social activity level. 
In particular, 
the latter contains information generated by the end-users when using social networks,
such as location and keywords, 
which may allow to better capture particular social events, 
such as concerts and football matches.
The correlation between these data sets and the network traffic is analyzed and used to improve the prediction accuracy. 
A novel deep learning based traffic prediction architecture is then proposed. 
This architecture can fuse the cross-domain data sets into a unified representation. 
Spatial, temporal, and external factors are then captured and processed by \acp{ConvLSTM}.
In order to consider the pattern diversity and similarity of the network traffic of different city functional zones, 
the authors have also proposed an algorithm for grouping city areas into different clusters. 
Then, inter-cluster transfer learning is proposed to capture regional similarities and differences.
The achieved results have shown that cross-domain data sets have high correlation with the network traffic, 
and thus the introduction of the aforementioned data sets benefits the prediction accuracy. In details, the authors show that, in the considered scenario, cross-domain data sets allowed to improve the \ac{MAE} by up to 14\%

\paragraph{Model re-usability}

Another important issue related to traffic prediction is that of re-usability.
Prediction algorithms generally lack of re-usability, 
which require them to be re-trained to learn a new representation of the spatio-temporal information, 
when adopted in a new
---or dramatically changing---
scenario.
The generalization problem of prediction algorithms has been recently discussed in~\cite{wang2020bsenet}, 
where the authors have proposed a model based on auto-encoders, 
which learns the embedding of \acp{BS} based on raw data. 
In this framework, 
the embeddings are vectors,
which contain spatio-temporal information of the \acp{BS}, 
and their size is much smaller than the raw data, 
which allows to improve the generalisation capabilities of the prediction algorithm, 
while also reducing its computational cost.
The architecture is composed of three main modules: an encoder, a spatial adder and a decoder. 
In more detail, 
the encoder is designed to extract information from the \ac{BS}, and infer its embedding.
The spatial adder is in charge of building the relation among different \acp{BS}, 
whereas the  decoder restores original data from the embeddings. 
In this way, 
the training phase makes the encoder learn how to generate an embedding that conforms to the spatial relation with neighboring \acp{BS}. 
After training the model, 
the encoder is able to use the raw data from the \ac{BS} itself to infer its embedding, 
which contains information about how this \ac{BS} influence other \acp{BS}. 
Numerical results have shown that this approach helps temporal models to achieve similar performance as spatio-temporal ones, 
at the cost of a small increase in the training time.

%% file: sections/ML_RL.tex
The wireless network environment is complex and stochastic by nature as already discussed earlier,
and in more detail,
traffic requirements, user mobility, interference and channel variations in time and space make system-wide optimisation a hard problem. 
In the past,
most of solutions proposed in the literature to configure mobile network parameters have not considered the
dynamic nature of wireless networks. 
More specifically,  
state-of-the-art algorithms, 
as many of the once already surveyed,
are typically based on perfect 
---or partial--- 
knowledge of the instantaneous system conditions, 
which requires to re-compute the solution of a problem whenever a notable change has occurred in the environment. 
With the complexity carried by such approaches,
they may lead to significant computation and signaling overhead. 
Thus, 
there is an urgent need for more light-weight, flexible and adaptive solutions with respect to environment dynamics to minimise the energy consumption of practical networks.

\begin{figure}
    \centering
    \includegraphics[scale=0.6]{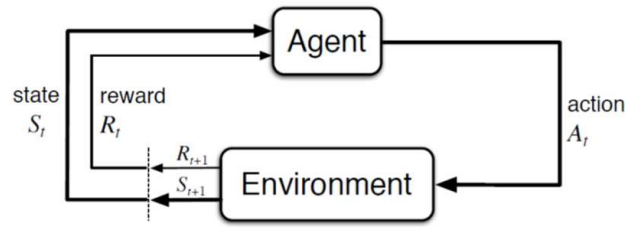}
    \caption{The perception-action-learning loop.}
    \label{fig:MDP_interactions}
\end{figure}

In the last decade, 
\ac{RL}, 
and more recently \ac{DRL}, 
have emerged as potential tools to pave the way for artificial intelligence driven optimisation in \ac{5G} systems and beyond.
For more details on the motivation,
refer to \cite{Klaine2017} and \cite{Luong2019} and the references therein. 
In \ac{RL}, the learning process is represented by the interaction between a learner and
decision maker, named agent, and the environment. The agent selects actions and the
environment responds to those actions by returning new states of the environment.
The environment also returns a numerical value named reward that the agent tries
to maximize over time. The agent interactions and learning process are represented for clarity in Fig.~\ref{fig:MDP_interactions}.

This problem is characterised by multiple challenges.
For instance, 
the environment is typically partially observable, 
i.e., an agent does not have full knowledge of the system state, 
but only a partial observation. 
In the context of wireless network optimisation, 
this is very likely, 
and it can represent the case where a \ac{BS} is not aware of the load of other \acp{BS}.
Moreover, 
the perception of the reward related to a state-action pair is often delayed, 
which makes hard to evaluate the effect of an action, 
and thus to learn the optimal policy. 
This effect is known as the temporal credit assignment problem~\cite{Sutton2018}.
Similarly, 
in a multi-agent system, 
a perceived reward depends on the actions of multiple agents behaving independently from each other, 
i.e., a cell throughput depends on how each neighbouring \ac{BS} schedules its own resources.

When the environment can be fully modelled, 
dynamic programming can be used to solve the learning problem through algorithms whose complexity is polynomial in the size of the set of states
~\cite{Sutton2018}.

The objective of \ac{RL} is to learn how to map experienced situation (i.e., the state
of the system) into action to take so as to maximise a numerical reward, through continuous interactions with the environment.
Through this learning by interaction loop, 
a exploration-exploitation trade-off arises,
i.e., exploiting the information collected so far to benefit the locally optimal decision, 
or exploring for achieving a better characterisation of the environment and achieve higher long-term gains.


Recently, 
in the context of energy efficiency,
new \ac{RL} schemes have been proposed to manage sleep modes in \ac{5G} \acp{BS}.
In \cite{Salem2019},
the authors have mapped the time-domain sleep mode control
---see Section~\ref{sec:TimeD}----
as a decision making problem in which a \ac{BS} sequentially sets the sleeping level length.
Specifically, 
when the cell becomes idle, 
this approach first puts the \ac{BS} in the deepest level of sleep, 
and then gradually switches it on. 
At each stage, 
the \ac{BS} decides the number of slots during which the current sleep mode status will be kept. 
If a \ac{UE} request arrives during a sleep period, 
the associated data is saved in a buffer until the \ac{BS} wakes up. 
Accordingly, 
the state set includes the possible states in which a \ac{BS} can operate, 
i.e., idle, active, or one of the sleep modes enabling energy saving. 
The action set includes the values of the possible number of time slots that can be associated to a given sleep mode. 
The reward is defined as the weighted sum of the energy saving gain due to the sleep mode and the additional latency experienced at the \acp{UE} due to the buffering of their traffic. 
In this way, 
different sleep mode policies can be defined according to whether the operator wants to trade end-users' \ac{QoS} for energy savings. 
The authors have used a popular \ac{RL} scheme,
named as Q-Learning, 
to find the optimal policy. 
This scheme is a model-free algorithm, which uses state transition experiences to iteratively construct an estimate of the optimal state-action function, also referred as Q-function. A learning parameter allows to control how the estimates are iteratively blended together over time. If each state-action pair is visited infinitely often, and the learning rate is decreased over time, the estimated Q-function converges to the optimum~\cite{Jaakkola1994}.
Note that the optimal state-action pairs are stored into a look-up-table.
The results in~\cite{Salem2019} have shown that, 
if delay is critical, 
the sleep mode should not be activated for a cell load larger than 30$\%$. 
In contrast, 
for very low loads, 
up to 55$\%$ of energy savings can be achieved,
even when prioritizing the end-users' \ac{QoS}.

A well known problem of \ac{RL} algorithms is the so-called curse of dimensionality, 
meaning that their computational requirements grow exponentially with the size of the state and action spaces~\cite{Sutton2018}. 
To deal with this challenge, 
function approximation can be used to approximate the state–action value function when the state and/or action spaces are large or continuous. 
Multiple function approximation methods have been investigated in the literature for \ac{RL},
e.g., linear functions or \acp{ANN}.

In \cite{DeDomenico2018}, 
the authors proposed a fuzzy Q-learning model to deal with complexity, 
where a network controller jointly optimises the \ac{DTX} of the underlying cells and backhaul nodes to minimise the energy consumption and satisfy the end-users' \ac{QoS}. 
Specifically, 
to reduce complexity, 
the controller maintains a distinct model for each cell. 
The state space of each cell characterises its buffer state in terms of rate and latency requirements, the \ac{BS} capacity, and the estimated spectral efficiency loss due to the interference of the nearby \acp{BS}, 
which are expected to be active. 
Then, in each time slot, 
the controller observes individually each state space, 
and decides in a distributed manner which cells (and associated backhaul nodes) to keep in energy saving mode or activate. 
The authors have associated to each state-action pair a cost function, 
which models the weighted sum of the \ac{BS} power consumption and the packet lost, 
either due to latency constraints or interference. 
As the state space is composed by continuous variables, 
this would prevent a classic \ac{RL} algorithm to converge to an optimal policy in a finite time. 
Accordingly, 
the authors have integrated a \ac{FIS} to their framework, 
and reduced the state space by mapping the state representation in fuzzy sets~\cite{ZADEH1975}. 
The authors have shown that the proposed framework is able to coordinate the activation and deactivation of neighbouring \acp{BS}, 
thus limiting the inter-cell interference. 
Moreover, 
this scheme takes advantage of the latency-energy trade-off, 
and achieves up to 38$\%$ of energy savings with respect to a baseline \ac{DTX},
which does not exploit data buffering.

To manage curse of dimensionality,
\acp{DNN} are currently widely used as a powerful global function approximator,
where a neural network is used to compress the Q-table~\cite{mnih2015humanlevel}.
However, the combination of \ac{DNN} and \ac{RL}, 
i.e., \ac{DRL} 
can lead to instability, 
and even divergence during the training process~\cite{Arulkumaran2017}. 
To address these issues,
recently, 
the authors of~\cite{mnih2015humanlevel} have proposed a \ac{DQL} framework that leverages two main ideas,
i.e.,
the usage of experience replay, 
and the introduction of the target network. 
Experience replay consists in the usage of a buffer, 
where tuples of experiences 
(i.e., interactions with the environment) 
are saved and continuously replayed to break the correlation across subsequent observations during training. 
Moreover, during training, 
two distinct deep networks are used.
One that is continuously updated, 
and another one, 
updated less frequently. 
These modifications make the algorithm training more stable.

These enhancements have led to continuous research innovations and \ac{DRL} architectures that can successfully deal with problems that were previously considered intractable. 
For instance, 
the authors of~\cite{Liu2018} have proposed a \ac{DQL} model to dynamically (de)activate \acp{BS}
based on traffic requests. 
This framework has introduced few enhancements with respect to the baseline \ac{DQL} in~\cite{mnih2015humanlevel}. 
First, 
they have observed that non-stationary traffic leads to oscillation between waking- and sleeping-dominating regimes. 
To break this correlated sequence of actions, 
the authors propose an \textit{action-wise} experience replay,
where experiences related to different actions are saved into distinct buffers, 
which are uniformly sampled during the training process. 

In the literature, 
other mechanisms have been proposed to improve the effectiveness of the experience replay process, 
e.g., the well known prioritized experience replay~\cite{Schaul2016}. 
Moreover,
and although the reward is clipped to $[-1; 1]$ in classic \ac{DQL}, 
to capture the strong variations characterising the wireless environment, 
this work has also proposed an adapting reward re-scaling scheme, 
which consists into dividing the instantaneous reward by a positive adapting scaling factor, 
and summing a saturation penalty to the \ac{DQL} loss function. 
In addition, 
the authors of this work have also used an interrupted Poisson process to model the traffic requests, 
and generate additional pseudo experiences, 
which, using the DynaQ framework~\cite{Sutton2018}, 
are periodically stored into the replay memory along with real experiences, 
and used indiscriminately for training. 
Empowered by these innovations, 
their \ac{DQL} algorithm attempts to learn the optimal policy that control the \ac{BS} status, 
based on a reward that takes into account the served requests, the queued or re-transmitted requests, and the failed ones.
The reward considers the cost to wake up the \ac{BS} and the one for changing the \ac{BS} status.
Their experiments have shown that modelling the traffic requests and generating pseudo experiences does not lead to large gains. In contrast, 
action-wise experience replay and adaptive reward scaling improve the stability and adaptability of the proposed framework. 
Overall, the proposed scheme achieves large gains with respect to a baseline Q-learning approach, 
in terms of energy saving and end-users' \ac{QoS}.

\begin{figure}
    \centering
    \includegraphics[scale=0.6]{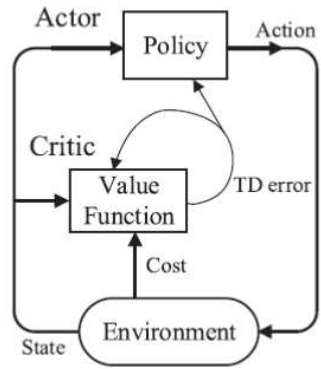}
    \caption{The actor-critic framework.}
    \label{fig:AC}
\end{figure}

Similarly, the authors of~\cite{Ye2020} have proposed a \ac{DRL} model to control the small cell (de)activation in dense \acp{HetNet}. 
In this framework,
the system state comprises the  status of each small cell and the estimate of its traffic arrival rate,
while the action set includes the (de)activation actions.
The cost function provides a qualitative description of the small cell network power consumption, the additional latency experienced due to deactivating \acp{BS}, and the switching cost due to the change of status of the small cells 
(i.e., from off to on and vice-versa). 
This work has improved the state-of-the-art solutions by considering an actor-critic \ac{DRL} scheme. 
In the actor-critic algorithm, 
the actor selects the action given the state of the environment, 
and the critic estimates the value function, 
given the state and the action.
Then, it delivers a feedback to the actor 
(see Fig.~\ref{fig:AC}). 
Importantly,
it should be noted that this type of \ac{DRL} based on actor-critic has emerged as a powerful solution to deal with continuous action spaces~\cite{lillicrap2019continuous}. 
Conventionally, 
the actor provides a probability distribution of the possible actions at a given state. 
In~\cite{Ye2020}, 
the action space has the size, $2^{N_{SC}}$, 
where ${N_{SC}}$ is the number of small cells in the network. 
Therefore, 
the output of the actor is defined as a single vector of continuous values. 
To compensate for the lack of exploration in the actor's side, 
this work proposed to add noise to the output action vector.
The noisy vector is then converted in a hard decision 
(i.e., the proto-action),
and then,
the algorithm explores the set of actions close to the proto-action, 
and selects the action with the minimum estimated cost.
For training the proposed model, the authors have used
a deep deterministic policy gradient framework,
in which the policy and value function are both approximated by \acp{DNN}.
The authors have shown for that this approach limits the cumulative network cost over time with respect to baseline \ac{RL} algorithms, achieving up to 30$\%$ of gain with respect to a Q-learning model,
and provides larger stability in case of non-stationary traffic. 
Moreover, they have indicated that the proposed action exploration method reduces the convergence time.

\bigskip

To conclude this section,
let us highlight that,
although \ac{DRL} has allowed great progresses in the context of system optimization in a stochastic environment, many challenges are still open, such as enabling distributed optimization in the context of multiple competitive or collaborative agents, or designing fast and low complex methods to update the learned policy after notable change in the system, which have not been observed during the training phase.
More challenges faced by \ac{RL} with respect to the energy efficiency are described in the next section. 

Table~\ref{tab:MLsummary} summarizes the main contributions presented in this section on \ac{ML} and data driven energy efficient optimization.

\begin{table}[!ht]
\caption{Summary of main \ac{ML} and data driven energy efficient optimization literature.}
\label{tab:MLsummary}
\scriptsize
\centering
  \begin{tabular}{|c||c|c|} 
  \hline
  Paper & Focus &Main contribution/idea \\
  \hline
\cite{zhang2020gated} & Cellular traffic forecasting & Encoder-decoder model based on gated dilated causal convolution \\ \hline
\cite{hossain2020deep} & Traffic congestion prediction & Tree-based model with convolutional layers to capture the spatial information \\ \hline
\cite{wang2017spatiotemporal} & Cellular traffic forecasting &  Hybrid model consisting of \ac{LSTM} unit and auto-encoder\\ & & to capture temporal and spatial information, respectively \\ \hline
\cite{zhang2018citywide} & Cellular traffic forecasting	& Densely connected \acp{CNN} used to capture the spatio-temporal characteristics of the traffic \\ \hline
\cite{zhang2018long} & Cellular traffic forecasting & Combination of ConvLSTM and \ac{CNN} to model both temporal and \\ & & long-distance spatial dependencies \\ \hline
\cite{zhang2020gated} & Cellular traffic forecasting & Decomposition of the total data traffic volume into in-tower and inter-tower traffic \\ & & to explicitly account for the spatial dependency \\ \hline
\cite{zhang2019deep} & Cellular traffic forecasting & Cross-domain data used to enhance the accuracy of traffic forecasting \\ \hline
\cite{wang2020bsenet} & Cellular traffic forecasting & Model re-usability enhanced by considering a model based on auto-encoders \\ \hline
\cite{Salem2019} & BS sleeping policy optimization & Q-learning method adopted to obtain the optimal sleeping time duration for the \acp{BS} \\ \hline
\cite{DeDomenico2018} & BS sleeping policy optimization & Fuzzy Q-learning method adopted to deal with the complexity of the \ac{DTX} optimization problem \\ \hline
\cite{Liu2018} & BS sleeping policy optimization & Action-wise experience replay proposed as a method \\ & & for breaking the correlation between actions \\ \hline
\cite{Ye2020} & BS sleeping policy optimization & Actor-critic \ac{DRL} method adopted to control the small cell (de)activation in dense HetNets \\ \hline
  \end{tabular}
\end{table}

%% file: sections/OpenResearchDirections.tex
In this section, 
we identify lines of research,
which according to the authors' understanding, 
still require further efforts to aid increasing the energy efficiency of \ac{5G} and future \acp{RAN}.

\subsection{Multi-cell Energy Efficiency Theoretical Modelling}

As discussed throughout the survey,
serving the end-users' \ac{QoS} demands with the minimum power consumption is key to energy savings,
and while the bounds and trade-offs to drive such optimisation in a single-cell case may be well understood,
the fundamental understanding of energy efficiency in multi-cell \acp{RAN} is still limited,
due to complexity issues. 

Large-scale multi-cell \acp{RAN} are intricate to model (see Section~\ref{sec:mMIMO_multi_cell}),
and as a result,
for tractability reasons,
most of the current theoretical understanding on energy efficiency for wide-area networks have been derived based on, 
for example, 
the performance of the average \ac{UE} in uniform networks with simplistic channel, operational and \ac{BS} power consumption models~\cite{Bjornson2016}~\cite{Xin2016}.
These models do not generally capture, however, all relevant features, 
such as \ac{BS} and \ac{UE} distributions, \ac{NLoS} and \ac{LoS} transmissions, directional channels, and antenna correlations.
Novel theoretical analyses embracing such complexities are thus required to characterise still unknown energy efficiency trade-offs, 
which may exist, 
and allow further technology breakthroughs in real deployments. 

In this line, 
the work in~\cite{Renzo2018,Cui2018,Wang2020} have represented a step forward, 
accounting for non-uniform \ac{BS} and \ac{UE} distributions in \acp{RAN},
while being able to estimate local performance,
i.e., not only the performance of the average \ac{UE}, 
but also its distribution. 
On the same note, 
the work in~\cite{Ghatak2018} has  characterised the cell load distribution for a given traffic density, 
and studied how this affects the network performance. 
These frameworks, however, are still in their infancy, 
and have been mostly applied, 
up to now, 
to the analysis of simpler single-antenna small cell networks.
Further research is needed for their application to more complex \acp{RAN} and advanced features,
such as \ac{mMIMO}. 

With regard to channel models,
to give another example,
it is widely accepted that most \ac{mMIMO} performance bounds used today work well when the useful signal coefficients behave almost deterministically, 
i.e., they have a non-zero mean and a small variance.
However, the research in~\cite{caire2018} has recently proven that under highly directional channels,
for instance, 
the channel hardening effect does not so clearly appear,
and that new bounds are thus required,
in terms of both capacity and energy efficiency, 
for a more accurate performance evaluation in these more realistic \acp{RAN} setups.  

It is also important to mention that there is a gap in the literature with respect to sophisticated energy efficiency performance analyses, 
via detailed numerical and/or system-level simulation tools, 
able to capture the complexity of large-scale multi-cell \acp{RAN}.
Gaining understanding of the interplay among complex features such as \ac{mMIMO}, \ac{CA}, and coordinated transmissions, 
together with their power consumption,
which is hard to derive through a pure theoretical analysis,
can provide new road maps to fundamental \acp{RAN} deployment and operation.
The intuition gained via these tools can also lead to new theoretical research avenues. 

\subsection{Energy Efficiency-driven Network Planning Tools}

Deploying and operating a large-scale multi-cell \acp{RAN} is expensive,
and thus requires careful network dimensioning and planning to ensure an optimum radio resource utilisation,
e.g., spectrum and bandwidth, number of \acp{BS}, their location, architecture and transmit power, number of antennas, and transceivers per \ac{BS},
to cite a few~\cite{Nawrocki2006,Laiho2006}.
Importantly,
\acp{RAN} planning tools must ensure that the deployed system has a sufficient amount of radio resources, 
and can use it in an effective manner,
to achieve the required level of network performance at the appropriate cost.
Unfortunately, however, 
such tools are mostly network capacity-driven as of today,
and not yet designed to derive optimal energy efficient deployments.  

Once the \acp{RAN} is planned and deployed,
such implementation also imposes hard constraints on the future network performance and its energy consumption.
It is thus of imperative importance to equip \acp{MNO} with sophisticated \acp{RAN} planning tools for wide-area network design with energy efficiency at heart.
For example,
should an \ac{MNO} deploy less \acp{BS} and \acp{CC} with larger \ac{mMIMO} arrays, 
or in contrast, 
use  more \ac{BS} and \acp{CC} with smaller \ac{mMIMO} arrays? 
The applicability of optimisation algorithms in the network planning phase to find such type of practical answers is crucial,
and should rely on accurate topological descriptions of the deployment scenario, knowledge of current site deployments and performance, as well as \ac{UE} and required traffic distributions and accurate \ac{BS} power consumption models,
among others~\cite{Mishra2018}.

It is also important to highlight that optimisation performance strictly depends on a reasonable trade-off between modelling accuracy and complexity,
and thus \acp{MNO} should chose their network models carefully on a per problem basis.
This reinforces the need for flexible and efficient numerical radio propagation and system-level simulation tools as well as tailored optimization theories and algorithms,
which have energy efficiency as key driver~\cite{Penttinen2019}.

More developments in this area are needed at a professional level to make sure future \acp{RAN} are optimally dimensioned, 
and energy waste is avoided. 

\subsection{Multi-carrier and Heterogeneous Network Analysis}

In most scenarios,
new \ac{3GPP} \ac{NR} \acp{RAN} deployments will coexist with existing ones,
e.g., \ac{3GPP} \ac{LTE}.
In some cases, 
these deployments may be orthogonal in frequency with, e.g., \ac{3GPP} \ac{NR} in the 3.5$\,$GHz band and \ac{3GPP} \ac{LTE} in the 2$\,$GHz one.
In some other cases, 
due to the scarcity of spectrum,
\ac{3GPP} \ac{NR} deployments will have to take place in the same spectrum already used by \ac{3GPP} \ac{LTE}~\cite{Book_Dahlman2018,Book_Holma2020}.
In the latter case, 
the fundamental tool to enable such \ac{3GPP} \ac{NR}/\ac{LTE} spectrum coexistence is the dynamic time scheduling of both \ac{3GPP} \ac{NR} and \ac{LTE},
for which the \ac{3GPP} \ac{NR} specification provides tools~\cite{Wan2019}. 

Given that \ac{3GPP} \ac{NR} and \ac{LTE} sites have very different characteristics (coverage, bandwidth, antennas, etc.),
leading to distinct performance and energy consumption,
it would also be desirable to inter-work and (de)activate these two technologies in a coordinated manner,
while satisfying end-users' \ac{QoS} demands with the minimum energy consumption.
In some cases, 
\ac{3GPP} \ac{LTE} may operate at lower carrier frequencies than \ac{3GPP} \ac{NR}, 
and thus be able to provide a better blanket coverage at a smaller energy consumption.
This may be the most energy efficient at low load periods.
In contrast,
at medium loads,
\ac{3GPP} \ac{NR} may be sufficient to provide the required capacity to the active \acp{UE},
and \ac{3GPP} \ac{LTE} can be deactivated.
If operated at different frequencies, 
at high loads, 
both technologies may be aggregated via dual connectivity~\cite{Yilmaz2019}. 

In scenarios with the mentioned frequency imbalance, 
and because the \acp{BS} can avail of larger transmit power than the \acp{UE},
it may also make sense to simultaneously activate both technologies, 
and use downlink/uplink split~\cite{Lopez2016},
i.e.,
\ac{3GPP} \ac{NR} for downlink transmissions and \ac{3GPP} \ac{LTE} for the uplink~\cite{Book_Holma2020}.
However, 
this should only be done where and when necessary,
under energy efficient conditions,
avoiding potential energy waste due to having both technologies activated. 

The optimisation of the inter-working between \ac{3GPP} \ac{NR} and \ac{3GPP} \ac{LTE} is also of high importance when \ac{3GPP} \ac{NR} appears in the form of small cells~\cite{Lopez2015} or millimetre wave~\cite{Giordani2019} access points. 
These types of cells have a much smaller coverage radius,
and can be (de)activated to provide boosted capacity where and when needed. 
Since some implementations of this type of cell, e.g., millimetre wave, may be power hungry,
the usage of coordinated \ac{5G} sleep modes across a large number of this type of cells is critical. 
This can be facilitated through the separation of the data plane and control plane \cite{Mohamed2016}, 
where the latter is continuously provided by underlaying macrocells to ensure robust connectivity and mobility support, 
while capacity cells (i.e., small cells) allow for enhanced capacity and high rate data transmissions, 
locally and on demand.

In general, 
there is a lack of studies covering the understanding and optimisation of these technology inter-working practical use cases from a energy efficiency point of view.

\subsection{Data-Driven Optimisation}


State-of-the-art data-driven approaches for network traffic prediction are mainly related to measurements biased by the observation point, 
i.e., the \acp{BS}, 
which usually does not report the effective traffic demand but rather the cell throughput or resource usage, 
which depend on the network deployment, interference, current \ac{RRM} parameters, and running energy saving schemes.
The use of this potentially biased measurements may be an issue when adopting prediction algorithms as enablers for improving the mobile network energy efficiency. 
In fact, the implementation of any energy saving scheme will impact the load distribution across the network, 
and make related forecasting unreliable. 
As a result, 
further research is encouraged with regard to the estimation and prediction of unbiased metrics, 
whose characterization is not affected by the algorithms implemented using these observations. 

Moreover, 
there is currently a lack of understanding on how prediction errors, 
e.g. traffic forecasting errors, 
affect the gains provided by energy saving schemes.
In this line,
most of the current literature focuses on measuring the performance of the prediction in the metric space,
(function of the difference between the predicted value and ground-truth,
e.g., \ac{PRB} usage, throughput).
However, 
it is generally not straightforward to derive how improvements in the prediction of such metrics help to minimise the \acp{RAN} power consumption. 
It is thus recommended that prediction accuracy is investigated also in terms of energy savings when developing data-driven optimisation schemes~\cite{Vallero2019}.

As discussed in Section \ref{sec:ML}, 
there is also a lack of end-to-end \ac{ML} frameworks that jointly use supervised learning and \ac{RL} to characterise and optimise the \ac{5G} system.
Specifically, 
supervised learning models can provide multi-step traffic predictions, 
achieving a comprehensive forecast of future status of the mobile network environment. 
Using this information can help \ac{RL} algorithms to converge faster to optimal operational policies,
and enhance the performance of the exploration phase, 
e.g., optimally deciding the moment to (de)activate \ac{BS} functionalities without affecting network performance.

Importantly, 
it should be stressed that recent progresses in the areas of computational processing and data storage, 
as well as the increased availability of big data, 
have made the use of \ac{AI} more practical than ever in many challenging fields. 
However, the acquisition of large data sets in \acp{RAN} is currently challenging, 
and their processing energy demanding,
which limits the opportunities for implement data-driven optimization in \ac{5G} \acp{RAN}. 
To address this issue, 
the use of joint data-driven and model-based approaches is being widely explored~\cite{shlezinger2020modelbased,Zappone2019}, 
and is becoming the foundation of new optimization mechanisms for large-scale multi-cell \acp{RAN}.
Moreover, these techniques can be used for bootstrapping \ac{ML} models, 
thus reducing their need for data, computational complexity, power consumption, and latency.

\subsection{Green \ac{AI}}

Most of the recent breakthroughs led by novel \ac{ML} solutions have been possible thanks to the ever-increasing computational capacity of dedicated hardware platforms. 
The work in~\cite{Schwartz2020} has highlighted that, 
in the last decade, 
while \ac{ML} models evolved from AlexNet
---an image recognition \ac{DNN} presented in 2012~\cite{Krizhevsky2012}--- 
to AlphaZero
---a \ac{RL} algorithm proposed in 2018~\cite{silver2018general}---, 
the associated computational cost trend increased by 300,000x.
In this line, 
the work in~\cite{Strubell2019} has also analyzed the energy consumption issues arising from the need of exponentially larger computational resources to continue marginally increasing the model accuracy,
and has estimated that the carbon footprint of the current brute force trend is environmentally unfriendly.

The training of \acp{DNN} on mobile devices in a distributed, computationally and energy efficient manner is also an ongoing research topic,
which can brake down the aforementioned complexity. 
Collaborative learning schemes, 
such as federated learning \cite{MAL-083}, 
should be considered to mitigate the energy inefficiencies resulting from traditional, centralised \ac{ML} approaches.
Moreover, notable efforts are being performed towards hardware design and software accelerators,
which make possible to also move part of the ML process to the \ac{UE} itself. 
The more computing is performed on the mobile device, the less data needs to be sent to the cloud and/or network, enabling a reduction of the energy consumption due to the minimization of data exchange (see \cite{deng2019deep, lee2019device} and references therein).
In addition, 
tailored \emph{early stopping} can also be used to terminate the training process when a near-optimal solution has been found thus reducing the required number of iterations 
---and the associated energy consumption--- 
needed to train the \ac{ML} model~\cite{Prechelt2012}.

However, to pave the way for the successful integration of \ac{ML} techniques to drive the modelling and optimization of \ac{5G} \acp{RAN} and beyond, 
it is key to consider, 
not only accuracy key performance indicators, 
but also computational and energy consumption aspects during the design, training and exploitation phases of such data-driven optimisation algorithms. 
To achieve this goal, 
while designing an \ac{ML} model,
in addition to accuracy and optimisation-related metrics, 
computational efficiency and energy consumption must also be considered.
Using these metrics, 
\ac{ML} architectures that converge faster and/or need to be updated less frequently can be designed to optimise \acp{RAN},
prioritising energy consumption over model accuracy. 
One possible approach towards this goal is to investigate yet undiscovered \ac{ML} paradigms. 
For instance, 
as discussed in~\cite{shlezinger2020modelbased,Zappone2019},
model-based \ac{RL} solutions, 
which exploit a-priori expert knowledge to characterise physically/mathematically the system evolution, 
can be integrated with model-free \ac{RL} architectures, 
which interact with the environment to identify the optimal policy. 
Similarly, 
as explained in Section~\ref{sec:RL_LP}, 
\ac{RL} can leverage information about the future status of the system obtained by supervised learning forecasting to speed up the training speed and converge towards improved operating conditions.